\documentclass[iop]{emulateapj}
\usepackage{graphicx}
\usepackage{epstopdf}
\usepackage{epsfig}
\usepackage{changebar}
\usepackage{natbib}

\bibpunct{(}{)}{;}{a}{}{,}

\slugcomment{Accepted to the ApJ}

\shorttitle{DOG \& SMG SFHs}
\shortauthors{Bussmann et al.}

\usepackage{amsmath}
\usepackage{amssymb}

\begin{document}

\title{The Star-formation Histories of $z \sim 2$ DOGs and SMGs}

\author{R. S. Bussmann\altaffilmark{1,2}, Arjun Dey\altaffilmark{3}, L. Armus
\altaffilmark{4}, M. J. I. Brown\altaffilmark{5}, V.  Desai\altaffilmark{4}, A.
H. Gonzalez\altaffilmark{6}, B.  T.  Jannuzi\altaffilmark{3}, J.
Melbourne\altaffilmark{7}, B. T.  Soifer\altaffilmark{4,7}}

\altaffiltext{1}{Submillimeter Array Fellow, Harvard-Smithsonian Center for
Astrophysics, 160 Concord Ave., Cambridge, MA 02138; rbussmann@cfa.harvard.edu}
\altaffiltext{2}{Steward Observatory, Department of Astronomy, University of
Arizona, 933 N. Cherry Ave., Tucson, AZ 85721}
\altaffiltext{3}{National Optical Astronomy Observatory, 950 N. Cherry Ave., Tucson, AZ 85719}
\altaffiltext{4}{Spitzer Science Center, California Institute of Technology, MS
220-6, Pasadena, CA 91125}
\altaffiltext{5}{School of Physics, Monash University, Clayton, Victoria 3800,
Australia}
\altaffiltext{6}{Department of Astronomy, University of Florida, Gainesville,
FL 32611}
\altaffiltext{7}{Division of Physics, Math and Astronomy, California Institute 
of Technology, Pasadena, CA 91125}



\begin{abstract}

The {\it Spitzer Space Telescope} has identified a population of ultra-luminous
infrared galaxies (ULIRGs) at $z\sim2$ that may play an important role in the
evolution of massive galaxies.  We measure the stellar masses ($M_*$) of two
populations of {\it Spitzer}-selected ULIRGs that have extremely red $R-[24]$
colors (dust-obscured galaxies, or DOGs) and compare our results with
sub-millimeter selected galaxies (SMGs).  One set of 39 DOGs have a local
maximum in their mid-infrared (mid-IR) spectral energy distribution (SED) at
rest-frame 1.6$\mu$m associated with stellar emission (``bump DOGs''), while
the other set of 51 DOGs have power-law mid-IR SEDs that are typical of
obscured AGN (``power-law DOGs'').  We measure $M_*$ by applying
Charlot~\&~Bruzual stellar population synthesis models to broad-band photometry
in the rest-frame ultra-violet, optical, and near-infrared of each of these
populations.   Assuming a simple stellar population and a Chabrier initial mass
function (IMF), we find that power-law DOGs and bump DOGs are on average a
factor of 2 and 1.5 larger than SMGs, respectively (median and inter-quartile
$M_*$ values for SMGs, bump DOGs and power-law DOGs are
log$(M_*/M_\sun)=10.42_{-0.36}^{+0.42},~10.62_{-0.32}^{+0.36},$~and~$10.71_{-0.34}^{+0.40}$,
respectively).  More realistic star-formation histories drawn from two
competing theories for the nature of ULIRGs at $z\sim2$ (major merger vs.
smooth accretion) can increase these mass estimates by up to 0.5~dex.  A
comparison of our stellar masses with the instantaneous star-formation rate
(SFR) in these $z\sim2$ ULIRGs provides a preliminary indication supporting
high SFRs for a given $M_*$, a situation that arises more naturally in major
mergers than in smooth accretion powered systems.

\end{abstract}


\section{Introduction} \label{sec:intro4} 

Ultra-luminous infrared galaxies (ULIRGs) are defined to have extremely high
infrared (IR) luminosities ($L_{\rm IR} > 10^{12}~L_{\sun}$).  These
luminosities require significant dust heating, usually thought to arise from
extreme episodes of star-formation ($\dot{M} > 100 ~M_{\sun}~$yr$^{-1}$) or
accretion onto super-massive black holes.  These objects are rare in the local
Universe, yet they have been associated with a critical phase of galaxy
evolution linking mergers \citep[e.g.,][]{1987AJ.....94..831A,
1996AJ....111.1025M} with quasars and red, dead elliptical galaxies
\citep{1988ApJ...325...74S, 1988ApJ...328L..35S}.  ULIRGs are more commonplace
in the distant Universe, to the extent that they contribute a significant
component of the bolometric luminosity density of the Universe at $z > 1$
\citep[e.g.][]{2001AandA...378....1F, 2005ApJ...632..169L, 2005ApJ...630...82P}.
This realization implies that ULIRGs may represent an important evolutionary
phase in the assembly history of massive galaxies and has inspired a host of
new techniques for identifying ULIRGs at $z > 1$.

The two most successful techniques for identifying high-redshift ULIRGs rely on
selection at either mid-infrared or far-infrared wavelengths.  Surveys at
24$\mu$m with the Multiband Imaging Photometer for Spitzer
\citep[MIPS;][]{2004ApJS..154...25R} instrument for the {\it Spitzer Space
Telescope} have been remarkably successful for the mid-IR identification of
ULIRGs  \citep{2004ApJS..154...60Y, 2005ApJ...622L.105H, 2006ApJ...651..101W,
2008ApJ...672...94F, 2008ApJ...677..943D, 2009ApJ...693..447F}.  In particular,
\citet{2008ApJ...677..943D} select sources from the 9~deg$^{2}$ NOAO Deep
Wide-Field Survey (NDWFS) Bo\"otes field that satisfy $R - [24] > 14$ (Vega
magnitudes; $\approx$$\, F_{\rm 24\mu m}/F_R > 1000$) and $F_{\rm 24\mu m} >
0.3 ~ $mJy.  These objects are called dust-obscured galaxies (DOGs), lie at $z
\approx 2 \pm 0.5$ \citep[][Soifer et al., in prep., 2011]{2005ApJ...622L.105H,
2006ApJ...653..101W, 2009ApJ...700.1190D}, have ULIRG luminosities
\citep[e.g.][]{2009ApJ...705..184B}, have a space density of $(2.82 \pm 0.05)
\times 10^{-5}~h_{70}^3~$Mpc$^{-3}$ \citep{2008ApJ...677..943D}, and inhabit
dark matter haloes of mass $M_{\rm DM} \sim 10^{12.3}~M_{\sun}$
\citep{2008ApJ...687L..65B}.  These results show that DOGs are undergoing a very
luminous, likely short-lived phase of activity associated with the growth of
the most massive galaxies.  

In addition, DOGs can be divided into two groups according to the nature of
their mid-IR spectral energy distribution (SED): those with a peak or bump at
rest-frame 1.6$\mu$m, likely produced by the photospheres of old stars
(``bump DOGs''), and those dominated by a power-law in the mid-IR (``power-law
DOGs'').  The SED shapes, as well as spectroscopy in the near-IR
\citep{2007ApJ...663..204B, 2008ApJ...683..659S} and mid-IR
\citep{2007ApJ...658..778Y, 2007ApJ...664..713S, 2008ApJ...677..957F,
2009ApJ...700.1190D, 2009ApJ...700..183H} indicate that the bolometric
luminosities of bump DOGs are dominated by star-formation, while those of
power-law DOGs are dominated by obscured active galactic nuclei (AGN).  This
implies that the phase of DOG activity is characterized by both vigorous
stellar bulge and nuclear black hole growth.  

Another method of selecting high redshift ULIRGs is imaging at sub-millimeter
(sub-mm) wavelengths.  The advent of the Sub-mm Common User Bolometer Array
\citep[SCUBA;][]{1999MNRAS.303..659H} has allowed wide-field surveys at
850$\mu$m which have identified hundreds of sub-millimeter selected galaxies
(SMGs).  These objects have similar redshifts ($z =2.2\pm0.5$), number
densities \citep[$n \sim 9 \times
10^{-6}~h_{70}^3~$Mpc$^{-3}$;][]{2005ApJ...622..772C}, and clustering
properties \citep[$M_{\rm DM} \sim 10^{12.2}~M_{\sun}$;][]{2004ApJ...611..725B}
as DOGs.  

The fact that SMGs and DOGs have similar properties suggests they might be
related in an evolutionary sequence similar to that of ULIRGs in the local
Universe \citep[e.g.][]{1988ApJ...325...74S}.  It has been hypothesized that
such a sequence does indeed exist \citep{2009ASPC..408..411D}, and that DOGs
function as an important intermediate stage between gas-rich major mergers and
quasars at $z \sim 2$ \citep[which have similar clustering properties as DOGs
and SMGs;][]{2006ApJS..163....1H,2008ApJ...687L..65B, 2009ApJ...697.1656S}.
One intriguing potential piece of support for this idea comes from measurements
of H$\alpha$ line strengths, which indicate that power-law DOGs have lower
star-formation rates (SFRs) by an order of magnitude compared to SMGs
\citep{2011AJ....141..141M}.

A theoretical understanding of how this evolutionary sequence might occur has
recently been advanced using $N$-body/smoothed particle hydrodynamic
simulations combined with 3D polychromatic dust radiative transfer models
\citep{2010MNRAS.407.1701N}.  In these models, simulations are used to follow
the evolution of the SED of both isolated disk galaxies and major mergers.
These authors find that simulated systems with $F_{\rm 24\mu m} > 0.3~$mJy are
associated with gas-rich ($f_{\rm g} \approx 0.4$) major mergers with a minimum
total baryonic mass of $M_{\rm b} \approx 3 \times 10^{11}~M_{\sun}$.  While
there is significant variation associated with different viewing angles,
initial orbital configurations, etc., the typical simulated major merger
achieves peak SFRs of $\sim 1000~M_{\sun}~$yr$^{-1}$ at the beginning of final
coalescence when tidal torques funnel large quantities of gas into the nucleus
of the system \citep{1996ApJ...464..641M}.  This period is also when the system
is brightest at sub-mm wavelengths and thus can be selected as an SMG.  

\begin{figure*}[!tbp] 
\includegraphics[width=1.0\textwidth]{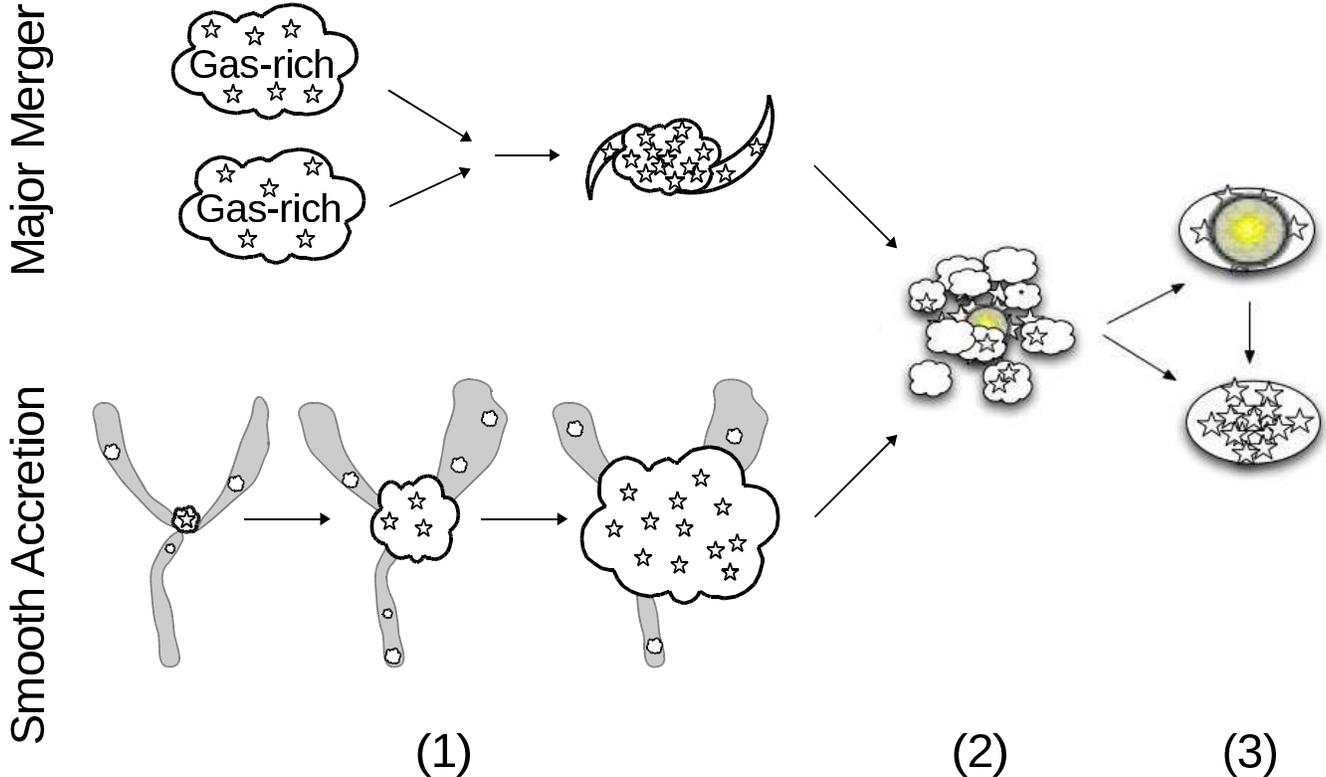}

\caption[Cartoon picture illustrating the two star-formation histories examined
in this paper]{Cartoon picture illustrating two possible evolutionary paths for
massive galaxies at $z \sim 2$ \citep[adapted from][]{2009ASPC..408..411D}.
{\it Top}: (1) A gas rich major merger leads to an intense, dust-enshrouded
phase of star-formation.  (2) Energetic feedback, possibly from the growth of a
central super-massive black hole, heats the dust and gas, cutting off
star-formation.  (3) Depending on the relative timescales of AGN fuelling, dust
dissipation, and star formation, the system may be briefly visible as a quasar
before settling on the red sequence.  {\it Bottom}: An alternative scenario in
which massive galaxies are assembled via smooth accretion of gas and small
satellites along filamentary structures (some mechanism is still needed to
quench star-formation; in this cartoon picture, steps (2) and (3) are assumed
to be the same as in the major merger driven scenario).  One goal of this paper
is to test the two different possibilities illustrated in step (1) of this
diagram using the stellar masses and star-formation rates of high redshift
ULIRGs.  \label{fig:cartoon}}

\end{figure*}

At the same time, central inflows begin to fuel the growth of a supermassive
black hole.  Approximately 100~Myr after the peak SFR, the black hole accretion
rate peaks (at about 1-2~$M_{\sun}$~yr$^{-1}$).  The simulations include a
prescription for active galactic nucleus (AGN) feedback that helps terminate
star-formation (along with consumption of the gas by star-formation).  In these
models, this period of AGN feedback coincides with the DOG phase ($F_{\rm 24\mu
m}/F_R > 1000$).  As the gas and dust are consumed by star-formation, optical
sightlines open up and the system can be optically visible as a quasar.  The
evolutionary progression in the simulations is driven by major mergers and
proceeds from SMG to DOG to quasar to red, dead, elliptical galaxy (illustrated
qualitatively in the top panel of Figure~\ref{fig:cartoon}).

Alternative theories for the formation of SMGs which do not involve major
mergers have also been advanced recently \citep{2010MNRAS.404.1355D}.  These
studies rely on numerical simulations of cosmological volumes and select SMGs
as the most actively star-forming systems that match the observed number
densities of SMGs.  The objects in the simulations that are designated as SMGs
have stellar masses in the range $M_* = (1 - 5) \times 10^{11} ~ M_{\sun}$ and
SFRs in the range 200-500~$M_{\sun}$~yr$^{-1}$.  These SFRs are a factor of 3
lower than what is inferred observationally in SMGs, which
\citet{2010MNRAS.404.1355D} attribute primarily to systematic effects in the
SFR calibration (in particular, a ``bottom-light'' initial mass function
requires lower SFRs to produce the observed IR luminosities of SMGs).  Because
the star-formation histories (SFHs) which produce these simulated SMGs do not
involve major mergers, they are referred to here as ``smooth accretion'' SFHs
(a qualitative illustration of this SFH is given in the bottom panel of
Figure~\ref{fig:cartoon}). 

Studies attempting to connect the mid-IR and far-IR selected ULIRG population
at high redshift have so far focused on their basic properties such as
bolometric luminosities \citep{2008ApJ...683..659S, 2008MNRAS.384.1597C,
2009ApJ...692..422L, 2009ApJ...705..184B, 2009AandA...508..117F}, clustering
strengths \citep{2004ApJ...611..725B, 2008ApJ...687L..65B}, and morphologies.
In particular, high-spatial resolution imaging \citep{2008ApJ...680..232D,
2008AJ....136.1110M, 2009AJ....137.4854M, 2009ApJ...693..750B,
2010MNRAS.405..234S} and dynamics \citep{2006ApJ...640..228T,
2008ApJ...680..246T, 2011AJ....141..141M} have shown no distinction in axial
ratio that might be suggestive of orientation effects; instead, these studies
have identified morphological trends which are consistent with an evolutionary
scenario driven by major mergers in which sources that show a bump in their
mid-IR SED (i.e., bump DOGs and most SMGs) evolve into those with a power-law
dominated mid-IR SED  \citep[i.e., power-law DOGs][]{2011ApJ...733...21B}.  To
test the origins of these sources further, it is imperative to use alternative,
complementary methods of constraining the SFHs of DOGs and SMGs at $z \sim 2$.

This paper is focused on one such technique: stellar population synthesis (SPS)
modeling of broad band photometry of DOGs and SMGs with known spectroscopic
redshifts.  The primary goal of this study is to place the tightest constraints
possible given the existing data on the stellar masses ($M_*$) and SFHs of bump
DOGs, power-law DOGs, and SMGs using a uniform SPS modeling analysis with
common model assumptions and fitting techniques for each ULIRG population.
There are several reasons to pursue this goal.  

First, constraints on the $M_*$ values and SFHs of {\it Spitzer}-selected
ULIRGs are limited to a few studies that have focused on bump sources
\citep{2007AandA...476..151B, 2009ApJ...692..422L}.  In contrast, the constraints
on $M_*$ and SFHs presented here for power-law DOGs are the first such results
for this potentially very important population of galaxies.  If power-law DOGs
do not have significantly different masses than SMGs or bump DOGs, this might
imply that the power-law phase occurs during the same time that most of the
mass in stars is being built up.  If the power-law is a signature of black hole
growth, then this would mean that the stellar mass and black hole mass are
likely being assembled during the same period of dust-obscured, intense
star-formation.  
A uniform analysis of all three populations is necessary to test this
hypothesis.

Second, while SPS modeling methods have become more sophisticated, stellar mass
results for a given population have not necessarily converged.  For example,
\citet{2005ApJ...635..853B} use {\it Spitzer}/Infrared Array Camera
\citep[IRAC;][]{2004ApJS..154...10F} data to infer average SMG stellar masses
of $M_* \approx 2.5 \times 10^{11}~M_{\sun}$.  More recently,
\citet{2008MNRAS.386.1107D} and \citet{2010AandA...514A..67M} have found median
stellar masses for SMGs of $M_* = 6.3 \times 10^{11} ~M_{\sun}$ and $3.5 \times
10^{11} ~ M_{\sun}$, respectively.  A new study by \citet{2011ApJ...740...96H}
using essentially the same data set as \citet{2010AandA...514A..67M} finds
significantly lower median SMG stellar masses of $M_* = (7 \pm 3) \times 10^{10} ~
M_{\sun}$.  Finally, measurements of the width of CO emission lines in 12
ULIRGs at $z \sim 2$ have provided a median dynamical mass estimate of $M_{\rm
dyn} \sim 2 \times 10^{11} \: M_\sun$ \citep{2010ApJ...724..233E}.  These
sources typically have high gas fractions of $\approx 0.5$, implying that the
stellar masses should be $M_* \leq 10^{11} \: M_\sun$.  This emphasizes the
significant systematics that affect stellar mass estimates based on SPS
modeling and underscores the need for a uniform analysis when comparing
different ULIRG populations.

Third, the disagreement in observed stellar masses has significant bearing on
theoretical models for the formation of high redshift ULIRGs.  As outlined
earlier, the cosmological hydrodynamical simulations of
\citet{2010MNRAS.404.1355D} predict that SMGs have large stellar masses that
are roughly consistent with the estimates of \citet{2005ApJ...635..853B} and
\citet{2010AandA...514A..67M}, but a factor of $\approx 4$ larger than the
estimates of \citet{2011ApJ...740...96H}.  The \citet{2011ApJ...740...96H} mass
estimates are also somewhat lower than what is expectated from merger
simulations \citep{2010MNRAS.407.1701N}, with the caveat that such expectations
are highly dependent on the stage of the merger, viewing angle, etc.  A
systematic, uniform comparison of the relative stellar mass distributions of
DOGs and SMGs with simulated SFHs from theoretical models for the evolution of
massive galaxies represents a significant component of this paper.

In section~\ref{sec:data4}, we present the data used in this analysis,
including DOG SEDs from rest-frame ultra-violet (UV) to near-IR.
Section~\ref{sec:models} outlines the general methodology and describes the SPS
libraries, initial mass functions (IMFs), and SFHs that are used in the
analysis.  We present our results in section~\ref{sec:results4}, including
constraints on stellar masses, visual extinctions, and stellar population ages.
In section~\ref{sec:disc4}, we compare our results with similar studies of SMGs
and other {\it Spitzer}-selected ULIRGs and explain the implications of the
results for models of galaxy evolution.  Conclusions are presented in
section~\ref{sec:conc4}.

Throughout this paper we assume a cosmology in which
$H_0=$70~km~s$^{-1}$~Mpc$^{-1}$, $\Omega_{\rm m} = 0.3$, and $\Omega_\Lambda =
0.7$.  All magnitudes are in the AB system.

\section{Data}\label{sec:data4}

The goal of this paper is to study the relative mass distributions of samples
of high-$z$ ULIRGs, specifically DOGs and SMGs, via population synthesis
modeling of their rest-frame UV through near-IR SEDs.  To minimize degeneracies
in the models, it is important to limit the analysis to sources with
spectroscopic redshifts.  Thus, the present sample consists of ULIRGs with
spectroscopic redshifts at $z > 1.4$ and broad-band photometry from the
rest-frame UV through near-IR.  The sample comprises three main sub-groups: two
selected with {\it Spitzer} at 24$\mu$m (DOGs), and one selected with the
Sub-mm Common User Bolometer Array (SCUBA) at 850$\mu$m (SMGs).

\subsection{DOGs}\label{sec:dogdata}

\subsubsection{Sample Selection}\label{sec:dogsample}

For the {\it Spitzer}-selected ULIRGs, a total of 2603 DOGs satisfying $R -
[24] > 14$ (Vega mag) and $F_{\rm 24\mu m} > 0.3 \:$mJy were identified in the
8.6 deg$^2$ NDWFS Bo\"otes field with deep {\it Spitzer}/MIPS 24$\mu$m coverage
\citep{2008ApJ...677..943D}.  This paper focuses on the subset of 90 of these
objects that have known spectroscopic redshifts at $z > 1.4$ either from
observations with the Keck telescope ($\approx 60\%$, Soifer et al., in prep.,
2011) or with the InfraRed Spectrometer \citep[IRS][]{2004ApJS..154...18H}
onboard {\it Spitzer} \citep{2005ApJ...622L.105H,2006ApJ...651..101W}.
Spectroscopic redshifts for our sample of DOGs are given in
Table~\ref{tab:positions}.

\begin{figure}[!tp] 
\begin{center}
\includegraphics[width=0.47\textwidth]{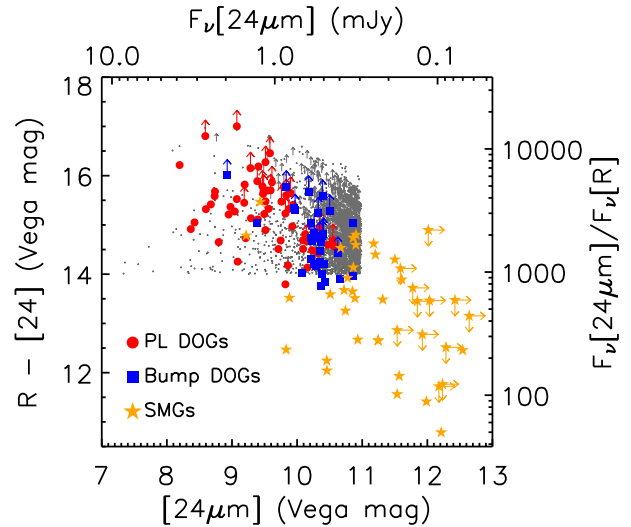}
\end{center}

\caption[$R - 24$ color vs. 24$\mu$m magnitude distribution for DOGs in the
NDWFS Bo\"{o}tes field]{$R - [24]$ color vs. 24$\mu$m magnitude distribution
for DOGs in the NDWFS Bo\"{o}tes field.  Gray dots and upward arrows show the
full sample of DOGs, with and without an $R$-band detection (2$\sigma$ limits),
respectively.  Highlighted are the subsamples with spectroscopic redshifts and
either a mid-IR power-law SED (PL DOGs, red circles) or a mid-IR bump SED (Bump
DOGs, blue squares).  Also shown are SMGs (orange stars) with spectroscopic
redshifts from \citet{2005ApJ...622..772C} and 24$\mu$m photometry from
\citet{2009ApJ...699.1610H}.  \label{fig:sample4}}

\end{figure}

\begin{figure}[!bp] 
\begin{center}
\includegraphics[width=0.47\textwidth]{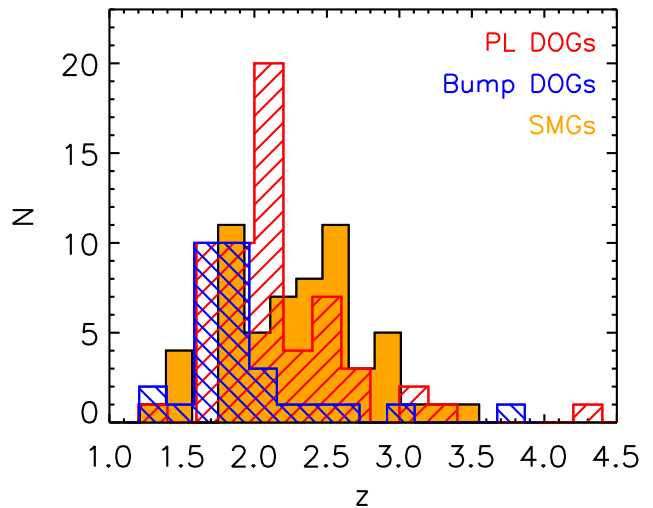}
\end{center}

\caption[Spectroscopic redshift distribution of DOGs in the Bo\"{o}tes
Field]{Redshift distribution of DOGs in the Bo\"{o}tes Field with spectroscopic
redshifts.  The redshift distribution of bump DOGs (blue hatched) is relatively
narrow due to selection effects \citep[for details see][]{2009ApJ...700.1190D},
while power-law DOGs (red hatched) are weighted towards slightly larger
redshifts.  Also shown is the redshift distribution of SMGs (orange filled
region) from \citet{2005ApJ...622..772C}.  \label{fig:zdist4}}

\end{figure}

Figure~\ref{fig:sample4} shows the $R - [24]$ color as a function of 24$\mu$m
magnitude for the subsample studied here (the ``spectroscopic sample'') in
comparison to the overall sample of DOGs in Bo\"otes.  To optimize the
spectroscopic detection rate, the spectroscopic sample is biased towards bright
24$\mu$m sources, although the full range of $R - [24]$ colors is sampled.  The
spectroscopic sample consists of 39 star-formation dominated ``bump'' sources
(those that show a peak at rest-frame 1.6$\mu$m) and 51 active galactic nucleus
(AGN) dominated ``power-law'' sources.  Bump and power-law DOGs are separated
according to the statistical criteria given in section~3.1.2 of
\citet{2008ApJ...677..943D}.  Also shown in this diagram are 53 sub-millimeter
galaxies (SMGs) with spectroscopic redshifts from \citet{2005ApJ...622..772C}
(see section~\ref{sec:smgdata4}).  The redshift distributions of these groups
of galaxies are shown in Figure~\ref{fig:zdist4}.  The positions, $R-[24]$
colors, and nature of mid-IR SED for each DOG in the sample are given in
Table~\ref{tab:positions}.

\subsubsection{Optical Photometry}\label{sec:ndwfs}

The NOAO Deep Wide Field Survey \citep[NDWFS;][]{1999ASPC..191..111J} is a
ground-based optical and near-IR imaging survey of two 9.3~deg$^2$ fields, one
in Bo\"otes and one in Cetus.  In this paper, we utilize the optical imaging of
the Bo\"otes field, conducted using the NOAO 4m telescope on Kitt Peak.  The
survey reaches 5$\sigma$ point-source depths in $B_W$, $R$, and $I$ of 27.1,
26.1, and 25.4 (Vega mag), respectively.  The NDWFS astrometry is tied to the
reference frame defined by stars from the United States Naval Observatory A-2
catalog.  NDWFS data products are publicly available via the NOAO science
archive \footnote{http://archive.noao.edu/nsa}.

Photometry for each DOG was measured in 4$\arcsec$ diameter apertures, centered
on the 3.6$\mu$m centroid position measured from the {\it Spitzer} Deep
Wide-field Survey \citep[SDWFS;][]{2009ApJ...701..428A} imaging data
\citep{2009ApJ...701..428A}.  Foreground and background objects were removed
using SExtractor segmentation maps, and the sky level was determined using an
annulus with an inner diameter of 6$\arcsec$ and a width of 5$\arcsec$.  The
background level and photometric uncertainty were computed by measuring the
sigma-clipped mean and RMS of fluxes measured in roughly fifty 4$\arcsec$
diameter apertures within 1$\arcmin$ of the target.  Aperture corrections were
derived using bright, non-saturated stars for each of the 27 sub-fields that
comprise the NDWFS.

\subsubsection{Near-Infrared Photometry}\label{sec:newfirm}

The NOAO Extremely Wide Field InfraRed iMager (NEWFIRM) has conducted a survey
at near-IR wavelengths of the full 9.3~deg$^2$ Bo\"otes field using the NOAO 4m
telescope on Kitt Peak during the spring semesters of 2008 and 2009.  The
nominal 5$\sigma$ limits of the survey within a 3$\arcsec$ diameter aperture in
$J$, $H$, and $Ks$ are 22.05, 21.3, and 19.8 (Vega mag), respectively.  All of
the survey data are publicly available (Gonzalez et al., in prep.).

Photometry was computed in the same manner as with the NDWFS images (see
section~\ref{sec:ndwfs}).  Aperture corrections were computed using bright,
non-saturated stars for each of the 52 sub-fields that comprise the NEWFIRM
survey of Bo\"otes.  Photometry in the optical and near-IR is presented in
Table~\ref{tab:phot1}.

\subsubsection{Mid-Infrared Photometry}\label{sec:sdwfs}

The SDWFS is a four-epoch survey of roughly 8.5~deg$^2$ of the Bo\"otes field
of the NDWFS.  The first epoch of the survey took place in 2004 January as part
of the IRAC Shallow Survey \citep{2004ApJS..154...48E}.  Subsequent visits to
the field as part of the SDWFS program reimaged the same area three times to
the same depth each time.  The final co-added images have 5$\sigma$ depths
(aperture-corrected from a 4$\arcsec$ diameter aperture) of 19.77, 18.83,
16.50, and 15.85 (Vega mag) at 3.6$\mu$m, 4.5$\mu$m, 5.8$\mu$m, and 8.0$\mu$m,
respectively.  All SDWFS data are publicly available.

Part of the SDWFS Data Release 1.1 includes band-matched catalogs created with
Source Extractor \citep[SExtractor,][]{1996AandAS..117..393B}.  Astrometry in
these catalogs is tied to 2MASS positions within 0$\farcs$2.  We identify DOGs
in these catalogs using a 3$\arcsec$ search radius, and use the values in these
catalogs for our flux density measurements of DOGs.  SExtractor underestimates
the true magnitude uncertainties because it assumes a Gaussian noise
distribution where noise is uncorrelated.  In place of the SExtractor-derived
values, we determine our own estimates of the uncertainty on each flux density
measurement using 4$\arcsec$ diameter apertures randomly placed within
1$\arcmin$ of each object of interest.  Photometry in the mid-IR is presented
in Table~\ref{tab:phot1}.

\subsection{SMGs}\label{sec:smgdata4}

\subsubsection{Sample Selection}\label{sec:smgsample}

For the SCUBA-selected SMGs, we use the sample of 53 objects with spectroscopic
redshifts at $z > 1.4$ (we have removed from the sample three sources with
extremely blue rest-frame ultra-violet colors as well as two sources which were
subsequently shown to be spurious detections by \citet{2011ApJ...740...96H})
from \citet{2005ApJ...622..772C}.  These are sources with precise positional
information derived from Very Large Array 1.4~GHz imaging and redshifts
obtained with optical ground-based spectroscopy with the Keck~I telescope.
Their clustering properties indicate they inhabit very massive dark matter
haloes \citep[$M_{\rm DM} \approx 10^{12} \:
M_{\sun}$;][]{2004ApJ...611..725B}, comparable to the dark matter halo masses
of DOGs \citep{2008ApJ...687L..65B}.

\subsubsection{SMG Photometry}\label{sec:smgimaging}

The broad-band photometry of SMGs used in this paper has been collected from a
variety of sources.  $B$- and $R$-band photometry were obtained with several
telescopes and were presented in \citet{2005ApJ...622..772C}.  $I$-, $J$-, and
$K$-band photometry also were obtained with several telescopes and were
presented in \citet{2004ApJ...616...71S}.  These photometry values were
derived with 4$\arcsec$ diameter apertures and have been aperture-corrected.
Mid-IR photometry of SMGs was obtained from \citet{2009ApJ...699.1610H}, who
compute aperture-corrected 4$\arcsec$ diameter aperture photometry using
SExtractor.

\section{Stellar Population Synthesis Models}\label{sec:models}

Stellar population synthesis (SPS) modeling offers a means of constraining the
mass and star-formation history of a galaxy's stellar population.  This section
contains a description of the technique adopted here to apply the SPS models to
the high-$z$ ULIRG photometry outlined in section~\ref{sec:data4}.
Additionally, details are provided regarding three SFHs and three initial mass
functions (IMFs) that are used in this paper for testing theories for the
formation of massive galaxies at high redshift.  Results from this analysis are
presented in section~\ref{sec:results4}.  A detailed analysis of the
differences in $M_*$ measurements obtained with four SPS libraries may be found
in Appendix~\ref{sec:spslibraries}.

\subsection{General Methodology}\label{sec:genmethod}

SPS models are parameterized at minimum by their luminosity-weighted age and
their stellar mass, $M_*$.  The attenuation of stellar light by dust adds a
third parameter, $A_V$.  In all models used here, the simplifying assumption of
a uniform dust screen ($A_V$ ranging from 0 to 3) is adopted which obscures the
intrinsic stellar light according to the reddening law for starbursts from
\citet{2000ApJ...533..682C} for wavelengths between $0.12 - 2.2 \: \mu$m and
that of \citet{2003ARAandA..41..241D} for longer wavelengths.  The available data
do not allow constraints to be placed on more complex models in which younger
stars have different dust obscuration prescriptions than older stars
\citep[e.g.,][]{2000ApJ...539..718C}.  

The broad band photometry used here is not sufficient to break the degeneracy
between age and $A_V$ (except under special assumptions).  For this reason, the
main goal here is to measure the relative $M_*$ values of three distinct
populations of high redshift ULIRGs (power-law DOGs, bump DOGs, and SMGs) using
a uniform, self-consistent analysis.  This will allow the stellar masses of
these objects to be measured in a relative sense and therefore minimize many of
the uncertainties discussed above (however, note that the masses of the
power-law DOGs in general are upper limits since the AGN contribution to the
3.6$\mu$m and 4.5$\mu$m IRAC channels is unknown).  Furthermore, competing
models of galaxy formation and evolution make different predictions about the
stellar mass properties of the most luminous galaxies at $z\sim2$.  The
distribution of stellar masses of populations of power-law DOGs, bump DOGs, and
SMGs is therefore (in principle) a viable tool with which to test these
competing models.


The approach used here is to apply SPS models of varying $A_V$ and age values
to generate a probility density function for the stellar mass of each galaxy,
$\phi (M_*,{\rm age}, A_V) $.  $\phi (M_*,{\rm age}, A_V)$ is computed directly
from the best-fit $\chi^2$ value for the given number of degrees of freedom,
$N_{\rm DOF}$.  Since we have 7 data points and 3 model parameters, $N_{\rm
DOF} = 4$.  For a few sources (SST24J~142648.9+332927, SMMJ030227.73+000653.5,
SMMJ123600.15+621047.2, SMMJ123606.85+621021.4, SMMJ131239.14+424155.7,
SMMJ163631.47+405546.9, SMMJ221735.15+001537.2, SMMJ221804.42+002154.4), no
models achieved statistically acceptable fits.  These systems are assumed to
have a uniform stellar mass probability density function between $10^{10} -
10^{12}~M_\sun$.  This has the effect of broadening the resulting stellar mass
constraints for a given galaxy population.  Each individual galaxy's $\phi
(M_*,{\rm age}, A_V) $  is normalized such that it contributes equally to the
final stellar mass probability density function for that population of galaxies
($\phi_{\rm PL DOG}$, $\phi_{\rm Bump DOG}$, and $\phi_{\rm SMG}$).  


The use of SPS models to determine intrinsic properties of galaxies assumes
that all of the observed flux is emitted by stars.  In fact, many of the
sources in this study have a significant contribution in the rest-frame near-IR
from obscured AGN (this is especially true for the power-law DOGs).  Some
authors add this component (in the form of a variable slope power-law) to their
SPS modelling efforts \citep[e.g.,][]{2011ApJ...740...96H}.  Alternatively, it
is possible to minimize the AGN contribution by considering only the first two
IRAC channels (i.e., up to observed-frame 4.5$\mu$m).  We adopt the latter
approach in this study.  For bump DOGs and most SMGs, this should
provide a reasonably reliable measurement of the stellar light from these
objects.  For power-law DOGs and those SMGs with power-law tails in the
near-IR, there still exists a significant possibility that the observed-frame
4.5$\mu$m light is contaminated by AGN, though it should be noted that
high-spatial resolution imaging with {\it HST}/NICMOS indicates that only
10-20\% of the rest-frame optical light is emitted by a point source in
power-law DOGs \citep{2009ApJ...693..750B}.  For this reason, the stellar mass
estimates of power-law DOGs should be regarded as upper limits on the true
stellar mass.  

The observed-frame $B_W$ photometry have been excluded from the fitting
process.  These data typically probe rest-frame 1500~\AA\ and as such are
highly sensitive to the youngest stellar populations and the detailed geometry
of the dust distribution surrounding them.  The most robust model fits were
obtained when the $B_W$ photometry were not used. 

Only solar metallicity models are tested in this study.  This is a reasonable
assumption, since high-redshift (median redshift of 2.4) dusty galaxies have
been found to have near-solar metallicities \citep{2004ApJ...617...64S}.
Moreover, our broad-band SED data do not provide the ability to constrain
metallicity.  The adoption of a single metallicity in SPS modeling typically
introduces uncertainties at the level of 10-20\%
\citep{2009ApJ...699..486C,2009ApJ...701.1839M}, which are insignificant
compared to systematic uncertainties related to the IMF, SFH, and age of the
stellar population.  


\subsection{SPS Star-formation Histories}\label{sec:spssfhs}

One of the most critical adjustable parameters in SPS modeling is the
star-formation history (SFH).  \citet{2011arXiv1108.6058M} suggest that
the use of multiple component SFHs, in which different stellar populations are allowed
to have distinct ages and obscuration, can lead to factors of 2-4 difference in best-fit stellar
mass.  The focus in this paper is placed on three distinct SFHs that broadly
encompass a reasonable range of parameter space while maintaining a level of
simplicity in accordance with the quality of the available data.

The first SFH adopted here is the simplest one possible: an infinitely short
burst of star-formation at time $t=0$ during which all the stars of the galaxy
are formed, followed thereafter by passive evolution.  This is called a simple
stellar population (SSP), and is used commonly in SPS modeling in the
literature.  If the objects under study here have recently had star-formation
shut off by some process (e.g., AGN feedback), then the SSP model provides
constraints on how long ago such an event ocurred.  Models used here have ages
spaced logarithmically from 10~Myr up to 1~Gyr.

The second SFH used in this paper is borrowed from a representative simulation
of a major merger which undergoes a very luminous sub-mm phase (SMG) as well as
a highly dust-obscured phase (DOG) before star-formation is shut off by AGN
feedback effects \citep{2010MNRAS.407.1701N}.  This SFH traces the
star-formation rate from the beginning of the simulation --- before the two
gas-rich ($f_g \sim 0.8$) disks begin to interact --- through the period of
final coalescence when the SFR peaks near 1000~$M_{\sun} \:$yr$^{-1}$, to the end
of the simulation and a red, dead, elliptical galaxy.  
Models used here have ages spaced roughly linearly from 10~Myr to 0.8~Gyr.

\begin{figure}[!tp] 
\begin{center}
\includegraphics[width=0.47\textwidth]{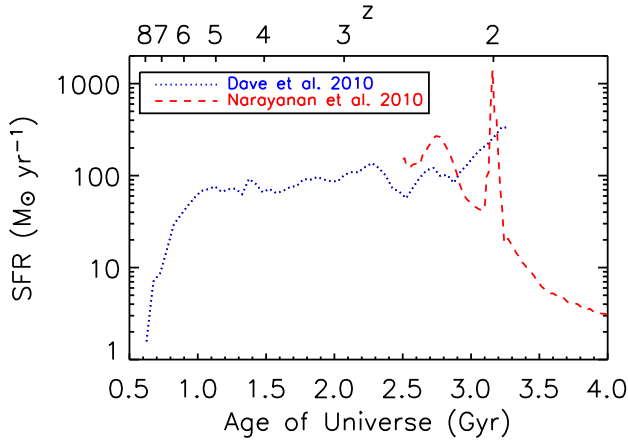}
\end{center}

\caption[Star-formation histories used in stellar population synthesis models]{
Star-formation histories used in stellar population synthesis models.  Dotted
line represents high-$z$ ULIRGs identified in cosmological hydrodynamical
simulations formed via smooth gas inflow and accretion of small satellites
\citep[Galaxy A from Figure~4 of][]{2010MNRAS.404.1355D}.  Dashed line
represents high-$z$ ULIRGs formed via major mergers of two gas-rich disks
\citep{2010MNRAS.407.1701N} and has been shifted in time so that the peak
star-formation rate occurs at $z \approx 2$.  \label{fig:sfhplot}}

\end{figure}

The third SFH adopted in this study comes from cosmological hydrodynamical
simulations in which SMGs are posited to correspond to the most rapidly
star-forming systems that match the observed number density of SMGs
\citep{2010MNRAS.404.1355D}.  In particular, the SFH and metallicity history of
the highest SFR simulated SMG are used.  This object has a SFR of
$\approx150~M_{\sun} \:$yr$^{-1}$ for most of the simulation but is boosted to
$\approx500~M_{\sun} \:$yr$^{-1}$ at $z=2$ and reaches a mass of $M_* = 2.8
\times 10^{11} ~ M_{\sun}$ by the same redshift.  As nearly all of the mass is
assembled in a quiescent mode, this SFH is nearly opposite to a SSP, in which
all stars are formed in a single infinitely short burst.  Models used here have
ages spaced roughly linearly over the full range of the SFH, from 10~Myr to
3~Gyr.  Figure~\ref{fig:sfhplot} shows the SFHs from
\citet{2010MNRAS.407.1701N} and \citet{2010MNRAS.404.1355D} that are used in
this analysis.

\subsection{Initial Mass Functions}\label{sec:spsimfs}

Another critical adjustable parameter involved in SPS modeling is the
IMF.  Despite its importance, the detailed nature of the IMF in galaxies at
high redshift is poorly constrained.  The relevant parameter space is
characterized here by three different forms: a Salpeter~IMF
\citep{1955ApJ...121..161S}, a Chabrier~IMF \citep{2003PASP..115..763C}, and a
bottom-light~IMF \citep[e.g.,][]{2008ApJ...674...29V, 2008MNRAS.385..147D}.  All
of these have a lower mass
cutoff of 0.1~$M_{\sun}$ and an upper mass cutoff of 100~$M_{\sun}$.  The
Chabrier~IMF has fewer low mass stars compared to a Salpeter~IMF (and hence a
lower mass-to-light ratio), while a bottom-light~IMF has even fewer low-mass
stars (and a correspondingly lower mass-to-light ratio).

%

The contribution of low mass stars to the bottom-light~IMF is governed by the
characteristic mass, $m_c$, which controls both the cutoff mass at which the
lognormal form dominates as well as the shape of the lognormal part of the IMF
itself.  In particular, \citet{2008ApJ...674...29V} use the color and
luminosity evolution of cluster ellipticals to infer $m_c \sim 2~M_{\sun}$ at
$z > 4$ \citep[however, see ][which argues instead for a steeper-than-Salpeter
IMF slope based upon spectral features that are strong in stars with $M_* < 0.3
\: M_\sun$ found in elliptical galaxies in the local
Universe]{2010Natur.468..940V}.  In this study, a characteristic mass of $m_c =
0.4 ~ M_{\sun}$ has been adopted, as this value matches both the (very rough)
estimates for SMGs at $z \sim2$ as well as theoretical expectations based on a
model in which the characteristic mass is a function of the CMB temperature:
$m_c \propto T_{\rm CMB}^{3.35}$.  The effect of such a change in the
characteristic mass is to produce a Salpeter-like slope at $M > 1~M_{\sun}$ and
a turnover at $M \approx 1 ~ M_{\sun}$.  This reduces the number of low-mass
stars relative to the high-mass ones, thereby lowering the mass-to-light ratio
relative to the Chabrier IMF (for intermediate age stars or younger).  

Since observational constraints on the IMF are not readily available, each IMF
has been tested with each SFH (see section~\ref{sec:spssfhs}).  In the case of
the simple stellar population (SSP), this provides a measure of the uncertainty
resulting from the unknown IMF.  However, for the purposes of testing the
self-consistency of more complicated SFHs of ULIRGs at high redshift, it is
necessary to select certain IMFs for each model.  The simulations of major
mergers tested here \citep{2010MNRAS.407.1701N} adopt a Kroupa~IMF for their
radiative transfer, so a Chabrier~IMF (which is very similar to a Kroupa~IMF)
is what is focused on here.  Meanwhile, the IMF is a free parameter in the
smooth accretion SFH \citep{2010MNRAS.404.1355D}.  A Chabrier~IMF is adopted in
this paper for this SFH (with an accompanying thorough discussion of the
implications of a more ``bottom-light'' IMF), since a Salpeter IMF overpredicts
the sub-mm fluxes of SMGs.  

\section{Results}\label{sec:results4}

This section presents measurements of the stellar masses ($M_*$) of bump DOGs,
power-law DOGs, and SMGs.  SEDs for each source may be found in
Appendix~\ref{sec:seds44}.  The nominal fiducial model chosen in this paper is
the CB07 SPS library with a SSP SFH and Chabrier IMF (meaning that we have
chosen this as the standard by which the other models will be compared), and is
presented in section~\ref{sec:SSPresults}.  In later sections, alternative SFHs
and IMFs are explored.  Although differences exist between various SPS
libraries in the treatment of aspects of stellar atmospheres and evolution,
these details are sub-dominant to the choice of SFH and IMF (for an explanation
of this, see Appendix~\ref{sec:spslibraries}).  For this reason, our modeling
process does not include marginalization over SPS library.  




\subsection{Simple Stellar Population}\label{sec:SSPresults}

The SSP represents a SFH in which all stars form in an infinitely short burst
of star-formation and evolve passively thereafter.  While this is an idealized
scenario for the formation of massive galaxies, it is worth studying
since SSPs form the building blocks of more complex SFHs and
can be used more directly to compare the effect of different SPS libraries and
IMFs (see section~\ref{sec:libraryresults} for more details on this last point).

\begin{figure}[!tp] 
\includegraphics[width=0.47\textwidth]{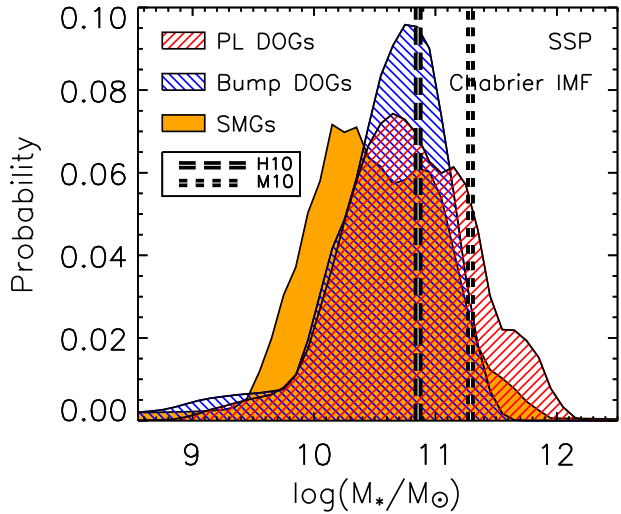}

\caption[Stellar mass probability density function of power-law DOGs, bump
DOGs, and SMGs derived using using the CB07 library, a Chabrier IMF, and a
simple stellar population]{ Stellar mass probability density function of
power-law DOGs (red), bump DOGs (blue), and SMGs (orange) derived using using
the CB07 library, a Chabrier IMF, and a simple stellar population SFH.  The
median $M_*$ value (corrected to a Chabrier~IMF) from studies by
\citet{2011ApJ...740...96H} and \citet{2010AandA...514A..67M} are given by the
long and short dashed lines, respectively.  The mass estimates presented here
indicate both types of DOGs have masses similar to SMGs and are closer to the
\citet{2011ApJ...740...96H} values than those of \citet{2010AandA...514A..67M}.
%
\label{fig:mstarSSP}}

\end{figure}

Figure~\ref{fig:mstarSSP} shows the stellar mass probability density function
resulting from fitting a SSP (computed with the CB07 SPS library and a Chabrier
IMF) to each power-law DOG, bump DOG, and SMG.  
All three populations have a similar range of acceptable $M_*$ values.
Power-law DOGs tend to be the most massive systems, followed by bump DOGs and
then SMGs.  However, their median stellar masses are separated by $\approx
0.1$~to~0.2~dex, while the spread in their distributions are $\approx 0.3$~dex.  This
implies that the differences in stellar mass between the populations are
suggestive rather than conclusive.  Perhaps the most interesting feature of
this result is that the masses of all three populations are not significantly
different.  This may imply that that the power-law phase occurs during the same
time that most of the mass in stars is being built up.  If the mid-IR power-law
is a signature of black hole growth, then this implies that the stellar mass
and black hole mass are being assembled during the same period of
dust-obscured, intense star-formation.  A low mass tail is present in each
population which is in fact a reflection of the fact that the constraints on
the stellar mass of a small percentage of each group are weak.
The median and inter-quartile range of stellar masses for this SPS model are
given in Table~\ref{tab:mstar}.

One feature of the fitting process that is not shown in
Figure~\ref{fig:mstarSSP} is the well-known significant degeneracy between
$A_V$ and stellar age -- the broad-band photometry of these high-$z$ ULIRGs can
be fit either by young (10~Myr) and dusty ($A_V \sim 1.5-2$) stellar
populations or intermediate age (500~Myr) and less dusty ($A_V \sim 0.0 - 0.5$)
stellar populations.  Given the large quantities of dust that are known to
exist in these systems based on observations at longer wavelengths
\citep[e.g.,][]{2006ApJ...650..592K, 2008MNRAS.384.1597C,
2009ApJ...693..750B,2009ApJ...692..422L,2010ApJ...717...29K}, it is unlikely
that $A_V < 1$ solutions are acceptable.  Indeed, mid-IR spectra of {\it
Spitzer}-selected ULIRGs generally show strong silicate absorption features
indicative of highly obscured sources \citep{2007ApJ...664..713S}.
Furthermore, measurements of H$\alpha$ and H$\beta$ in a handful of sources
find strong Balmer decrements implying $A_V > 1$ \citep{2007ApJ...663..204B}.  

Assuming $A_V = E(B-V) / R_V$ (where $R_V = 3.1$) and the relation between
$E(B-V)$ and the hydrogen column density ($N_{\rm H}$) from
\citet{1978ApJ...224..132B}, $A_V \sim 1$ implies $N_{\rm H} \sim 2 \times
10^{21}~$cm$^{-2}$.  Under the assumption of a spherical shell around the
source with radius equal to the effective radius ($R_{\rm eff}$), the dust mass
can be estimated from $N_{\rm H}$ using:

\begin{equation}
    M_{\rm d} = \frac{1}{f_{\rm gd}} \mu_{\rm p} N_{\rm H} 4 \pi R_{\rm eff}^2,
\end{equation}

\noindent where $f_{\rm gd}$ is the gas-to-dust mass ratio \citep[assumed to be
60, the value found appropriate for SMGs;][]{2006ApJ...650..592K} and $\mu_{\rm
p}$ is the mean molecular weight of the gas (assumed to be 1.6 times the mass
of a proton).  Morphological measurements indicate these objects have typical
effective radii of 3-8~kpc
\citep{2008ApJ...680..232D,2009ApJ...693..750B,2010ApJ...719.1393D}.  All
together this implies $M_{\rm d} \sim (0.5 - 3) \times 10^8~M_{\sun}$,
depending on the size of $R_{\rm eff}$.  In fact, based on 350$\mu$m
observations, \citet{2010ApJ...717...29K} find dust masses of $M_{\rm d}
\approx (5-10) \times 10^8~M_{\sun}$ for {\it Spitzer}-selected ULIRGs with a
mid-IR bump feature.  This suggests that $A_V>1$ and hence age $ < 200$~Myr
models should be preferred.  Note however that for any given galaxy, we do not
have independent constraints on $A_V$ and hence have applied no priors on this
quantity in the fitting process.

\subsection{Merger-Driven Star-Formation History}\label{sec:dtnresults}

One of the major goals of this paper is to go beyond instantaneous burst SFHs
(SSPs) and test the self-consistency of more complicated SFHs.  Two in
particular that are tested here are a SFH driven by a major merger
\citep{2010MNRAS.407.1701N} and a SFH driven mainly by smooth accretion of gas
and nearby small satellites \citep{2010MNRAS.404.1355D}.  The merger-driven SFH
is described here, while the smooth accretion SFH is described in
section~\ref{sec:radresults}.

\begin{figure*}[!tp] 
\includegraphics[width=0.5\textwidth]{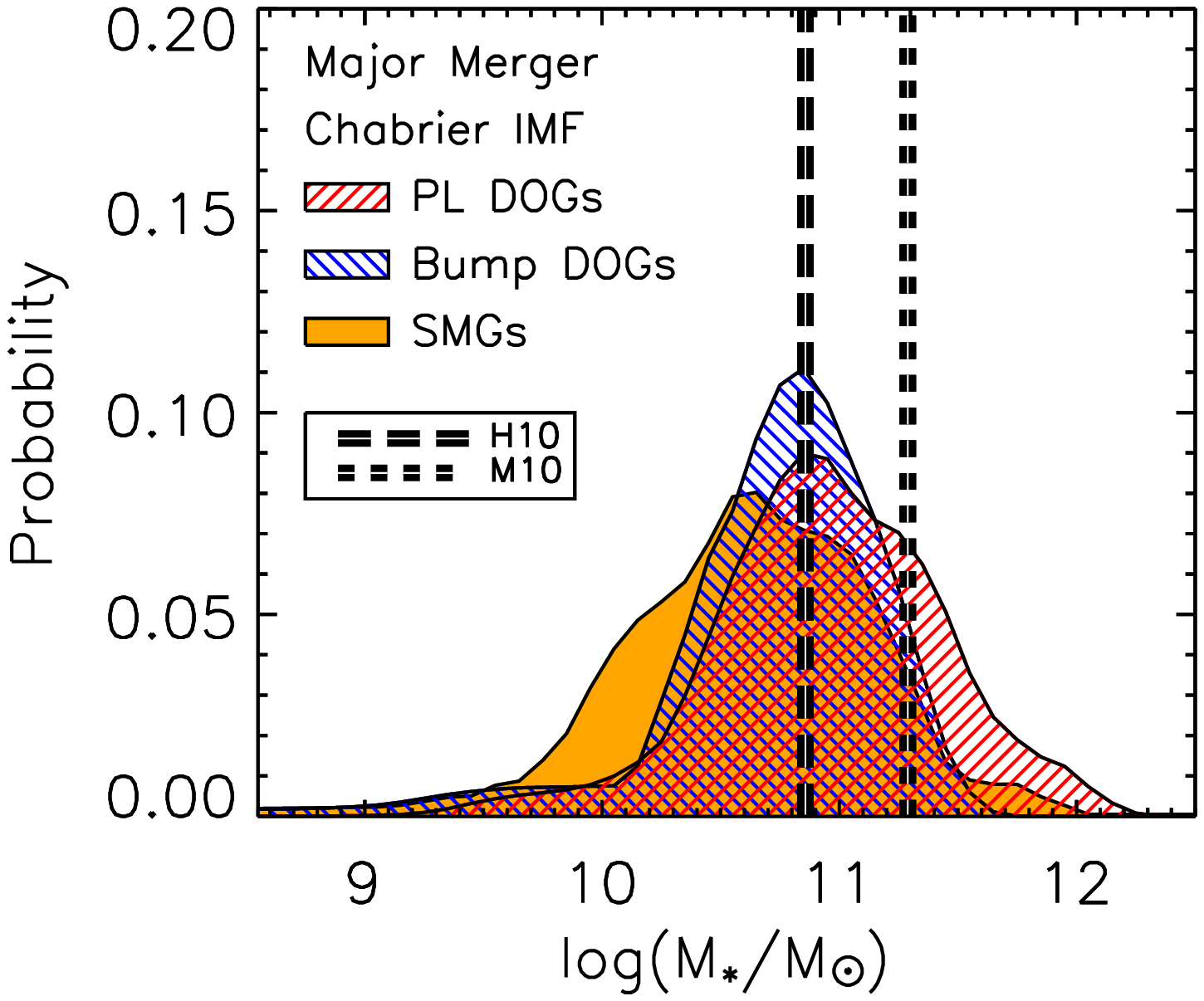} 
\includegraphics[width=0.5\textwidth]{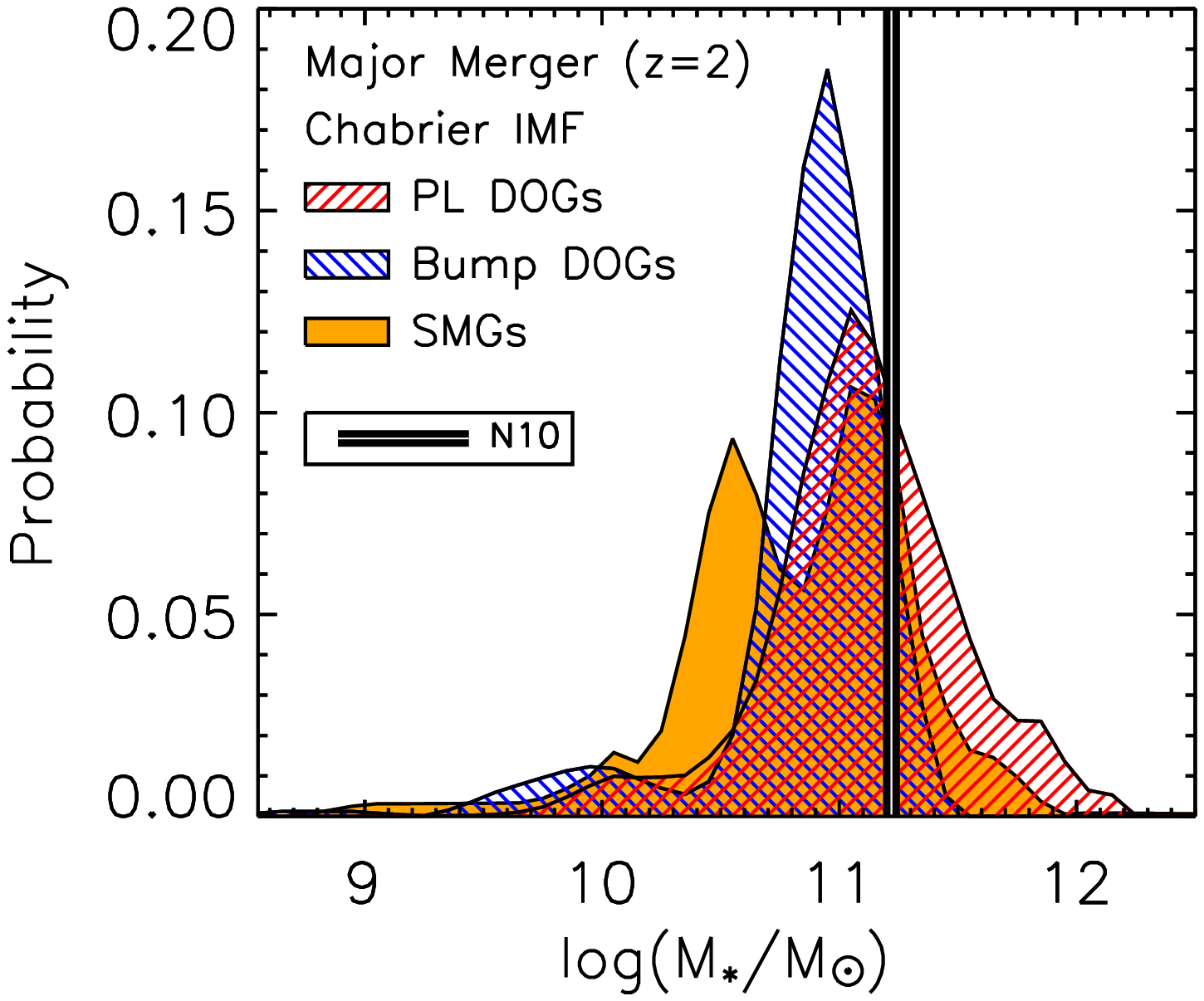}

\caption[Similar to previous figure, but assuming a major merger SFH]{ Similar
to Figure~\ref{fig:mstarSSP}, but assuming a major merger SFH \citep[see
Figure~\ref{fig:sfhplot} of this paper and][]{2010MNRAS.407.1701N}.  {\it
Left:} Stellar masses obtained when marginalizing over the full age range of
the model. Mass estimates from \citet{2011ApJ...740...96H} and
\citet{2010AandA...514A..67M} (dashed and dotted lines, respectively) have been
corrected to a Chabrier~IMF. {\it Right:} Stellar masses obtained when
marginalizing over only the $z = 2$ period of the model (i.e., the timestep
during which the SFR peaks).  When only the $z = 2$ period is considered, the
inferred stellar masses increase by 0.1-0.3~dex (see Table~\ref{tab:mstar})
thanks to the increased contribution from old stars with high mass-to-light
ratios.  The median stellar masses of these $z \sim 2$ ULIRGs are still about
0.2-0.3~dex lower than expected from the \citet{2010MNRAS.407.1701N} models
(solid line).  

\label{fig:mstardtn}}

\end{figure*}

Figure~\ref{fig:mstardtn} (left panel) shows the stellar mass probability
density function for power-law DOGs, bump DOGs, and SMGs derived using a
merger-driven SFH \citep[from][]{2010MNRAS.407.1701N} with the CB07 SPS library
and a Chabrier~IMF.  The median and inter-quartile range of $M_*$ values are
given for this SFH in Table~\ref{tab:mstar} and are about 0.1-0.2~dex larger
than the same values derived using a SSP and a Chabrier~IMF (again the trend in
masses is that power-law DOGs are the most massive and SMGs the least massive,
with bump DOGs falling in between).  Multi-component SFHs in general produce
higher mass-to-light ratios than SSPs because even a modest amount of
rest-frame UV emission will strongly constrain the age of the SSP to be less
than a few hundred million years.  Such a young stellar population will have a
low mass-to-light ratio.  In contrast, a multi-component SFH can have a low
mass young stellar component (which reproduces the rest-frame UV emission) as
well as an old stellar component which boosts the mass-to-light ratio.  

This point is made more clearly in the right panel of
Figure~\ref{fig:mstardtn}, which shows the stellar mass probability density
function for power-law DOGs, bump DOGs, and SMGs derived from the merger-driven
SFH but focusing on the portion of the SFH when the system is expected to be in
its ULIRG phase (i.e., maximum SFR).  By this stage (about 0.7~Gyr into the
SFH), the presence of a significant amount of low mass stars increases the
inferred stellar masses by 0.1-0.3~dex (relative to the masses derived from the
SSP SFH).  These mass estimates are also reported in Table~\ref{tab:mstar}.
The increase in our estimates of $M_*$ is actually mitigated somewhat because
the SFR is so high that the fraction of very massive stars relative to all
other stars is higher than at other times in the SFH and because we have made
the simplest possible assumption for the dust geometry of a uniform dust
screen.  In reality, the youngest stars should experience greater extinction
than the older stars.  This effect is likely to be amplified by the merger, in
which the peak SFR occurs when all the gas and dust have been dumped into the
central, most obscured regions.  For this reason, we expect that our
measurements of $M_*$ for the merger SFH during this period are likely to
underestimate the true stellar masses.

It is somewhat interesting that a bimodal distribution in SMG stellar masses
appears when one focuses on the period of peak SFR in the merger simulation.
This bimodality is smoothed out in the left panel of Figure~\ref{fig:mstardtn},
which shows the superposition of all ages during the SFH.  The origin of the
bimodality is not entirely clear, but is likely due to the presence of a
significant number of SMGs that are rest-frame UV-bright and therefore are
found to have relatively low stellar masses.  In contrast, DOGs are selected to
be rest-frame UV-faint and do not show this bimodality in stellar masses.  

 
\subsection{Smooth Accretion Star-Formation History}\label{sec:radresults}

In the cosmological hydrodynamical simulations of \citet{2010MNRAS.404.1355D},
SMGs are posited to be the maximally star-forming galaxies whose number
densities match the observed number density of SMGs.  This results in the
typical simulated SMG having a SFH described by a relatively constant SFR of
100-200~$M_{\sun}~$yr$^{-1}$ over a period of 3~Gyr and leads to a stellar mass
in these systems in the range $M_{\sun} \approx (1-5) \times 10^{11} ~
M_{\sun}$.  \citet{2010MNRAS.404.1355D} note that their simulated SFRs are a
factor of $\sim 3$ lower than the typical values observationally inferred for
SMGs, and hypothesize that a ``bottom-light'' IMF such as that proposed by
\citet{2008ApJ...674...29V} and \citet{2008MNRAS.385..147D} could explain this
discrepancy.  This type of IMF would also have the consequence of modifying the
$M_*/L_V$ of the galaxy, meaning that at a given $L_V$, the inferred stellar
mass will be lower than for other IMFs such as Chabrier or Salpeter.  It is for
this reason that the constraints on the stellar masses of the high-$z$ ULIRGs
with this SFH are of particular interest.  

Figure~\ref{fig:mstarrad} (left panel) shows the stellar mass probability
density function for power-law DOGs, bump DOGs, and SMGs derived using a SFH
driven mainly by smooth accretion of gas and nearby satellites (with the CB07
SPS library and a Chabrier~IMF).  The median and inter-quartile range of $M_*$
estimates are provided in Table~\ref{tab:mstar}.  In this case, the median
stellar masses of the three populations are separated by $\approx 0.15$~dex,
with power-law DOGs being the most massive and SMGs being the least massive
(note that this is still well below the typical inter-quartile range in the
stellar mass estimates of $\approx 0.3$~dex).  In comparison to the SSP SFH,
the smooth accretion mass estimates are $\approx 0.2$~dex larger, for similar
reasons as those outlined at the end of section~\ref{sec:dtnresults}.

Restricting the age range of the SFH for the smooth accretion model to coincide
with the period during which the simulated systems are expected to be ULIRGs
(i.e., at $z \sim 2-3$) leads to inferred stellar masses that are larger by
0.3-0.4~dex compared to the SSP SFH (Table~\ref{tab:mstar} and
Figure~\ref{fig:mstarrad}, right panel).  As described in
section~\ref{sec:dtnresults}, this is a result of a greater contribution from
older stars that have higher mass-to-light ratios than younger stars.  

\begin{figure*}[!tp] 
\includegraphics[width=0.5\textwidth]{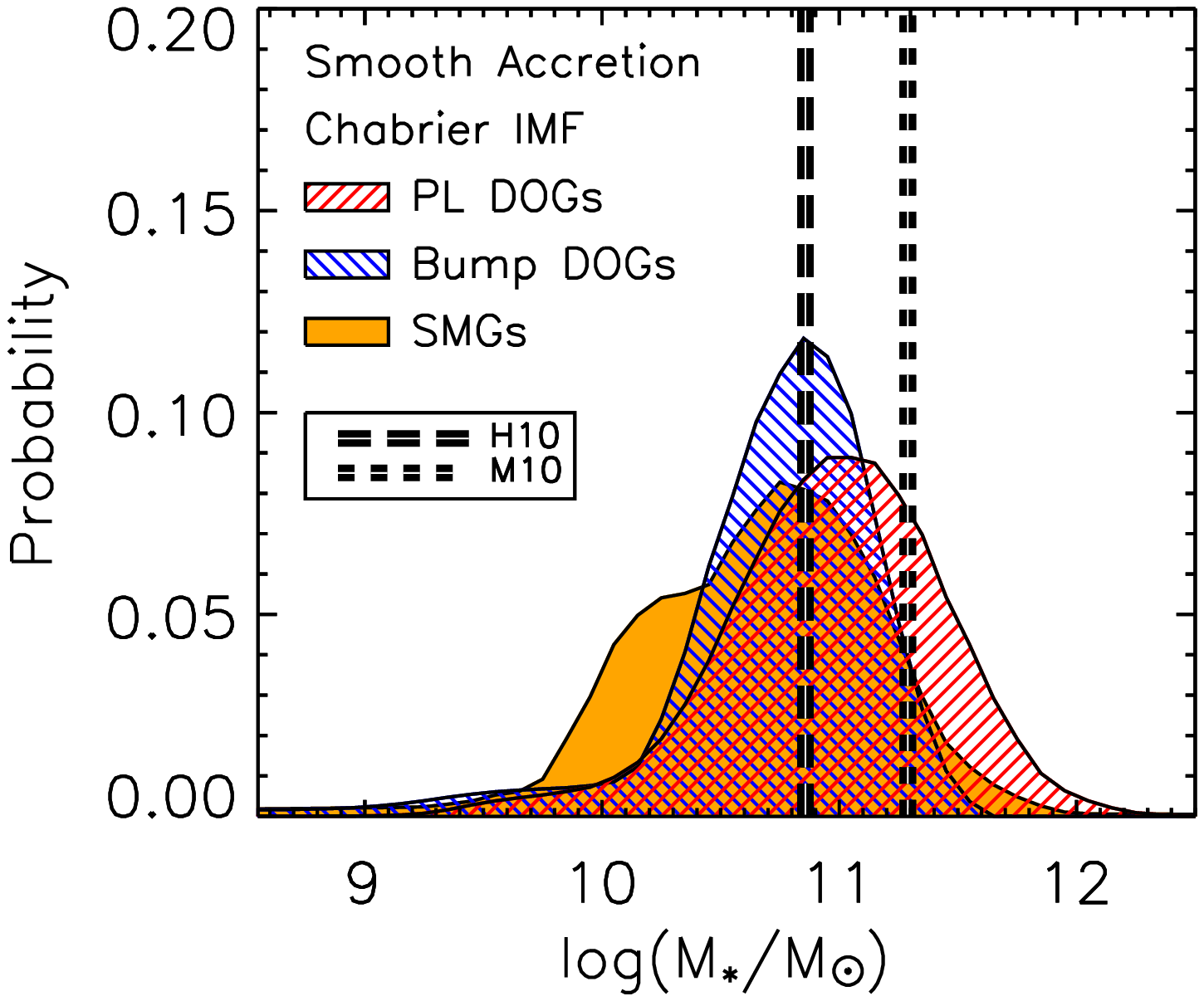}
\includegraphics[width=0.5\textwidth]{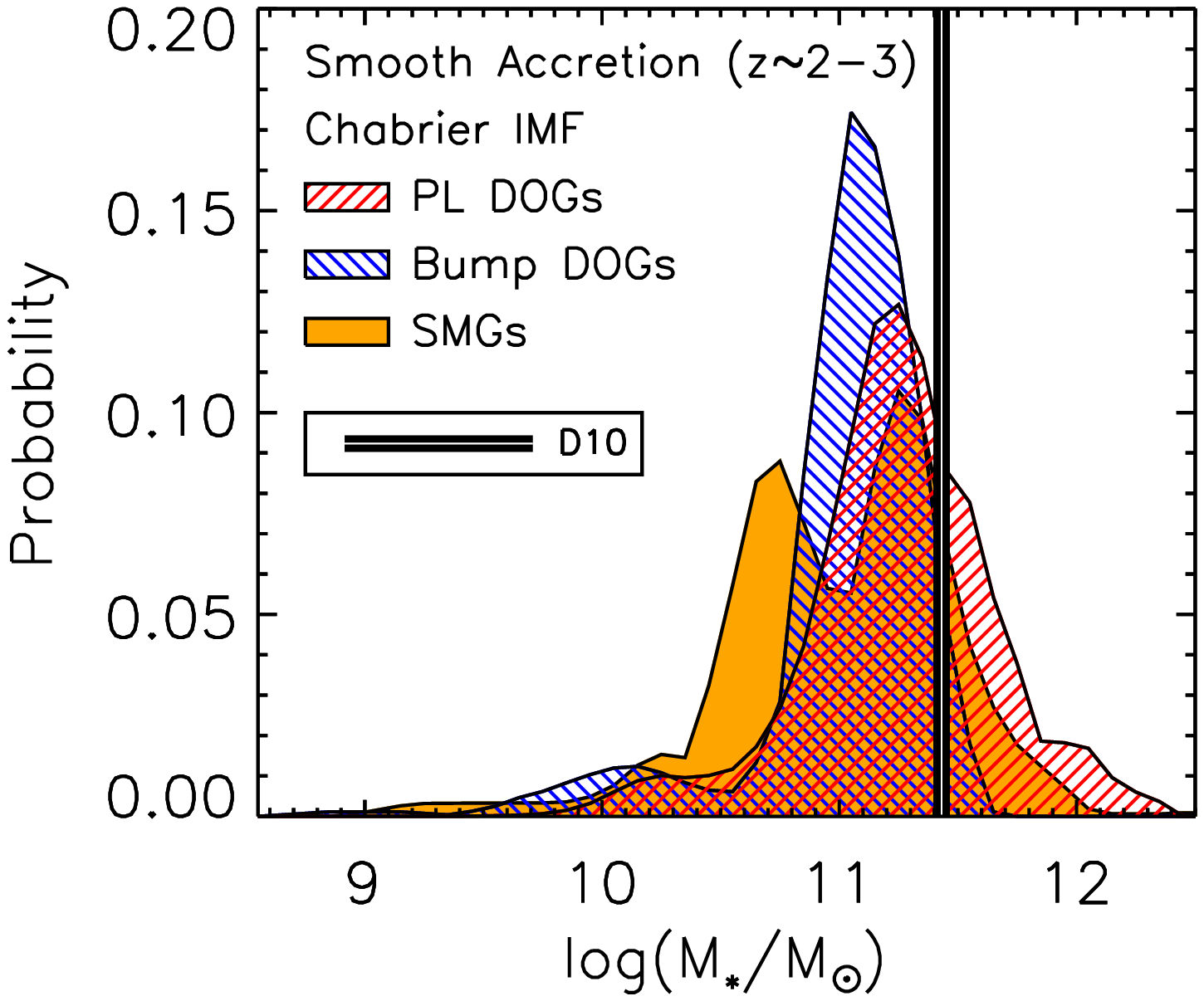}

\caption[Similar to previous figure, but assuming a smooth accretion SFH]{
Similar to Figure~\ref{fig:mstarSSP}, but assuming a smooth accretion SFH
\citep[see Figure~\ref{fig:sfhplot} of this paper and][]{2010MNRAS.404.1355D}.
{\it Left:} Stellar masses obtained when marginalizing over the full age range
of the model.  {\it Right:} Stellar masses obtained when marginalizing over
only the $z \sim 2-3$ period of the model.  Here, the stellar mass values
increase by 0.3-0.4~dex when only the $z \sim 2$ epoch of the SFH is considered
due to the increased contribution of low-mass stars which have high
mass-to-light ratios.  The median stellar masses of these $z \sim 2$ ULIRGs are
still about 0.3-0.4~dex lower than expected from the
\citet{2010MNRAS.404.1355D} models (solid line).  \label{fig:mstarrad}}

\end{figure*}

\subsection{Variation with IMF}
\label{sec:libraryresults}

In SPS modeling, the IMF affects primarily the mass-to-light ratio of the
synthesized stellar population.  \citet{2003MNRAS.344.1000B} showed that the
$B-V$ and $V-K$ colors of SPS models distinguished only by their IMFs (Chabrier
vs.  Salpeter) are very similar.  On the other hand, the Salpeter IMF gives
mass-to-light ratios that are $\approx 0.2$ dex larger than the Chabrier IMF.
Bottom-light IMFs \citep[such as that advocated by][]{2008ApJ...674...29V} have
more complicated mass-to-light ratios that depend on both the characteristic
mass ($m_c$) and the age of the stellar population.
\citet{2008ApJ...674...29V} find that for $m_c = 0.4 ~ M_{\sun}$ (as adopted
here) and ages $< 1~$Gyr, the mass-to-light ratio is lower by 0.2-0.3~dex
compared to a Chabrier IMF.  The results of this study are consistent with this
finding: assuming a SSP SFH and this bottom-light IMF, the stellar masses of
bump DOGs are in the range $M_* = (0.1 - 0.6) \times 10^{11} ~ M_{\sun}$, or
about 0.3-0.4~dex lower than those inferred using a Chabrier IMF.  A similar
reduction in $M_*$ occurs when using the bottom-light IMF in conjunction with
more complicated SFHs such as the merger-driven SFH and the smooth accretion
SFH detailed in sections~\ref{sec:dtnresults} and \ref{sec:radresults}.

\section{Discussion}\label{sec:disc4}

The focus of this section is to build upon the constraints on the stellar
masses and star-formation histories of bump DOGs, power-law DOGs, and SMGs
presented in section~\ref{sec:results4}.  Estimates of $M_*$ presented here are
compared with estimates of other dust-obscured high-redshift ULIRGs.  In
addition, implications for models of galaxy evolution are presented based upon
a comparison of the two theoretical SFHs considered in this study (major merger
and smooth accretion).

\subsection{Comparing Stellar Mass Estimates of ULIRGs at $z \sim
2$}\label{sec:mstarcompare}

Studies of other {\it Spitzer}-selected ULIRGs with a bump in the
observed-frame mid-IR SED have found median stellar masses of $M_* \approx
10^{11} ~ M_{\sun}$ \citep[for a Chabrier~IMF;][]{2007AandA...476..151B,
2009ApJ...692..422L, 2009ApJ...700..183H}.  This is a little more than
$1\sigma$ higher than the median stellar mass found for bump DOGs here.  The
small difference in stellar mass estimates can be fully accounted for by the
choice of star-formation history as well as the use in this study of the new
CB07 SPS libraries, which have redder near-IR colors and hence tend towards
lower inferred stellar masses \citep[see section~\ref{sec:libraryresults} and
also][]{2009ApJ...701.1839M}.

Two recent studies of SMGs using stellar population synthesis modeling have
come to differing conclusions regarding their median $M_*$.  While
\citet{2010AandA...514A..67M} find a median stellar mass of $M_* \approx 2 \times
10^{11} ~ M_{\sun}$ \citep[using SEDs from][and after converting to a
Chabrier~IMF]{2007ApJ...670..279I}, \citet{2011ApJ...740...96H} find $M_* = 7
\times 10^{10} ~ M_{\sun}$ \citep[assuming a Chabrier IMF and models
from][]{2005MNRAS.362..799M}.  \citet{2011ApJ...740...96H} argue that models
which do not consider the contribution of an obscured AGN in the mid-IR
(particularly in the 5.8$\mu$m and 8.0$\mu$m channels of IRAC) can bias stellar
mass estimates of SMGs upwards by a factor of $\approx 2$.  Our analysis (which
excludes these two IRAC channels to minimize the contribution from an obscured
AGN) indicates stellar masses that are closer to those of
\citet{2011ApJ...740...96H}, with median $M_* = 4.4 \times 10^{10} ~ M_{\sun}$
(for a smooth accretion SFH without constraints on the age of the stellar
population, which most closely resembles the SSP and constant star formation
histories adopted by \citet{2011ApJ...740...96H}).  Inclusion of the additional
two IRAC channels increases our stellar mass estimates by 50\% (median $M_* = 6
\times 10^{10} ~ M_{\sun}$).  A similar increase is seen when including the
5.8$\mu$m and 8.0$\mu$m channels of IRAC to models which focus on the age of
the smooth accretion SFH corresponding to the ULIRG phase (i.e., $z \sim 2$).
This is somewhat of a smaller effect than found by \citet{2011ApJ...740...96H},
possibly suggesting that there may be some contamination from AGN in the
3.6$\mu$m and 4.5$\mu$m IRAC channels that we are not accounting for as well.

\subsection{Implications for Galaxy Evolution at $z \sim 2$}
\label{sec:implications4}

Observational evidence indicates that ULIRGs in the local Universe are the
product of major mergers \citep{1987AJ.....94..831A} and that they are
connected in an evolutionary sense with quasars \citep{1988ApJ...325...74S,
1988ApJ...328L..35S}.  It is tempting to postulate a similar major-merger
origin for high-redshift ULIRGs. However, conclusive evidence linking variously
selected ULIRG populations to each other and to quasars at high redshift
requires measurements that challenge current observational capabilities.
Nevertheless, some tantalizing hints exist that suggest these diverse
populations are indeed linked.  First, the clustering strength of DOGs is
comparable to that of both the SMGs and QSOs at similar redshifts
\citep{2008ApJ...687L..65B}.  Second, the quantitative morphologies of DOGs and
SMGs are consistent with an evolutionary picture in which the SMG phase
precedes the bump DOG phase, which in turn precedes the PL DOG phase
\citep{2011ApJ...733...21B}.  However, such morphological studies are
challenging because of surface brightness dimming and dust-obscuration effects,
which prevent a straightforward merger identification based on imaging
\citep{2008ApJ...680..232D, 2008AJ....136.1110M, 2009ApJ...693..750B,
2009AJ....137.4854M, 2011ApJ...730..125Z, 2011ApJ...733...21B}.  

This study offers an independent means of testing both the evolutionary
hypothesis as well as the merger hypothesis via SPS modeling of broad-band
imaging in the rest-frame UV through near-IR.  The approach followed in this
paper is to test the self-consistency of two distinct SFHs.  One is
characterized by a gas-rich major merger which reaches a peak SFR of $\approx
1000~M_\sun\;$yr$^{-1}$ \citep{2010MNRAS.407.1701N}.  The other is
characterized by smooth accretion of gas and small satellites that typically
reaches SFRs of $\approx 200-300~M_\sun\;$yr$^{-1}$ at $z \sim 2$
\citep{2010MNRAS.404.1355D}.

First, it is worth noting that in the model of \citet{2010MNRAS.407.1701N}, an
evolutionary progression exists in which SMGs evolve into bump DOGs which
evolve into power-law DOGs.  This process occurs on a short time-scale ($\sim
50-100$~years), but if it is true then we should expect the SMGs to have the
youngest stellar population, followed by bump DOGs and then power-law DOGs.  In
this case, the relative differences in the inferred stellar masses for the
three populations become more significant and in the expected direction for the
evolutionary scenario outlined above.  In comparison, the model of
\citet{2010MNRAS.404.1355D} does not yet include radiative transfer
calculations and so cannot make a prediction for an evolutionary scenario
between these three populations.  Because of the short timescales involved in
the merger simulations and the nature of our seven filter broadband photometry
dataset, we do not pursue this point in a more quantitative manner, but we
nevertheless believe that it deserves mentioning.

Second, the stellar masses reported here are factors of 2-2.5 lower than expected from
the both the merger and smooth accretion models tested in this paper.  This
reflects the large uncertainties inherent in absolute measurements of stellar
mass and indicates that stellar masses alone are unlikely to provide a
definitive reason to favor either model over the other.
\cite{2011arXiv1108.6058M} show that the use of multi-component SFHs (i.e.,
multiple stellar populations with varying ages, extinctions, and masses) in SPS
modeling can
lead to higher inferred total stellar masses by virtue of using the young
stellar component to match the rest-frame UV flux and the old stellar component
to match the rest-frame near-IR flux.  We do not believe the data we have in
hand (broad-band photometry in seven filters) is sufficient to warrant such complex
models, but it is nevertheless important to recognize that such models are
indeed capable of implying larger stellar masses than the models we have
adopted in this paper.  

In addition, the unknown form of the IMF can potentially insert another factor
of 2-4 uncertainty in the absolute stellar mass measurements.  However, it must
be emphasized that modifications in the assumed IMF will affect not only the
inferred stellar masses, but also the inferred instantaneous SFRs.  Thus, the
effect of the IMF can be minimized by comparing the stellar masses of $z \sim
2$ ULIRGs to their star-formation rates.  Although SFRs are not yet well known
in {\it Spitzer}-selected ULIRGs, early evidence indicates that bump sources
may have similar SFRs as SMGs \citep[$\sim
1000~M_\sun~$yr$^{-1}$][]{2009ApJ...692..422L, 2010ApJ...717...29K}, whereas
power-law sources may have much lower SFRs \citep[e.g.  $\sim
100~M_\sun~$yr$^{-1}$;][]{2011AJ....141..141M}.  

A galaxy with a mass of $M_* = 1 \times 10^{11} \: M_\sun$ and a SFR of
$1000~M_\sun~$yr$^{-1}$ has a specific SFR of ${\rm sSFR} = 1 \times
10^{-8}~{\rm yr}^{-1}$.  A galaxy with a SFR lower by a factor 10 will have a
sSFR that is also lower by a factor of 10.  Thus the range for DOGs and SMGs in
sSFR is likely to be of order $1-10~{\rm Gyr}^{-1}$.  In comparison,
simulations of major mergers that produce DOG and SMG behavior tend to have
${\rm sSFR} = 6~{\rm Gyr}^{-1}$.  On the other hand, in smooth accretion driven
simulations, SMGs have ${\rm sSFR} = 0.7~{\rm Gyr}^{-1}$.  Even if we adopted
assumptions regarding the SFHs and dust geometry that led to stellar masses
that were a factor of 2-4 larger and were thus consistent with those found by
e.g. \citet{2010AandA...514A..67M}, the range in sSFR values for DOGs and SMGs
would still be higher than the expectation from the smooth accretion model.
This is merely a consequence of the fact that mergers provide a more ready
mechanism to obtain high sSFR values than smooth accretion models.



\section{Conclusions}\label{sec:conc4}

In this paper, we analyze the broad-band SEDs of a large sample of mid-IR
selected (bump and power-law DOGs) and far-IR selected (SMGs) ULIRG populations
with known spectroscopic redshifts and use stellar population synthesis models
to estimate self-consistently the stellar masses of these three populations.
We compare our mass estimates with predictions from two competing theories for
the formation of these systems and examine the implications for galaxy
evolution.  We list our findings below.


\begin{itemize}

    \item The median and inter-quartile range of stellar masses for SMGs, bump
	DOGs and power-law DOGs are log$(M_*/M_\sun) = 10.71^{+0.40}_{-0.34}$,
	$10.62^{+0.26}_{-0.32}$, and $10.42^{+0.42}_{-0.36}$, respectively,
	assuming a simple stellar population SFH, a Chabrier IMF, and the CB07
	stellar libraries.  The overlap in $M_*$ values between all three
	populations is consistent with the picture in which they represent a
	brief but important phase in massive galaxy evolution, with tentative
	evidence supporting a scenario in which SMGs evolve into bump DOGs
	which evolve into power-law DOGs.
	


    \item The use of more realistic SFHs in the SPS modeling in which both old
	and young stars contribute to the observed broad-band photometry can
	increase mass estimates significantly.  We show that using a major
	merger driven SFH during its peak SFR period (when it is expected to be
	identified as a ULIRG at $z \sim 2$) leads to median and inter-quartile
	stellar mass estimates for power-law DOGs, bump DOGs, and SMGs of
	log$(M_*/M_\sun) = 11.06^{+0.24}_{-0.21}$, $10.88^{+0.14}_{-0.13}$, and
	$10.86^{+0.24}_{-0.37}$, respectively.  Using a smooth accretion driven
	SFH (focusing on the predictions at $z \sim 2$) these values become
	log$(M_*/M_\sun) = 11.20^{+0.23}_{-0.20}$, $11.03^{+0.15}_{-0.14}$, and
	$11.02^{+0.25}_{-0.37}$, respectively. 

    \item The stellar masses we measure are inconsistent with those predicted
	by both numerical simulations we have tested (being lower by a factor
	of 2-2.5).  This indicates that either the simulations over-predict the
	stellar masses of high-$z$ ULIRGs, or that one (or more) of the
	assumptions in our SPS models is incorrect.  In either case, the
	stellar mass data presented here are by themselves insufficient to
	favor one model over another.  However, we note that the use of a
	bottom-light rather than a Chabrier~IMF may be needed for the SFRs of
	the smooth accretion model to match those that are observed.  Such a
	change would decrease our mass estimates by a factor of roughly 2
	(depending on the exact shape of the bottom-light IMF).  This line of
	reasoning suggests that, at least for the most luminous sources, the
	smooth accretion model has difficulty reproducing the observed far-IR
	emission (i.e., instantaneous SFR) without overestimating the observed
	optical and near-IR emission (i.e., stellar mass).

\end{itemize}

Estimates of the stellar masses of dust-obscured galaxies at high-redshift are
highly dependent on the age of the stellar populations within those galaxies.
The use of multiple component SFHs with different ages can lead to significant
variations in the inferred stellar mass \citep[e.g.,][]{2011arXiv1108.6058M}.
We do not consider the broad-band photometry in seven filters used here to be
sufficient to explore such complex SFHs.  However, in the near future,
wide-field medium-band photometry surveys in the near-IR \citep[e.g., the
NEWFIRM Medium-Band Survey, NMBS;][]{2009PASP..121....2V} will provide a finer
sampling of the rest-frame Balmer and 4000~\AA\ break and significantly improve
constraints on the stellar population age in DOGs and SMGs.  Further in the
future, the advent of the {\it James Webb Space Telescope} will provide
high-spatial resolution imaging in the mid-IR and provide improved constraints
on the amount of stellar emission vs.  AGN emission in ULIRGs at high redshift.
This is critical information especially for power-law DOGs, but holds
significance for bump DOGs and SMGs as well.

This work is based in part on observations made with the {\it Spitzer Space
Telescope}, which is operated by the Jet Propulsion Laboratory, California
Institute of Technology under NASA contract 1407.  {\it Spitzer}/MIPS
guaranteed time observing was used to image the Bo\"otes field at 24$\mu$m and
is critical for the selection of DOGs.  

We wish to acknowledge our referee, Laura Hainline, whose comments have helped
greatly to improve the clarity of the paper.  We thank the SDWFS team
(particularly Daniel Stern and Matt Ashby) for making the IRAC Legacy data
products available to us.  We are grateful to the expert assistance of the
staff of Kitt Peak National Observatory where the Bo\"{o}tes field observations
of the NDWFS were obtained.  The authors thank NOAO for supporting the NOAO
Deep Wide-Field Survey. In particular, we thank Ed Ajhar, Jenna Claver, Alyson
Ford, Tod Lauer, Lissa Miller, Glenn Tiede and Frank Valdes and the rest of the
NDWFS survey team, for their assistance with the execution of the NDWFS.  We
also thank the staff of the W.~M.~Keck Observatory, where many of the galaxy
redshifts were obtained.

RSB gratefully acknowledges financial assistance from HST grants GO10890 and
GO11195, without which this research would not have been possible.  Support for
Programs HST-GO10890 and HST-GO11195 was provided by NASA through a grant from
the Space Telescope Science Institute, which is operated by the Association of
Universities for Research in  Astronomy, Incorporated, under NASA contract
NAS5-26555.  The research activities of AD and BTJ are supported by NOAO, which
is operated by the Association of Universities for Research in Astronomy (AURA,
inc.) under a cooperative agreement with the National Science Foundation.  

Facilities used: {\it Spitzer Space Telescope}; Kitt Peak National Observatory
Mayall 4m telescope; W.~M.~Keck Observatory; Gemini-North Observatory.

\appendix

\section{SPS Libraries}\label{sec:spslibraries}

Four SPS libraries have been tested in this analysis of the SEDs of DOGs and
SMGs.  The first SPS library used in this paper is from the
\citet{2003MNRAS.344.1000B} population synthesis library.  It uses the
isochrone synthesis technique \citep{1991ApJ...367..126C} and the Padova~1994
evolutionary tracks \citep{1996AandAS..117..113G} to compute the spectral
evolution of stellar populations at ages between 10$^5$ and $2\times
10^{10}$~yr.  The STEllar LIBrary \citep[STELIB][]{2003AandA...402..433L} of
stellar spectra offer a median resolving power of 2000 over the wavelength
range 3200 to 9500~\AA.  Outside this wavelength range, the BaSeL~3.1 libraries
\citep{2002AandA...381..524W} are used and offer a median resolving power of 300
from 91~\AA\ to 160$\mu$m.  

The second SPS library used here is an updated version of the
\citet{2003MNRAS.344.1000B} population synthesis library (Charlot \& Bruzual,
private communication, hereafter CB07).  The primary improvement included in
these models is a new prescription for the thermally pulsing asymptotic giant
branch (TP-AGB) evolution of low- and intermediate-mass stars
\citet{2007AandA...469..239M} and \citet{2008AandA...482..883M}.  This has the
effect of producing significantly redder near-IR colors for young and
intermediate-age stellar populations, which leads to younger inferred ages and
lower inferred masses for a given observed near-IR color.  These new models
otherwise still rely on the Padova~1994 evolutionary tracks and the combination
of BaSeL~3.1 and STELIB spectral libraries.

The third SPS library employed in this paper is called a Flexible Stellar
Population Synthesis library
\citep[FSPS;][]{2009ApJ...699..486C,2010ApJ...708...58C,2010ApJ...712..833C}.
This library uses the isochrone synthesis technique as well, but with updated
evolutionary tracks
\citep[Padova~2008][]{2007AandA...469..239M,2008AandA...482..883M}.  FSPS adopts
the BaSeL~3.1 spectral library \citep{2002AandA...381..524W} but includes TP-AGB
spectra from a compilation of more than 100 optical/near-IR spectra spanning
the wavelength range 0.5 – 2.5$\mu$m
\citep{2000AandAS..146..217L,2002AandA...393..167L}.  One feature of this library
that is not available in the others is the ability to input a custom IMF (e.g.,
a ``bottom-light'' IMF).

The fourth and final SPS library used here is from \citet{2005MNRAS.362..799M}.
This library adopts the ``fuel-consumption'' approach, in which the integration
variable is the amount of hydrogen or helium consumed by nuclear burning during
a given post-main-sequence phase (unlike the isochrone synthesis approach, in
which the integration variable is the stellar mass).  This library features a
strong contribution from TP-AGB stars ($\approx 40\%$ of the bolometric light)
for age ranges of 0.2 - 2~Gyr.  A comparison between this library and that of
\citet{2003MNRAS.344.1000B} found that the near-IR colors of $z \sim2$ galaxies
were better fit by the former \citep{2006ApJ...652...85M}, highlighting the
importance of a proper treatment of the TP-AGB phase for intermediate age
stellar populations.

\section{SEDs}\label{sec:seds44}

Since every source in this study has a known spectroscopic redshift, it is
possible to construct SEDs for each source showing the luminosity per unit
frequency ($L_\nu$) as a function of rest-frame wavelength ($\lambda_{\rm
rest}$).  Figures~\ref{fig:pldogsed},~\ref{fig:bumpdogsed},~\ref{fig:smgsed}
show the SEDs for power-law DOGs, bump DOGs, and SMGs, respectively.  Also
shown in this diagram is the best-fit synthesized stellar population model
(CB07, simple stellar population, Chabrier~IMF). Inset in each diagram is the
stellar mass probability density function.  In a few cases
(SST24J~142648.9+332927, SMMJ030227.73+000653.5, SMMJ123600.15+621047.2,
SMMJ123606.85+621021.4, SMMJ131239.14+424155.7, SMMJ163631.47+405546.9,
SMMJ221735.15+001537.2, SMMJ221804.42+002154.4), no acceptable model was found
within the probed region of parameter space.  In the subsequent analysis, these
systems are assumed to have a uniform stellar mass probability density function
between $10^{10} - 10^{12}~M_\sun$.

Power-law DOGs have the brightest rest-frame near-IR luminosities, with
luminosities at 3$\mu$m approaching $\nu L_\nu = 10^{12}~L_{\sun}$.  This
represents a near-IR excess of a factor of 3-5 compared to bump DOGs and SMGs.
Such an excess is an indicator of thermal emission from an obscured nuclear
source \citep[i.e., obscured AGN;][]{1978ApJ...226..550R}.  Meanwhile, bump
DOGs and SMGs have rest-frame optical and near-IR SEDs that qualitatively match
the shape of the synthezied stellar population shown in
Figures~\ref{fig:bumpdogsed}~and~\ref{fig:smgsed}.  This is consistent with the
notion that this part of the SED of these objects is dominated by stellar
light.  

Relative to their rest-frame near-IR luminosities, SMGs show a rest-frame UV
excess compared to bump DOGs and power-law DOGs.  This is likely the result of
a selection effect, but the physical implications are unclear.  Possible
explanations include a difference in dust obscuration or in the luminosity
weighted-age of the stellar population.  Resolving this issue may require deep,
high spatial resolution imaging of SMGs in the rest-frame UV, optical, and
near-IR \citep[currently, only UV and optical imaging is available and only for
a handful of sources; e.g.][]{2003ApJ...596L...5C,2010MNRAS.405..234S}.  



    \tabletypesize{\tiny}
\clearpage
\begin{deluxetable*}{lccccc}
    \tablecolumns{6}
    \tablewidth{0pt}
    \tablecaption{Basic DOG Spectroscopic Sample Properties}
    \tablehead{
    \colhead{ID} &
    \colhead{R.A. (J2000)} &
    \colhead{Dec. (J2000)} &
    \colhead{$z$\tablenotemark{a}} &
    \colhead{Bump/Power-law} &
    \colhead{$R-[24]$}
    }
    \startdata
SST24 J142538.2+351855 &  216.4089050 &   35.3156586 & 2.26  &  Power-law & $ > $ 15.6 \\
SST24 J142541.3+342420 &  216.4219513 &   34.4056931 & 2.194  &  Power-law &   14.7 \\
SST24 J142554.9+341820 &  216.4792328 &   34.3057480 & 4.412  &  Power-law &   15.5 \\
SST24 J142607.8+330425 &  216.5326385 &   33.0739212 & 2.092  &  Power-law &   14.4 \\
SST24 J142622.0+345249 &  216.5918884 &   34.8804398 & 2.00  &  Bump      &   15.0 \\
SST24 J142626.4+344731 &  216.6102295 &   34.7919617 & 2.13  &  Power-law & $ > $ 15.7 \\
SST24 J142637.3+333025 &  216.6558075 &   33.5071220 & 3.200  &  Power-law & $ > $ 14.7 \\
SST24 J142644.3+333051 &  216.6846313 &   33.5143967 & 3.312  &  Power-law &   14.9 \\
SST24 J142645.7+351901 &  216.6904144 &   35.3169899 & 1.75  &  Power-law & $ > $ 16.3 \\
SST24 J142648.9+332927 &  216.7039337 &   33.4908333 & 2.00  &  Power-law &   15.7 \\
SST24 J142652.5+345506 &  216.7188568 &   34.9181824 & 1.91  &  Bump      &   15.0 \\
SST24 J142653.2+330221 &  216.7218781 &   33.0391388 & 1.86  &  Power-law &   15.8 \\
SST24 J142724.9+350824 &  216.8541260 &   35.1399765 & 1.70  &  Bump      & $ > $ 14.8 \\
SST24 J142748.4+344851 &  216.9518738 &   34.8142471 & 2.200  &  Power-law &   14.6 \\
SST24 J142759.8+351243 &  216.9991150 &   35.2118530 & 2.100  &  Power-law & $ > $ 15.4 \\
SST24 J142800.6+350455 &  217.0028992 &   35.0819473 & 2.223  &  Power-law &   14.7 \\
SST24 J142804.1+332135 &  217.0172119 &   33.3596916 & 2.34  &  Bump      & $ > $ 15.8 \\
SST24 J142810.5+352509 &  217.0439453 &   35.4192238 & 1.845  &  Power-law &   14.8 \\
SST24 J142814.2+352245 &  217.0593109 &   35.3795052 & 2.387  &  Power-law &   14.2 \\
SST24 J142815.4+324720 &  217.0640869 &   32.7887993 & 2.021  &  Power-law &   15.1 \\
SST24 J142827.9+334550 &  217.1163635 &   33.7639198 & 2.772  &  Power-law &   15.4 \\
SST24 J142832.4+340849 &  217.1351166 &   34.1473694 & 1.84  &  Bump      &   13.8 \\
SST24 J142842.9+342409 &  217.1790771 &   34.4030418 & 2.180  &  Power-law &   15.1 \\
SST24 J142846.6+352701 &  217.1942139 &   35.4504471 & 1.727  &  Bump      & $ > $ 15.3 \\
SST24 J142901.5+353016 &  217.2565460 &   35.5044174 & 1.789  &  Power-law & $ > $ 14.7 \\
SST24 J142920.1+333023 &  217.3341827 &   33.5063858 & 2.02  &  Bump      &   14.0 \\
SST24 J142924.8+353320 &  217.3533783 &   35.5559425 & 2.73  &  Power-law & $ > $ 15.9 \\
SST24 J142928.5+350841 &  217.3685455 &   35.1448898 & 1.855  &  Bump      & $ > $ 14.4 \\
SST24 J142931.3+321828 &  217.3808136 &   32.3076057 & 2.20  &  Power-law & $ > $ 15.7 \\
SST24 J142934.2+322213 &  217.3932343 &   32.3701096 & 2.278  &  Power-law &   15.2 \\
SST24 J142941.0+340915 &  217.4209595 &   34.1542397 & 1.90  &  Bump      & $ > $ 14.7 \\
SST24 J142951.1+342042 &  217.4629822 &   34.3447685 & 1.77  &  Bump      & $ > $ 14.7 \\
SST24 J142958.3+322615 &  217.4930878 &   32.4376068 & 2.64  &  Power-law &   15.6 \\
SST24 J143001.9+334538 &  217.5076904 &   33.7603149 & 2.46  &  Power-law &   16.2 \\
SST24 J143020.4+330344 &  217.5855865 &   33.0622444 & 1.482  &  Bump      & $ > $ 15.2 \\
SST24 J143022.5+330029 &  217.5941925 &   33.0080185 & 3.15  &  Power-law & $ > $ 15.7 \\
SST24 J143025.7+342957 &  217.6072998 &   34.4992828 & 2.545  &  Power-law &   15.4 \\
SST24 J143028.5+343221 &  217.6188049 &   34.5392456 & 2.178  &  Power-law &   15.1 \\
SST24 J143102.2+325152 &  217.7593689 &   32.8645210 & 2.00  &  Power-law & $ > $ 15.8 \\
SST24 J143109.7+342802 &  217.7908020 &   34.4673615 & 2.10  &  Power-law &   15.7 \\
SST24 J143135.2+325456 &  217.8971863 &   32.9158325 & 1.48  &  Power-law &   14.7 \\
SST24 J143137.1+334501 &  217.9042053 &   33.7503319 & 1.77  &  Bump      &   14.8 \\
SST24 J143152.3+350030 &  217.9683838 &   35.0082169 & 1.52  &  Bump      &   14.6 \\
SST24 J143201.8+340408 &  218.0076141 &   34.0688477 & 1.857  &  Power-law &   14.5 \\
SST24 J143216.8+335231 &  218.0702515 &   33.8754730 & 1.76  &  Bump      & $ > $ 14.8 \\
SST24 J143225.3+334716 &  218.1057739 &   33.7878914 & 2.00  &  Power-law & $ > $ 15.9 \\
SST24 J143242.5+342232 &  218.1771698 &   34.3757019 & 2.16  &  Power-law & $ > $ 15.5 \\
SST24 J143251.8+333536 &  218.2159729 &   33.5932732 & 1.78  &  Power-law & $ > $ 15.3 \\
SST24 J143312.7+342011 &  218.3028564 &   34.3364716 & 2.119  &  Power-law &   15.3 \\
SST24 J143315.1+335628 &  218.3133240 &   33.9411583 & 1.766  &  Power-law &   14.2 \\
SST24 J143318.8+332203 &  218.3284149 &   33.3674889 & 2.175  &  Power-law &   14.6 \\
SST24 J143321.8+342502 &  218.3410492 &   34.4173508 & 2.09  &  Bump      &   14.2 \\
SST24 J143324.3+334239 &  218.3508911 &   33.7109337 & 1.93  &  Bump      &   14.3 \\
SST24 J143325.8+333736 &  218.3575897 &   33.6268959 & 1.90  &  Power-law &   15.4 \\
SST24 J143330.0+342234 &  218.3752289 &   34.3762436 & 2.082 &  Power-law &   15.2 \\
SST24 J143331.9+352027 &  218.3831787 &   35.3409195 & 1.92 &  Bump      &   14.3 \\
SST24 J143332.5+332230 &  218.3855133 &   33.3750801 & 2.778 &  Bump      & $ > $ 15.3 \\
SST24 J143335.9+334716 &  218.3996735 &   33.7877769 & 2.355 &  Power-law &   14.5 \\
SST24 J143349.5+334601 &  218.4567871 &   33.7671394 & 1.87 &  Bump      & $ > $ 14.7 \\
SST24 J143353.7+343155 &  218.4738007 &   34.5321503 & 1.406 &  Bump      &   14.0 \\
SST24 J143358.0+332607 &  218.4916382 &   33.4355431 & 2.414 &  Power-law & $ > $ 16.5 \\
SST24 J143407.4+343242 &  218.5311125 &   34.5451361 & 3.791 &  Bump      & $ > $ 15.7 \\
SST24 J143410.6+332641 &  218.5445557 &   33.4447975 & 2.263 &  Power-law &   14.1 \\
SST24 J143411.0+331733 &  218.5457833 &   33.2924194 & 2.656  &  Power-law &   13.8 \\
SST24 J143424.4+334543 &  218.6019135 &   33.7619972 & 2.263 &  Power-law & $ > $ 15.2 \\
SST24 J143447.7+330230 &  218.6988373 &   33.0417976 & 1.78 &  Power-law & $ > $ 17.0 \\
SST24 J143458.9+333437 &  218.7454834 &   33.5770416 & 2.150 &  Bump      &   14.2 \\
SST24 J143502.9+342658 &  218.7622208 &   34.4496611 & 2.10 &  Bump      &   14.2 \\
SST24 J143503.2+340243 &  218.7635042 &   34.0454417 & 1.97 &  Bump      &   15.3 \\
SST24 J143504.1+354743 &  218.7672272 &   35.7955055 & 2.13 &  Power-law &   16.2 \\
SST24 J143508.4+334739 &  218.7854614 &   33.7942467 & 2.10 &  Power-law &   15.3 \\
SST24 J143509.7+340137 &  218.7904500 &   34.0269583 & 2.080 &  Power-law &   14.6 \\
SST24 J143518.8+340427 &  218.8285065 &   34.0741196 & 1.996 &  Bump      &   13.9 \\
SST24 J143520.7+340602 &  218.8361969 &   34.1007767 & 1.730 &  Bump      &   13.8 \\
SST24 J143520.7+340418 &  218.8364868 &   34.0716324 & 1.790 &  Power-law &   15.8 \\
SST24 J143523.9+330706 &  218.8497772 &   33.1186829 & 2.59 &  Power-law &   15.3 \\
SST24 J143539.3+334159 &  218.9140167 &   33.6998062 & 2.62 &  Power-law & $ > $ 16.8 \\
SST24 J143545.1+342831 &  218.9378204 &   34.4752998 & 2.50 &  Bump      & $ > $ 16.0 \\
SST24 J143631.8+350210 &  219.1326141 &   35.0360146 & 1.689 &  Bump      &   15.0 \\
SST24 J143632.7+350515 &  219.1362610 &   35.0877495 & 1.743 &  Power-law &   14.3 \\
SST24 J143634.3+334854 &  219.1430206 &   33.8151054 & 2.267 &  Power-law &   14.9 \\
SST24 J143641.0+350207 &  219.1708542 &   35.0353083 & 1.948 &  Bump      &   14.0 \\
SST24 J143641.6+342752 &  219.1735382 &   34.4644394 & 2.752 &  Power-law &   14.9 \\
SST24 J143644.2+350627 &  219.1842804 &   35.1075211 & 1.95 &  Power-law &   15.6 \\
SST24 J143701.9+344630 &  219.2582875 &   34.7751167 & 3.04 &  Bump      & $ > $ 15.6 \\
SST24 J143725.1+341502 &  219.3548889 &   34.2506104 & 2.50 &  Power-law & $ > $ 16.2 \\
SST24 J143740.1+341102 &  219.4176636 &   34.1841354 & 2.197 &  Power-law &   14.5 \\
SST24 J143742.5+341424 &  219.4276276 &   34.2403145 & 1.901 &  Power-law &   15.0 \\
SST24 J143808.3+341016 &  219.5347443 &   34.1708908 & 2.50 &  Power-law &   15.5 \\
SST24 J143816.6+333700 &  219.5695038 &   33.6167984 & 1.84 &  Bump      &   14.5 \\
\tablenotetext{a}{Redshifts are from either {\it Spitzer}/IRS
\citep[two-decimal point precision;][]{2005ApJ...622L.105H, 2006ApJ...653..101W}
or Keck (three decimal point precision; Soifer et al., in prep.) spectroscopy.
}
    \enddata                                         
    \label{tab:positions}                            
\end{deluxetable*}                                    
                                                     
\clearpage                                           
                                                     
    \tabletypesize{\tiny}                            
\begin{deluxetable*}{lccccccccccc}                    
    \tablecolumns{12}                                
    \tablewidth{0pt}                                 
    \tablecaption{Optical, Near-IR, and Mid-IR Photometry of
    DOGs.  All flux densities given in units of $\mu$Jy.  Note that all measurements and their uncertainties
    are reported, regardless of whether the measurement is statistically
    significant.}                                    
    \tablehead{                                      
    \colhead{ID} &                                   
    \colhead{$F_{B_W}$ } &
    \colhead{$F_{R}$ } &
    \colhead{$F_{I}$ } &
    \colhead{$F_{J}$ } &
    \colhead{$F_{H}$ } &
    \colhead{$F_{Ks}$ } &
    \colhead{$F_{3.6\mu {\rm m}}$ } &
    \colhead{$F_{4.5\mu {\rm m}}$ } &
    \colhead{$F_{5.8\mu {\rm m}}$ } &
    \colhead{$F_{8.0\mu {\rm m}}$ } &
    \colhead{$F_{24\mu {\rm m}}$ }
    }
    \startdata
SST24 J142538.2+351855 & -0.05$\pm$0.04 &  0.05$\pm$0.10 & -0.01$\pm$0.11 &   1.4$\pm$ 0.8 &   2.3$\pm$ 1.3 &   9.2$\pm$ 2.5 &  19.4$\pm$2.4 &  26.7$\pm$ 3.4 &  30.9$\pm$10.2 &  44.0$\pm$ 8.0 &  850$\pm$ 85 \\
SST24 J142541.3+342420 &  0.19$\pm$0.06 &  0.37$\pm$0.15 &  0.46$\pm$0.09 &   0.9$\pm$ 0.8 &   1.7$\pm$ 1.4 &   2.1$\pm$ 2.5 &  15.0$\pm$2.3 &  30.5$\pm$ 3.7 &  80.9$\pm$14.6 & 164.5$\pm$13.0 &  670$\pm$ 67 \\
SST24 J142554.9+341820 &  0.19$\pm$0.04 &  0.30$\pm$0.13 &  0.54$\pm$0.12 &   1.2$\pm$ 0.8 &   2.2$\pm$ 1.3 &   3.8$\pm$ 3.0 &   9.4$\pm$1.9 &  13.7$\pm$ 2.6 &  11.2$\pm$ 7.9 &  51.2$\pm$ 9.2 & 1140$\pm$114 \\
SST24 J142607.8+330425 &  0.12$\pm$0.05 &  0.38$\pm$0.05 &  0.73$\pm$0.12 &   3.9$\pm$ 0.7 &  10.7$\pm$ 1.6 &  14.8$\pm$ 2.8 &  32.0$\pm$3.3 &  44.3$\pm$ 4.5 &  75.8$\pm$13.8 & 131.1$\pm$12.3 &  540$\pm$ 54 \\
SST24 J142622.0+345249 &  0.44$\pm$0.05 &  0.52$\pm$0.12 &  0.63$\pm$0.16 &   0.8$\pm$ 1.0 &  -4.3$\pm$ 3.2 &   5.2$\pm$ 2.6 &   4.3$\pm$1.4 &   4.1$\pm$ 1.7 &   0.0$\pm$ 5.6 &  37.0$\pm$ 7.7 & 1290$\pm$129 \\
SST24 J142626.4+344731 &  0.04$\pm$0.05 &  0.00$\pm$0.13 & -0.15$\pm$0.21 &  -0.5$\pm$ 0.9 &   2.3$\pm$ 1.3 &   7.0$\pm$ 2.8 &  18.3$\pm$2.7 &  25.2$\pm$ 3.4 &  39.8$\pm$12.1 &  39.3$\pm$ 8.3 & 1170$\pm$117 \\
SST24 J142637.3+333025 &  0.10$\pm$0.05 &  0.14$\pm$0.18 &  0.28$\pm$0.11 &  -0.9$\pm$ 0.6 &   2.0$\pm$ 1.5 &   2.1$\pm$ 4.3 &   4.4$\pm$1.4 &  11.9$\pm$ 2.5 &  34.8$\pm$11.3 &  89.1$\pm$11.1 &  640$\pm$ 64 \\
SST24 J142644.3+333051 &  0.08$\pm$0.06 &  0.52$\pm$0.19 &  0.91$\pm$0.10 &   3.2$\pm$ 0.8 &   4.2$\pm$ 2.1 &  21.8$\pm$ 5.4 &  62.3$\pm$4.6 &  93.1$\pm$ 6.3 & 164.4$\pm$19.8 & 384.9$\pm$18.7 & 1140$\pm$114 \\
SST24 J142645.7+351901 &  0.04$\pm$0.03 &  0.07$\pm$0.07 &  0.30$\pm$0.13 &   2.2$\pm$ 0.7 &   5.0$\pm$ 1.1 &   5.3$\pm$ 1.5 &  32.5$\pm$3.4 &  52.7$\pm$ 4.8 &  84.3$\pm$14.7 & 156.5$\pm$12.5 & 1140$\pm$114 \\
SST24 J142648.9+332927 &  0.34$\pm$0.06 &  0.52$\pm$0.21 &  0.83$\pm$0.11 &   2.2$\pm$ 0.6 &   4.1$\pm$ 1.8 &   3.2$\pm$ 5.0 &  57.4$\pm$4.5 & 180.4$\pm$ 8.8 & 497.8$\pm$33.1 & 952.7$\pm$28.6 & 2330$\pm$233 \\
SST24 J142652.5+345506 &  0.09$\pm$0.04 &  0.24$\pm$0.11 &  0.30$\pm$0.18 &   1.0$\pm$ 0.7 &   4.5$\pm$ 1.6 &   2.3$\pm$ 1.6 &  22.0$\pm$0.7 &  30.0$\pm$ 1.1 &  28.0$\pm$ 5.9 &  22.9$\pm$ 6.8 &  598$\pm$ 50 \\
SST24 J142653.2+330221 &  0.10$\pm$0.04 &  0.18$\pm$0.08 &  0.42$\pm$0.16 &   0.8$\pm$ 0.8 &   5.3$\pm$ 1.4 &   1.7$\pm$ 2.9 &  19.2$\pm$2.6 &  29.6$\pm$ 3.7 &  34.5$\pm$11.2 &  64.5$\pm$ 9.2 &  880$\pm$ 88 \\
SST24 J142724.9+350824 &  0.09$\pm$0.04 &  0.15$\pm$0.13 &  0.52$\pm$0.23 &   3.7$\pm$ 1.1 &   8.7$\pm$ 3.0 &   4.3$\pm$ 2.5 &  43.6$\pm$3.6 &  57.4$\pm$ 4.6 &  72.3$\pm$12.9 &  65.1$\pm$ 9.1 &  510$\pm$ 51 \\
SST24 J142748.4+344851 &  1.66$\pm$0.06 &  1.26$\pm$0.13 &  0.80$\pm$0.27 &   2.7$\pm$ 0.8 &   5.7$\pm$ 1.3 &   5.9$\pm$ 2.7 &  15.4$\pm$2.4 &  50.5$\pm$ 4.8 & 162.6$\pm$20.2 & 473.0$\pm$20.8 & 2210$\pm$221 \\
SST24 J142759.8+351243 &  0.34$\pm$0.04 &  0.40$\pm$0.21 &  0.47$\pm$0.32 &   2.9$\pm$ 0.6 &   5.9$\pm$ 1.0 &   6.5$\pm$ 1.4 &  48.5$\pm$4.7 &  78.6$\pm$ 6.9 & 181.1$\pm$23.6 & 333.9$\pm$21.0 & 1540$\pm$154 \\
SST24 J142800.6+350455 &  0.40$\pm$0.05 &  0.51$\pm$0.14 &  0.70$\pm$0.26 &   4.7$\pm$ 0.8 &  12.6$\pm$ 1.1 &  12.4$\pm$ 1.7 &  57.2$\pm$4.4 &  85.9$\pm$ 6.1 & 163.8$\pm$19.4 & 300.2$\pm$16.5 &  920$\pm$ 92 \\
SST24 J142804.1+332135 &  0.00$\pm$0.05 & -0.01$\pm$0.09 & -0.14$\pm$0.14 &  -1.1$\pm$ 0.6 &   2.6$\pm$ 2.0 &   1.5$\pm$ 4.7 &   5.6$\pm$1.5 &   8.5$\pm$ 2.1 &   0.0$\pm$ 7.0 &   9.0$\pm$ 7.1 &  850$\pm$ 85 \\
SST24 J142810.5+352509 &  0.14$\pm$0.03 &  0.32$\pm$0.11 &  0.73$\pm$0.09 &   3.3$\pm$ 1.0 &   6.6$\pm$ 1.4 &   9.9$\pm$ 2.6 &  27.3$\pm$3.1 &  39.7$\pm$ 4.1 &  66.4$\pm$12.9 & 125.2$\pm$11.8 &  650$\pm$ 65 \\
SST24 J142814.2+352245 &  0.20$\pm$0.03 &  0.50$\pm$0.11 &  0.87$\pm$0.10 &   3.3$\pm$ 0.9 &   6.3$\pm$ 1.5 &   9.0$\pm$ 2.2 &  30.1$\pm$3.2 &  57.4$\pm$ 4.9 & 107.1$\pm$16.3 & 182.1$\pm$13.4 &  570$\pm$ 57 \\
SST24 J142815.4+324720 &  0.33$\pm$0.04 &  0.51$\pm$0.07 &  0.85$\pm$0.14 &   2.0$\pm$ 1.5 &   4.1$\pm$ 2.0 &   8.5$\pm$ 3.0 &  19.6$\pm$2.5 &  24.5$\pm$ 3.2 &  47.0$\pm$10.8 &  86.3$\pm$11.5 & 1400$\pm$140 \\
SST24 J142827.9+334550 &  0.20$\pm$0.04 &  0.22$\pm$0.09 &  0.37$\pm$0.14 &   2.4$\pm$ 1.2 &  10.2$\pm$ 1.6 &  18.7$\pm$ 2.7 &  51.0$\pm$4.2 &  79.8$\pm$ 5.9 & 153.0$\pm$19.1 & 292.0$\pm$17.1 &  770$\pm$ 77 \\
SST24 J142832.4+340849 &  0.29$\pm$0.02 &  0.68$\pm$0.15 &  1.17$\pm$0.14 &   4.3$\pm$ 1.1 &   5.7$\pm$ 1.3 &   8.5$\pm$ 2.3 &  35.9$\pm$3.5 &  43.7$\pm$ 4.3 &  49.8$\pm$11.6 &  34.5$\pm$ 7.8 &  520$\pm$ 52 \\
SST24 J142842.9+342409 &  1.12$\pm$0.06 &  1.23$\pm$0.17 &  2.66$\pm$0.15 &  13.3$\pm$ 1.6 &  16.4$\pm$ 1.5 &  24.4$\pm$ 2.9 & 126.2$\pm$5.2 & 200.7$\pm$ 7.8 & 393.4$\pm$26.6 & 695.7$\pm$23.8 & 3110$\pm$311 \\
SST24 J142846.6+352701 &  0.10$\pm$0.05 &  0.18$\pm$0.12 &  0.31$\pm$0.15 &   3.3$\pm$ 1.6 &   7.1$\pm$ 1.4 &  10.9$\pm$ 1.7 &  42.1$\pm$3.8 &  68.6$\pm$ 5.4 & 120.0$\pm$17.1 & 169.9$\pm$13.2 &  750$\pm$ 75 \\
SST24 J142901.5+353016 &  0.39$\pm$0.04 &  0.24$\pm$0.12 &  0.70$\pm$0.13 &   2.8$\pm$ 1.5 &   3.4$\pm$ 1.5 &   7.0$\pm$ 1.6 &  25.3$\pm$3.0 &  50.5$\pm$ 4.7 &  94.1$\pm$15.4 & 194.9$\pm$13.9 &  440$\pm$ 44 \\
SST24 J142920.1+333023 &  0.22$\pm$0.06 &  0.53$\pm$0.09 &  0.68$\pm$0.10 &   3.3$\pm$ 0.7 &   3.3$\pm$ 1.4 &   8.1$\pm$ 2.8 &  19.1$\pm$2.7 &  24.8$\pm$ 3.5 &  36.6$\pm$11.6 &  16.2$\pm$ 8.7 &  510$\pm$ 51 \\
SST24 J142924.8+353320 &  0.08$\pm$0.04 &  0.09$\pm$0.10 &  0.15$\pm$0.08 &  -0.2$\pm$ 1.8 &   0.4$\pm$ 2.5 &  -0.2$\pm$ 2.1 &   6.1$\pm$1.6 &  10.7$\pm$ 2.3 &  21.5$\pm$ 8.7 &  71.1$\pm$10.6 & 1040$\pm$104 \\
SST24 J142928.5+350841 & -0.01$\pm$0.07 &  0.14$\pm$0.14 & -0.02$\pm$0.29 &   2.3$\pm$ 0.8 &   2.9$\pm$ 2.3 &   3.5$\pm$ 2.1 &  27.2$\pm$2.9 &  32.6$\pm$ 3.6 &  29.6$\pm$10.7 &  30.0$\pm$ 8.2 &  410$\pm$ 41 \\
SST24 J142931.3+321828 & -0.13$\pm$0.07 & -0.09$\pm$0.12 &  0.39$\pm$0.24 &   0.0$\pm$ 0.0 &   0.0$\pm$ 0.0 &   0.0$\pm$ 0.0 &   9.8$\pm$1.9 &  12.7$\pm$ 2.5 &  23.0$\pm$10.1 &  65.3$\pm$ 8.9 & 1060$\pm$106 \\
SST24 J142934.2+322213 &  0.61$\pm$0.04 &  0.39$\pm$0.06 &  0.57$\pm$0.12 &   0.9$\pm$ 0.8 &   6.5$\pm$ 1.9 &  11.2$\pm$ 3.9 &  18.3$\pm$2.5 &  29.6$\pm$ 3.8 &  75.5$\pm$14.8 & 152.5$\pm$14.2 & 1160$\pm$116 \\
SST24 J142941.0+340915 &  0.05$\pm$0.04 & -0.07$\pm$0.16 &  0.29$\pm$0.11 &   2.1$\pm$ 1.2 &   3.7$\pm$ 1.6 &  11.2$\pm$ 2.6 &  31.4$\pm$3.2 &  42.1$\pm$ 4.2 &  47.9$\pm$11.5 &  41.5$\pm$ 8.4 &  590$\pm$ 59 \\
SST24 J142951.1+342042 &  0.24$\pm$0.04 &  0.26$\pm$0.16 &  0.82$\pm$0.11 &   1.6$\pm$ 1.0 &   4.8$\pm$ 1.1 &   9.5$\pm$ 2.6 &  42.6$\pm$3.4 &  54.9$\pm$ 4.3 &  60.4$\pm$12.3 &  42.8$\pm$ 7.5 &  600$\pm$ 60 \\
SST24 J142958.3+322615 &  0.20$\pm$0.04 &  0.27$\pm$0.09 &  0.31$\pm$0.13 &   1.3$\pm$ 0.9 &   0.1$\pm$ 1.5 &  10.2$\pm$ 3.4 &  28.9$\pm$3.2 &  48.0$\pm$ 4.6 & 111.2$\pm$16.5 & 219.0$\pm$14.4 & 1180$\pm$118 \\
SST24 J143001.9+334538 &  0.28$\pm$0.06 &  0.52$\pm$0.12 &  0.28$\pm$0.17 &   1.0$\pm$ 1.1 &   1.9$\pm$ 1.3 &   3.9$\pm$ 3.0 &  13.1$\pm$2.5 &  26.0$\pm$ 3.6 & 113.4$\pm$18.7 & 459.8$\pm$21.7 & 3840$\pm$384 \\
SST24 J143020.4+330344 &  0.06$\pm$0.05 &  0.17$\pm$0.09 &  0.57$\pm$0.15 &   3.7$\pm$ 0.5 &   9.8$\pm$ 1.4 &   7.9$\pm$ 2.5 &  34.9$\pm$3.6 &  44.1$\pm$ 4.5 &  54.2$\pm$12.6 &  47.1$\pm$ 9.1 &  540$\pm$ 54 \\
SST24 J143022.5+330029 &  0.01$\pm$0.05 &  0.16$\pm$0.09 &  0.05$\pm$0.11 &   1.2$\pm$ 0.8 &   6.5$\pm$ 1.8 &   8.1$\pm$ 3.2 &  39.3$\pm$3.7 &  48.0$\pm$ 4.5 &  89.1$\pm$14.8 & 196.8$\pm$13.9 &  800$\pm$ 80 \\
SST24 J143025.7+342957 &  0.46$\pm$0.04 &  0.70$\pm$0.12 &  1.15$\pm$0.13 &  -0.9$\pm$ 2.0 &   3.9$\pm$ 1.5 &   6.0$\pm$ 3.0 &  21.1$\pm$2.8 &  53.5$\pm$ 4.9 & 164.0$\pm$20.0 & 527.8$\pm$21.8 & 2470$\pm$247 \\
SST24 J143028.5+343221 &  0.35$\pm$0.06 &  0.47$\pm$0.10 &  0.66$\pm$0.14 &   3.6$\pm$ 1.1 &   6.4$\pm$ 1.5 &   7.9$\pm$ 2.5 &  28.0$\pm$3.2 &  47.6$\pm$ 4.7 & 120.9$\pm$17.0 & 288.4$\pm$16.4 & 1270$\pm$127 \\
SST24 J143102.2+325152 & -0.04$\pm$0.04 &  0.16$\pm$0.12 &  0.69$\pm$0.17 &  -2.0$\pm$ 1.1 &   0.7$\pm$ 2.2 &  -3.2$\pm$ 3.0 &   3.9$\pm$1.4 &   5.9$\pm$ 1.8 &   0.0$\pm$ 7.7 &  53.2$\pm$ 8.3 & 1190$\pm$119 \\
SST24 J143109.7+342802 &  0.02$\pm$0.05 &  0.24$\pm$0.09 &  0.23$\pm$0.17 &   2.5$\pm$ 0.9 &   3.0$\pm$ 1.3 &  15.4$\pm$ 5.5 &   7.5$\pm$1.7 &  10.1$\pm$ 2.4 &  27.4$\pm$ 9.1 &  62.6$\pm$ 9.7 & 1110$\pm$111 \\
SST24 J143135.2+325456 &  0.41$\pm$0.04 &  0.80$\pm$0.10 &  1.55$\pm$0.21 &   6.3$\pm$ 1.4 &   9.1$\pm$ 3.3 &  23.4$\pm$ 5.0 &  70.9$\pm$4.9 & 137.4$\pm$ 7.6 & 268.4$\pm$24.6 & 494.9$\pm$21.2 & 1510$\pm$151 \\
SST24 J143137.1+334501 &  0.17$\pm$0.06 &  0.28$\pm$0.12 &  0.80$\pm$0.15 &   1.5$\pm$ 1.2 &   4.7$\pm$ 1.9 &   8.2$\pm$ 3.3 &  29.4$\pm$3.0 &  40.4$\pm$ 3.9 &  43.2$\pm$11.1 &  35.6$\pm$ 8.2 &  570$\pm$ 57 \\
SST24 J143152.3+350030 &  0.14$\pm$0.03 &  0.30$\pm$0.09 &  0.66$\pm$0.10 &   4.8$\pm$ 0.8 &  10.0$\pm$ 1.2 &  14.3$\pm$ 2.9 &  49.0$\pm$4.0 &  63.1$\pm$ 5.1 &  63.3$\pm$12.7 &  51.7$\pm$ 8.9 &  520$\pm$ 52 \\
SST24 J143201.8+340408 &  0.43$\pm$0.05 &  0.43$\pm$0.18 &  1.26$\pm$0.17 &   4.8$\pm$ 1.0 &  12.6$\pm$ 1.6 &  17.5$\pm$ 2.5 &  44.8$\pm$3.9 &  72.3$\pm$ 5.5 & 121.2$\pm$16.8 & 230.3$\pm$14.7 &  670$\pm$ 67 \\
SST24 J143216.8+335231 &  0.11$\pm$0.05 &  0.23$\pm$0.12 &  0.39$\pm$0.12 &   3.4$\pm$ 0.8 &   5.6$\pm$ 1.5 &  11.9$\pm$ 2.6 &  32.4$\pm$0.7 &  41.4$\pm$ 1.1 &  46.6$\pm$ 5.7 &  42.1$\pm$ 6.5 &  502$\pm$ 44 \\
SST24 J143225.3+334716 &  0.07$\pm$0.04 &  0.13$\pm$0.12 &  0.14$\pm$0.13 &  -0.1$\pm$ 1.1 &   3.0$\pm$ 1.6 &   5.3$\pm$ 3.0 &  39.1$\pm$3.7 &  76.2$\pm$ 5.8 & 167.9$\pm$19.7 & 350.0$\pm$18.0 & 1280$\pm$128 \\
SST24 J143242.5+342232 &  0.05$\pm$0.04 &  0.20$\pm$0.12 &  0.30$\pm$0.18 &   3.4$\pm$ 1.1 &   3.9$\pm$ 1.7 &  12.4$\pm$ 2.9 &  36.6$\pm$3.6 &  59.3$\pm$ 5.2 & 127.8$\pm$18.0 & 225.0$\pm$15.1 &  910$\pm$ 91 \\
SST24 J143251.8+333536 &  0.06$\pm$0.05 &  0.13$\pm$0.13 &  0.34$\pm$0.14 &   3.3$\pm$ 0.6 &   4.2$\pm$ 1.4 &  10.9$\pm$ 2.1 &  41.5$\pm$3.7 &  55.2$\pm$ 4.8 &  69.3$\pm$13.1 & 110.4$\pm$10.9 &  820$\pm$ 82 \\
SST24 J143312.7+342011 &  0.53$\pm$0.04 &  0.57$\pm$0.13 &  0.85$\pm$0.12 &   2.9$\pm$ 1.0 &   4.7$\pm$ 1.5 &  10.6$\pm$ 2.7 &  27.9$\pm$3.2 &  35.1$\pm$ 4.0 &  65.5$\pm$13.4 & 106.3$\pm$11.5 & 1760$\pm$176 \\
SST24 J143315.1+335628 &  0.42$\pm$0.07 &  0.73$\pm$0.08 &  0.86$\pm$0.12 &   3.2$\pm$ 0.8 &   7.2$\pm$ 1.4 &  12.6$\pm$ 2.5 &  35.3$\pm$3.6 &  55.8$\pm$ 5.0 & 102.7$\pm$16.2 & 164.4$\pm$13.5 &  830$\pm$ 83 \\
SST24 J143318.8+332203 &  0.28$\pm$0.05 &  0.27$\pm$0.06 &  0.15$\pm$0.10 &   0.6$\pm$ 0.7 &   5.0$\pm$ 1.5 &  -1.4$\pm$ 2.3 &  11.5$\pm$2.0 &  18.6$\pm$ 2.8 &  31.0$\pm$ 9.4 &  56.1$\pm$ 9.1 &  430$\pm$ 43 \\
SST24 J143321.8+342502 &  0.18$\pm$0.07 &  0.50$\pm$0.09 &  0.91$\pm$0.11 &   5.0$\pm$ 0.8 &   9.1$\pm$ 1.4 &  14.4$\pm$ 2.6 &  32.8$\pm$3.3 &  41.3$\pm$ 4.2 &  56.2$\pm$12.7 &  48.5$\pm$ 9.2 &  560$\pm$ 56 \\
SST24 J143324.3+334239 &  0.24$\pm$0.06 &  0.44$\pm$0.12 &  1.04$\pm$0.10 &   4.5$\pm$ 1.2 &   7.3$\pm$ 1.4 &  13.8$\pm$ 2.5 &  41.5$\pm$3.5 &  54.0$\pm$ 4.7 &  50.4$\pm$11.2 &  52.9$\pm$ 8.8 &  530$\pm$ 53 \\
SST24 J143325.8+333736 &  0.20$\pm$0.06 &  0.55$\pm$0.10 &  0.92$\pm$0.13 &   7.8$\pm$ 0.7 &  11.2$\pm$ 1.4 &  20.7$\pm$ 2.0 &  62.0$\pm$4.6 &  81.3$\pm$ 6.0 & 118.0$\pm$16.5 & 141.3$\pm$12.1 & 1870$\pm$187 \\
SST24 J143330.0+342234 &  0.43$\pm$0.05 &  0.66$\pm$0.12 &  0.59$\pm$0.18 &   0.8$\pm$ 1.0 &   1.7$\pm$ 1.4 &   3.5$\pm$ 3.0 &   7.0$\pm$1.7 &  12.3$\pm$ 2.6 &  17.7$\pm$ 8.2 &  64.7$\pm$ 9.8 & 1920$\pm$192 \\
SST24 J143331.9+352027 &  0.18$\pm$0.03 &  0.47$\pm$0.06 &  0.78$\pm$0.10 &   3.0$\pm$ 1.5 &   5.1$\pm$ 1.6 &   8.0$\pm$ 3.9 &  26.5$\pm$3.1 &  35.4$\pm$ 4.0 &  41.4$\pm$11.0 &  26.0$\pm$ 7.7 &  600$\pm$ 60 \\
SST24 J143332.5+332230 &  0.09$\pm$0.05 &  0.14$\pm$0.07 &  0.17$\pm$0.13 &   1.0$\pm$ 0.8 &  -2.3$\pm$ 2.0 &  -1.0$\pm$ 3.4 &   4.6$\pm$1.4 &   2.4$\pm$ 1.5 &   0.0$\pm$ 7.1 &  13.5$\pm$ 7.2 &  460$\pm$ 46 \\
SST24 J143335.9+334716 &  0.36$\pm$0.06 &  0.39$\pm$0.14 &  0.49$\pm$0.12 &   1.5$\pm$ 1.1 &   1.3$\pm$ 2.7 &  14.5$\pm$ 3.9 &  30.1$\pm$3.2 &  41.6$\pm$ 4.2 &  64.5$\pm$12.7 &   0.0$\pm$ 9.2 &  590$\pm$ 59 \\
SST24 J143349.5+334601 &  0.12$\pm$0.06 &  0.26$\pm$0.14 &  0.50$\pm$0.15 &   4.5$\pm$ 1.1 &   9.2$\pm$ 1.5 &  12.0$\pm$ 3.2 &  37.2$\pm$3.8 &  42.0$\pm$ 4.9 &  62.2$\pm$13.6 &  32.0$\pm$ 8.1 &  530$\pm$ 53 \\
SST24 J143353.7+343155 &  0.33$\pm$0.04 &  0.69$\pm$0.13 &  1.10$\pm$0.17 &   8.6$\pm$ 0.8 &   9.0$\pm$ 1.5 &  14.3$\pm$ 2.5 &  32.2$\pm$3.2 &  37.6$\pm$ 4.1 &  43.6$\pm$11.7 & 100.5$\pm$10.6 &  680$\pm$ 68 \\
SST24 J143358.0+332607 &  0.05$\pm$0.04 &  0.03$\pm$0.06 &  0.04$\pm$0.09 &   1.1$\pm$ 0.7 &   5.0$\pm$ 1.6 &   4.8$\pm$ 2.3 &  13.4$\pm$2.4 &  19.2$\pm$ 3.2 &  42.2$\pm$10.9 &  88.8$\pm$10.7 & 1070$\pm$107 \\
SST24 J143407.4+343242 &  0.07$\pm$0.04 &  0.12$\pm$0.07 &  0.28$\pm$0.15 &   0.1$\pm$ 1.0 &  -0.7$\pm$ 1.8 &  -1.6$\pm$ 2.8 &   0.0$\pm$1.2 &   0.0$\pm$ 1.5 &   0.0$\pm$ 6.4 &   0.0$\pm$ 7.6 &  620$\pm$ 62 \\
SST24 J143410.6+332641 &  0.72$\pm$0.06 &  0.58$\pm$0.06 &  0.93$\pm$0.10 &   4.1$\pm$ 0.8 &  10.5$\pm$ 1.4 &  23.2$\pm$ 2.6 &  50.9$\pm$4.2 &  80.7$\pm$ 5.9 & 148.9$\pm$18.9 & 271.3$\pm$15.7 &  630$\pm$ 63 \\
SST24 J143411.0+331733 &  0.84$\pm$0.04 &  1.08$\pm$0.05 &  1.20$\pm$0.13 &   2.5$\pm$ 1.0 &   6.2$\pm$ 1.5 &   3.7$\pm$ 2.8 &  20.4$\pm$2.6 &  26.3$\pm$ 3.4 &  49.7$\pm$16.9 &  76.9$\pm$15.0 &  860$\pm$ 51 \\
SST24 J143424.4+334543 &  0.09$\pm$0.06 &  0.04$\pm$0.14 &  0.31$\pm$0.31 &   2.0$\pm$ 1.1 &   3.4$\pm$ 1.5 &   5.8$\pm$ 2.5 &  14.8$\pm$2.3 &  23.5$\pm$ 3.3 &  73.0$\pm$14.2 & 156.4$\pm$13.9 &  860$\pm$ 86 \\
SST24 J143447.7+330230 &  0.00$\pm$0.05 &  0.03$\pm$0.06 & -0.01$\pm$0.11 &   1.2$\pm$ 0.7 &   1.5$\pm$ 1.7 &   3.8$\pm$ 3.1 &  21.2$\pm$2.7 &  32.3$\pm$ 3.8 &  42.9$\pm$11.9 &  87.8$\pm$10.7 & 1710$\pm$171 \\
SST24 J143458.9+333437 &  0.20$\pm$0.04 &  0.49$\pm$0.12 &  0.66$\pm$0.11 &   4.2$\pm$ 0.8 &   5.5$\pm$ 1.4 &  13.2$\pm$ 2.3 &  40.0$\pm$3.7 &  48.6$\pm$ 4.6 &  60.5$\pm$13.0 &  53.9$\pm$ 8.4 &  570$\pm$ 57 \\
SST24 J143502.9+342658 &  0.28$\pm$0.04 &  0.43$\pm$0.13 &  0.46$\pm$0.15 &   2.5$\pm$ 1.1 &   2.9$\pm$ 1.6 &  11.6$\pm$ 2.4 &  44.7$\pm$3.4 &  47.2$\pm$ 4.3 &  46.2$\pm$12.5 &  44.0$\pm$ 8.4 &  500$\pm$ 50 \\
SST24 J143503.2+340243 &  0.03$\pm$0.06 &  0.23$\pm$0.11 &  0.38$\pm$0.12 &   2.6$\pm$ 1.1 &   6.5$\pm$ 1.6 &  11.3$\pm$ 2.8 &  34.3$\pm$3.5 &  46.2$\pm$ 4.6 &  54.5$\pm$13.0 &  45.1$\pm$ 9.0 &  760$\pm$ 76 \\
SST24 J143504.1+354743 &  0.01$\pm$0.04 &  0.17$\pm$0.07 &  0.02$\pm$0.10 &   0.0$\pm$ 0.0 &   0.0$\pm$ 0.0 &   0.0$\pm$ 0.0 &  21.0$\pm$2.7 &  33.8$\pm$ 4.0 &  50.8$\pm$12.1 &  86.6$\pm$10.8 & 1260$\pm$126 \\
SST24 J143508.4+334739 &  0.45$\pm$0.05 &  0.82$\pm$0.11 &  0.83$\pm$0.11 &   3.2$\pm$ 0.9 &   5.1$\pm$ 1.6 &  11.0$\pm$ 4.1 &  14.4$\pm$2.4 &  16.6$\pm$ 2.9 &  34.9$\pm$10.4 & 175.3$\pm$14.0 & 2650$\pm$265 \\
SST24 J143509.7+340137 &  0.07$\pm$0.06 &  0.28$\pm$0.11 &  0.55$\pm$0.15 &   1.2$\pm$ 0.8 &   2.5$\pm$ 1.4 &   8.4$\pm$ 2.9 &  13.1$\pm$1.9 &  15.8$\pm$ 2.5 &  32.6$\pm$ 8.8 &  53.0$\pm$ 9.9 &  470$\pm$ 47 \\
SST24 J143518.8+340427 &  0.13$\pm$0.05 &  0.45$\pm$0.11 &  0.81$\pm$0.19 &   0.9$\pm$ 1.3 &   6.0$\pm$ 1.5 &   7.5$\pm$ 2.0 &  23.4$\pm$2.8 &  31.8$\pm$ 3.8 &  53.9$\pm$12.2 &  48.2$\pm$ 8.9 &  400$\pm$ 40 \\
SST24 J143520.7+340602 &  0.40$\pm$0.04 &  0.59$\pm$0.09 &  1.03$\pm$0.13 &   2.8$\pm$ 1.2 &   8.3$\pm$ 1.8 &  11.3$\pm$ 2.1 &  29.8$\pm$3.2 &  35.1$\pm$ 4.0 &  40.5$\pm$11.1 &  25.2$\pm$ 8.2 &  490$\pm$ 49 \\
SST24 J143520.7+340418 &  0.37$\pm$0.06 &  0.30$\pm$0.13 &  0.67$\pm$0.22 &   0.0$\pm$ 1.0 &   2.0$\pm$ 1.5 &   5.7$\pm$ 2.1 &   5.8$\pm$1.5 &   7.1$\pm$ 2.0 &  15.1$\pm$ 8.5 &   7.4$\pm$ 7.5 & 1530$\pm$153 \\
SST24 J143523.9+330706 &  0.03$\pm$0.03 &  0.33$\pm$0.08 &  0.53$\pm$0.14 &   2.1$\pm$ 0.9 &   4.9$\pm$ 1.7 &   3.7$\pm$ 3.5 &  17.7$\pm$2.6 &  34.1$\pm$ 4.1 &  93.5$\pm$16.0 & 250.3$\pm$16.4 & 1090$\pm$109 \\
SST24 J143539.3+334159 &  0.10$\pm$0.07 &  0.16$\pm$0.10 &  0.39$\pm$0.16 &   1.0$\pm$ 1.1 &   0.6$\pm$ 2.4 &   3.7$\pm$ 3.8 &  14.1$\pm$2.3 &  23.9$\pm$ 3.4 &  65.8$\pm$13.6 & 249.5$\pm$15.7 & 2670$\pm$267 \\
SST24 J143545.1+342831 &  0.22$\pm$0.06 &  0.13$\pm$0.16 &  0.40$\pm$0.12 &   2.6$\pm$ 1.2 &   6.1$\pm$ 2.0 &   7.1$\pm$ 2.3 &  16.4$\pm$2.5 &  18.1$\pm$ 3.0 &  27.0$\pm$ 9.4 &  95.0$\pm$10.4 & 1960$\pm$196 \\
SST24 J143631.8+350210 &  0.00$\pm$0.05 &  0.13$\pm$0.06 & -0.16$\pm$0.14 &   1.3$\pm$ 0.8 &   2.8$\pm$ 1.2 &   7.8$\pm$ 4.3 &  25.4$\pm$2.8 &  31.5$\pm$ 3.4 &  33.0$\pm$10.0 &  20.7$\pm$ 6.6 &  330$\pm$ 33 \\
SST24 J143632.7+350515 &  1.31$\pm$0.03 &  1.39$\pm$0.08 &  1.71$\pm$0.15 &   6.9$\pm$ 0.9 &  10.5$\pm$ 1.5 &  16.7$\pm$ 3.9 &  53.2$\pm$4.2 &  92.2$\pm$ 6.2 & 172.8$\pm$20.1 & 348.1$\pm$17.9 & 1690$\pm$169 \\
SST24 J143634.3+334854 &  1.02$\pm$0.05 &  1.47$\pm$0.06 &  2.27$\pm$0.14 &  10.9$\pm$ 1.0 &  23.4$\pm$ 2.4 &  49.0$\pm$ 2.9 &  91.9$\pm$5.6 & 170.1$\pm$ 8.4 & 350.5$\pm$27.9 & 680.3$\pm$24.2 & 3280$\pm$328 \\
SST24 J143641.0+350207 &  0.29$\pm$0.04 &  0.36$\pm$0.05 &  1.56$\pm$0.18 &   2.0$\pm$ 0.9 &   7.0$\pm$ 1.4 &  15.6$\pm$ 4.5 &  20.6$\pm$2.4 &  26.0$\pm$ 3.2 &  30.6$\pm$ 9.4 &  43.4$\pm$ 8.2 &  330$\pm$ 33 \\
SST24 J143641.6+342752 &  0.30$\pm$0.03 &  0.23$\pm$0.12 &  0.44$\pm$0.11 &   1.8$\pm$ 1.4 &   5.3$\pm$ 1.6 &   8.5$\pm$ 2.5 &  23.8$\pm$2.9 &  38.8$\pm$ 4.1 &  77.9$\pm$14.0 & 162.1$\pm$13.2 &  530$\pm$ 53 \\
SST24 J143644.2+350627 &  0.39$\pm$0.04 &  0.57$\pm$0.06 &  0.79$\pm$0.16 &   3.4$\pm$ 1.0 &   6.4$\pm$ 1.5 &  10.5$\pm$ 2.5 &  37.9$\pm$3.6 & 103.3$\pm$ 6.6 & 308.7$\pm$26.4 & 734.2$\pm$25.1 & 2340$\pm$234 \\
SST24 J143701.9+344630 & -0.00$\pm$0.06 &  0.09$\pm$0.06 & -0.12$\pm$0.19 &  -0.1$\pm$ 0.9 &  -2.5$\pm$ 1.5 &   6.4$\pm$ 2.7 &  18.0$\pm$0.8 &  17.9$\pm$ 1.3 &  13.2$\pm$ 8.8 &  37.0$\pm$ 7.7 &  508$\pm$ 60 \\
SST24 J143725.1+341502 &  0.10$\pm$0.03 &  0.08$\pm$0.10 &  0.06$\pm$0.10 &   1.9$\pm$ 1.0 &   6.3$\pm$ 1.7 &  12.9$\pm$ 4.5 &  52.9$\pm$4.3 &  87.9$\pm$ 6.1 & 167.5$\pm$19.6 & 283.4$\pm$16.3 & 1410$\pm$141 \\
SST24 J143740.1+341102 &  0.43$\pm$0.04 &  0.62$\pm$0.12 &  0.50$\pm$0.12 &   5.1$\pm$ 1.0 &  13.4$\pm$ 1.7 &  21.6$\pm$ 3.9 &  52.3$\pm$4.2 &  79.8$\pm$ 5.8 & 148.0$\pm$18.9 & 236.9$\pm$15.1 &  950$\pm$ 95 \\
SST24 J143742.5+341424 &  0.31$\pm$0.05 &  0.33$\pm$0.12 &  0.61$\pm$0.14 &   3.9$\pm$ 1.0 &   8.0$\pm$ 1.8 &  17.7$\pm$ 3.8 &  32.7$\pm$3.4 &  54.2$\pm$ 4.8 &  98.0$\pm$15.8 & 172.9$\pm$13.4 &  780$\pm$ 78 \\
SST24 J143808.3+341016 &  0.30$\pm$0.04 &  0.44$\pm$0.12 &  0.72$\pm$0.17 &   2.6$\pm$ 0.9 &   7.3$\pm$ 1.8 &  13.0$\pm$ 2.6 &  35.9$\pm$3.5 &  73.2$\pm$ 5.6 & 193.7$\pm$21.1 & 411.9$\pm$19.5 & 1710$\pm$171 \\
SST24 J143816.6+333700 &  0.18$\pm$0.04 &  0.36$\pm$0.10 &  0.71$\pm$0.14 &   1.6$\pm$ 1.5 &   4.2$\pm$ 1.1 &   7.0$\pm$ 2.5 &  24.1$\pm$0.7 &  29.4$\pm$ 1.1 &  31.2$\pm$ 6.1 &  19.8$\pm$ 6.4 &  530$\pm$ 36 \\
    \enddata
    \label{tab:phot1}
\end{deluxetable*}
\clearpage

\clearpage

    \tabletypesize{\small}
\begin{deluxetable}{lcccccc}
    \tablecolumns{7}
    \tablewidth{0pt}

    \tablecaption{Median and inter-quartile $M_*$ values for
PL DOGs, bump DOGs, and SMGs using the CB07 library and Chabrier~IMF.  SFH as noted.}

    \tablehead{
    \colhead{} &
    \colhead{PL DOGs}  &
    \colhead{Bump DOGs}  &
    \colhead{SMGs}  \\
    \colhead{SFH} &
    \colhead{log$(M_*/M_{\sun})$} &
    \colhead{log$(M_*/M_{\sun})$} &
    \colhead{log$(M_*/M_{\sun})$}
    }
    \startdata
Instantaneous Burst\tablenotemark{a} & 10.71$^{+0.40}_{-0.34}$ & 10.62$^{+0.26}_{-0.32}$ & 10.42$^{+0.42}_{-0.36}$ \\
Major Merger\tablenotemark{b}	     & 10.90$^{+0.32}_{-0.30}$ & 10.74$^{+0.23}_{-0.26}$ & 10.59$^{+0.34}_{-0.36}$  \\
Major Merger\tablenotemark{c}	     & 11.06$^{+0.24}_{-0.21}$ & 10.88$^{+0.14}_{-0.13}$ & 10.86$^{+0.24}_{-0.37}$  \\
Smooth Accretion\tablenotemark{d}    & 10.94$^{+0.29}_{-0.30}$ & 10.75$^{+0.20}_{-0.25}$ & 10.64$^{+0.31}_{-0.39}$ \\
Smooth Accretion\tablenotemark{e}    & 11.20$^{+0.23}_{-0.20}$ & 11.03$^{+0.15}_{-0.14}$ & 11.02$^{+0.25}_{-0.37}$ \\
\tablenotetext{a}{Simple stellar population}
\tablenotetext{b}{SFH from \citet{2010MNRAS.407.1701N}}
\tablenotetext{c}{SFH from \citet{2010MNRAS.407.1701N} and restricting the time range to $z=2$ (i.e., the peak SFR period)}
\tablenotetext{d}{SFH from \citet{2010MNRAS.404.1355D}}
\tablenotetext{e}{SFH from \citet{2010MNRAS.404.1355D} and restricting the time
range to the $z \sim 2-3$ range}
    \enddata
    \label{tab:mstar}
\end{deluxetable}

\clearpage

%
%

%
%


\begin{figure}[!tp] 
\begin{center}
\includegraphics[height=0.75\textheight]{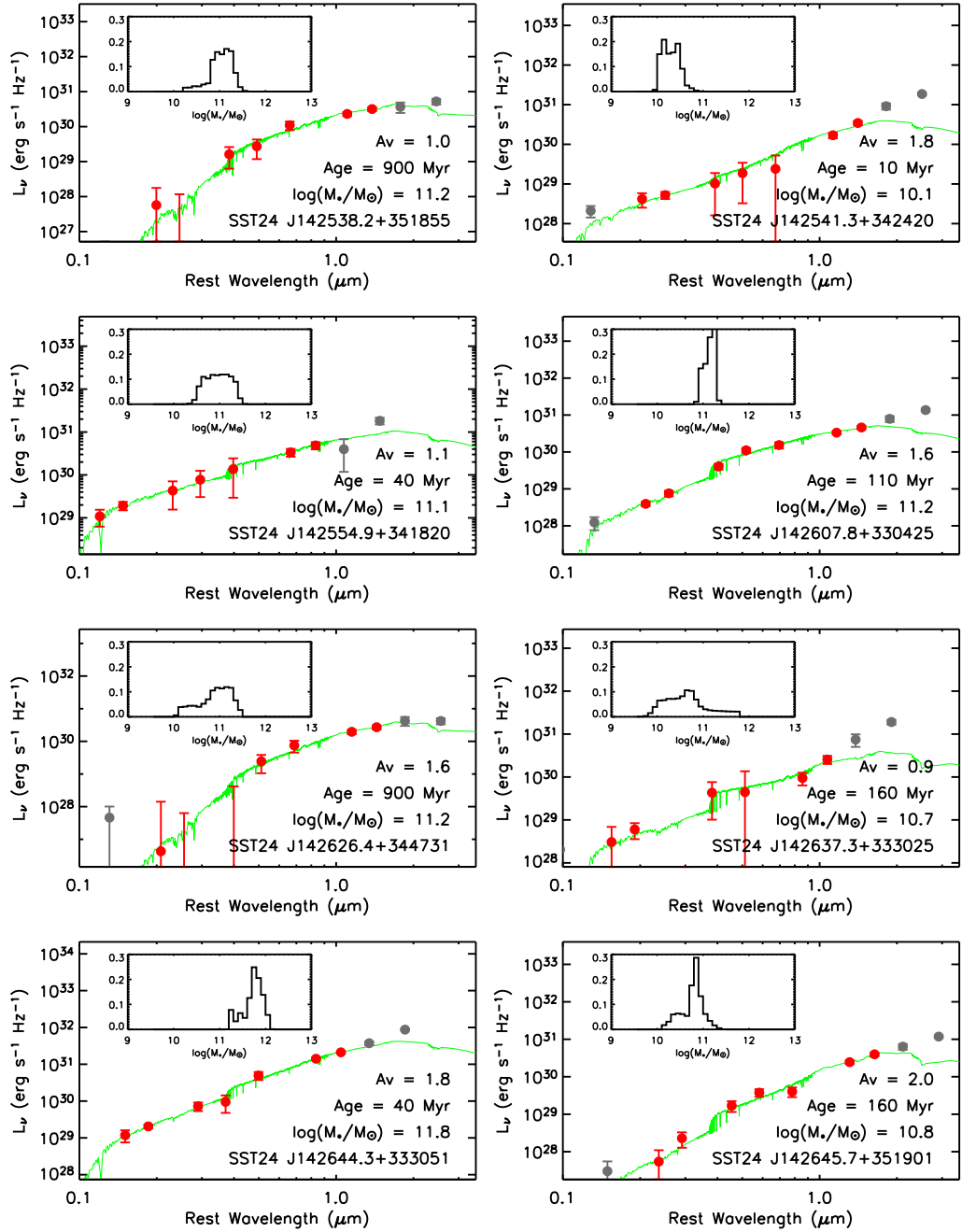}
\end{center}

\caption[Rest-frame UV through near-IR SEDs of power-law DOGs]{Luminosity per
unit frequency as a function of rest-frame wavelength for power-law DOGs (red
circles).  Gray circles indicate that $B_W$ and IRAC 5.8$\mu$m and 8.0$\mu$m
data are not used to constrain the SPS models (see
section~\ref{sec:genmethod}).  The best-fit CB07 synthesized stellar population
(assuming a Chabrier~IMF and a simple stellar population SFH) is shown in green
with the best-fit parameters printed in the bottom right of each panel.  The
inset shows the stellar mass probability density function (marginalizing over
model age and $A_V$) for each source.

\label{fig:pldogsed}}
\addtocounter{figure}{-1}
\end{figure}

\begin{figure}[!tp] 
\begin{center}
\includegraphics[height=0.8\textheight]{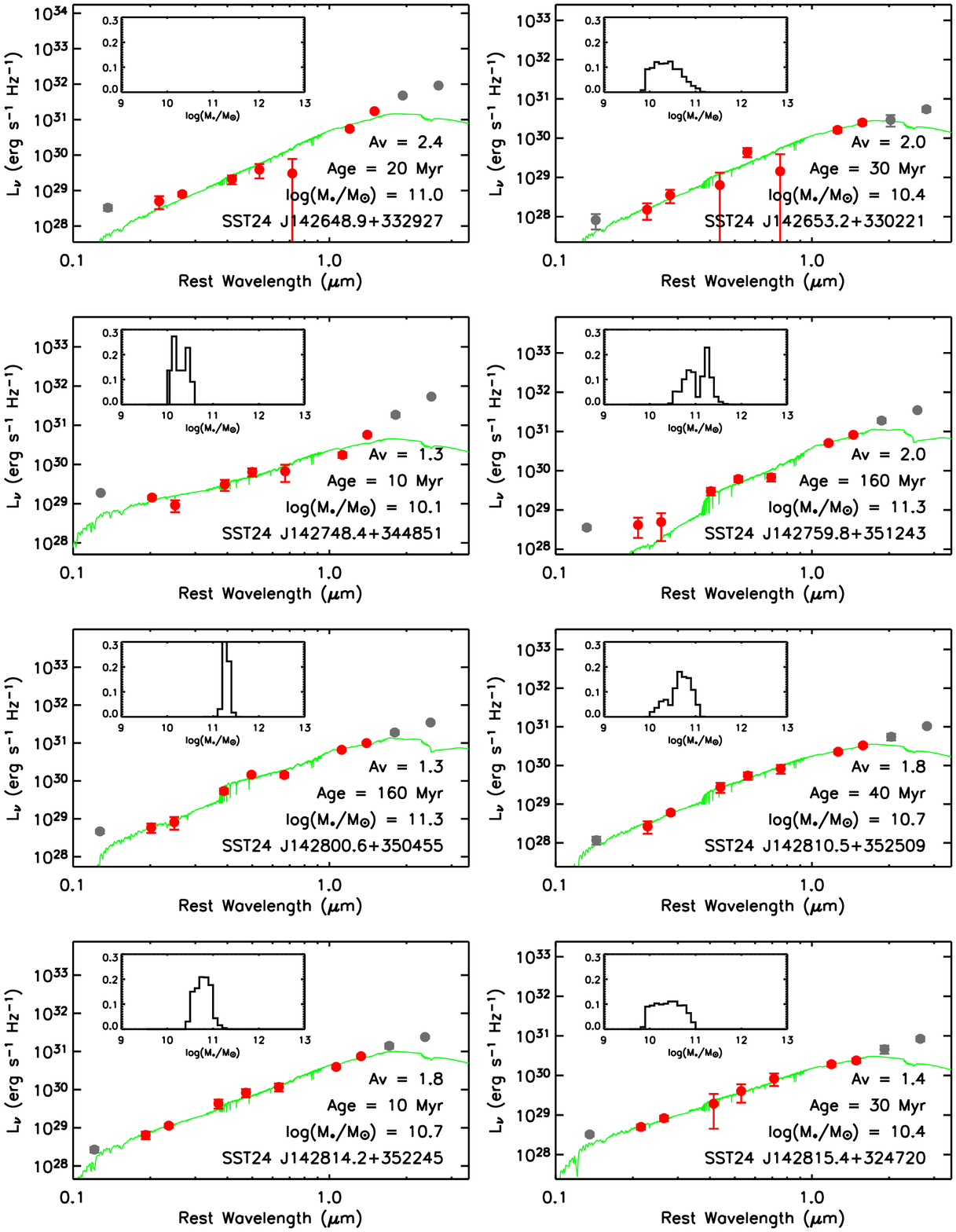}
\end{center}
\caption{Continued.}
\addtocounter{figure}{-1}
\end{figure}

\begin{figure}[!tp] 
\begin{center}
\includegraphics[height=0.8\textheight]{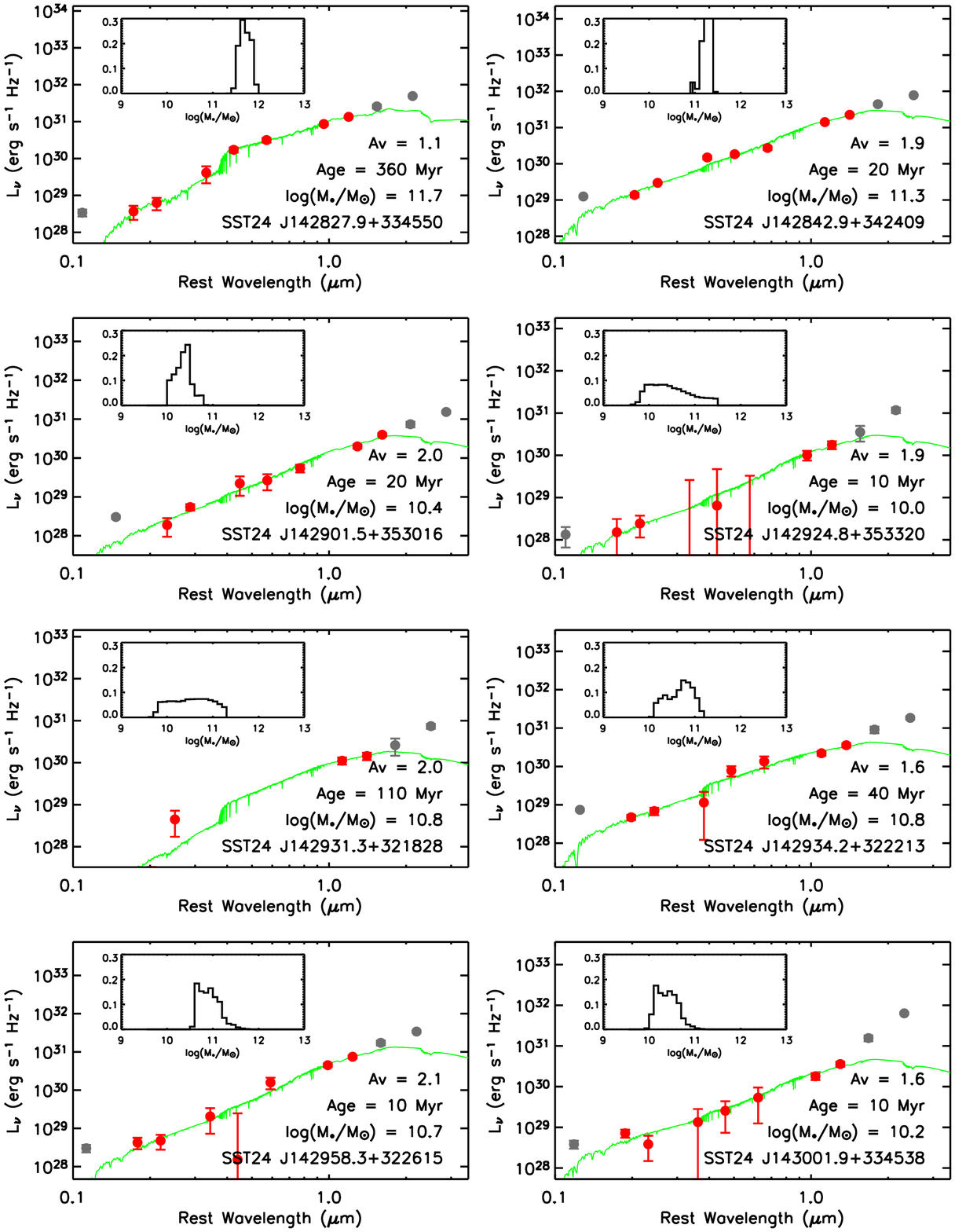}
\end{center}
\caption{Continued.}
\addtocounter{figure}{-1}
\end{figure}

\begin{figure}[!tp] 
\begin{center}
\includegraphics[height=0.8\textheight]{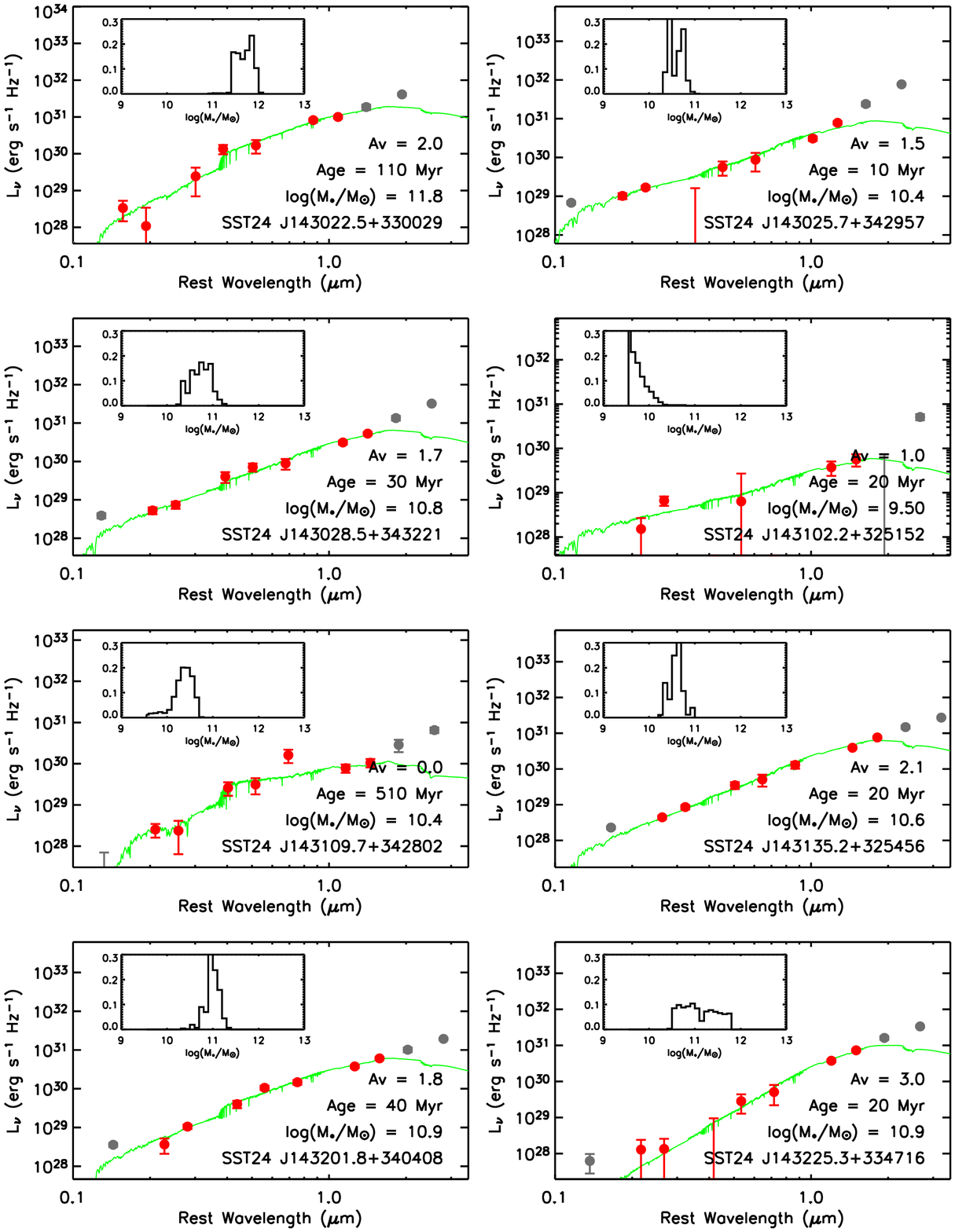}
\end{center}
\caption{Continued.}
\addtocounter{figure}{-1}
\end{figure}

\begin{figure}[!tp] 
\begin{center}
\includegraphics[height=0.8\textheight]{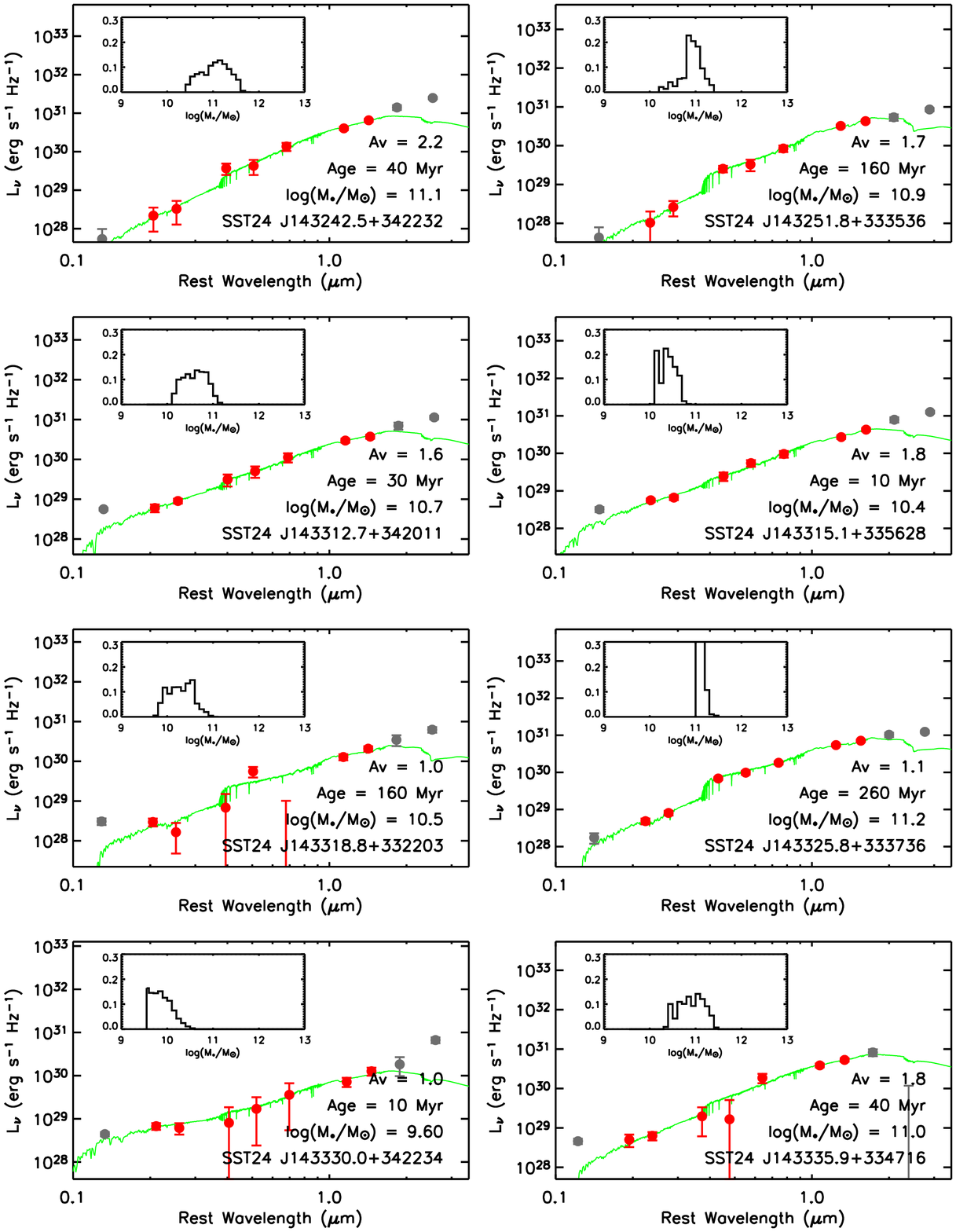}
\end{center}
\caption{Continued.}
\addtocounter{figure}{-1}
\end{figure}

\begin{figure}[!tp] 
\begin{center}
\includegraphics[height=0.8\textheight]{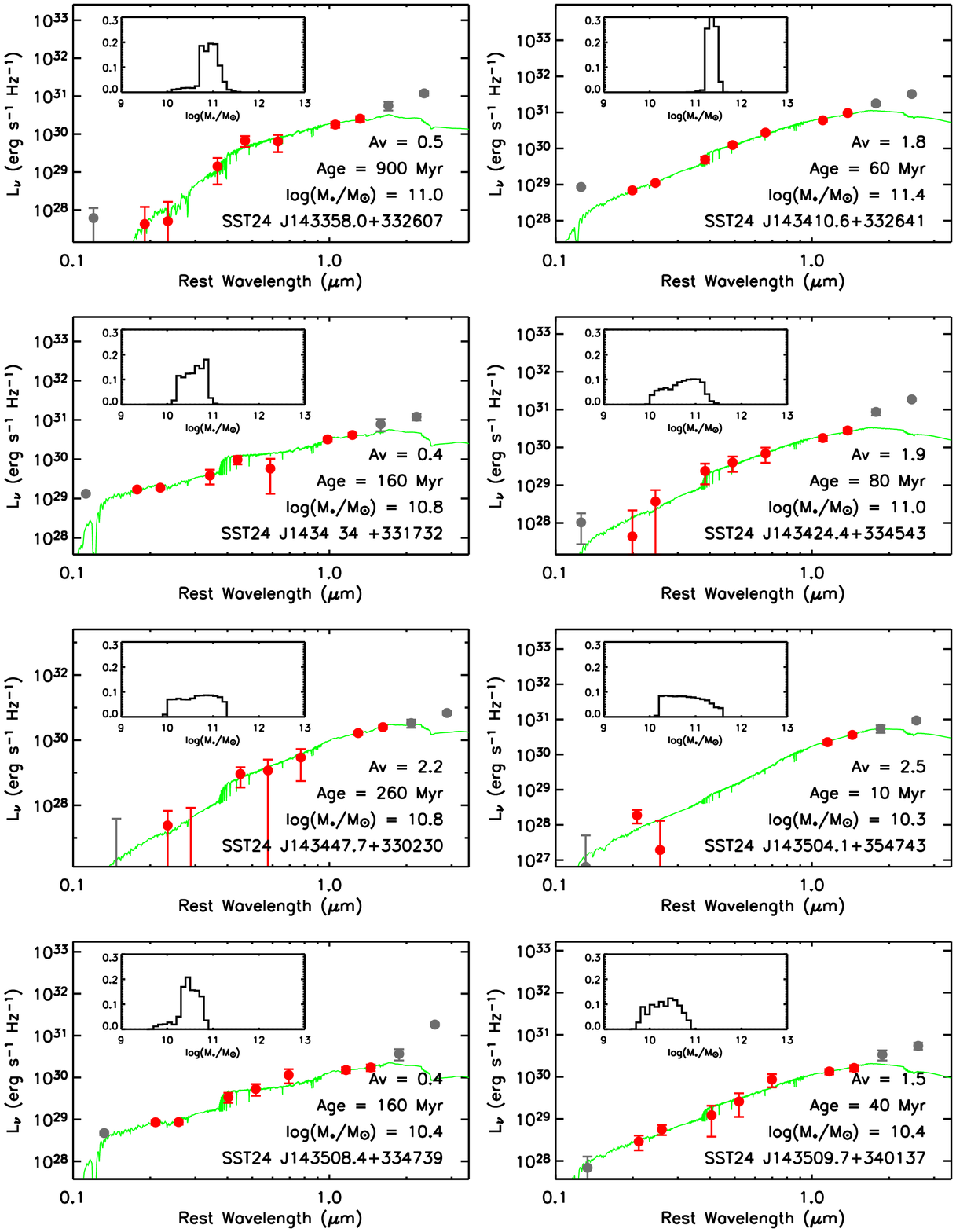}
\end{center}
\caption{Continued.}
\addtocounter{figure}{-1}
\end{figure}

\begin{figure}[!tp] 
\begin{center}
\includegraphics[height=0.8\textheight]{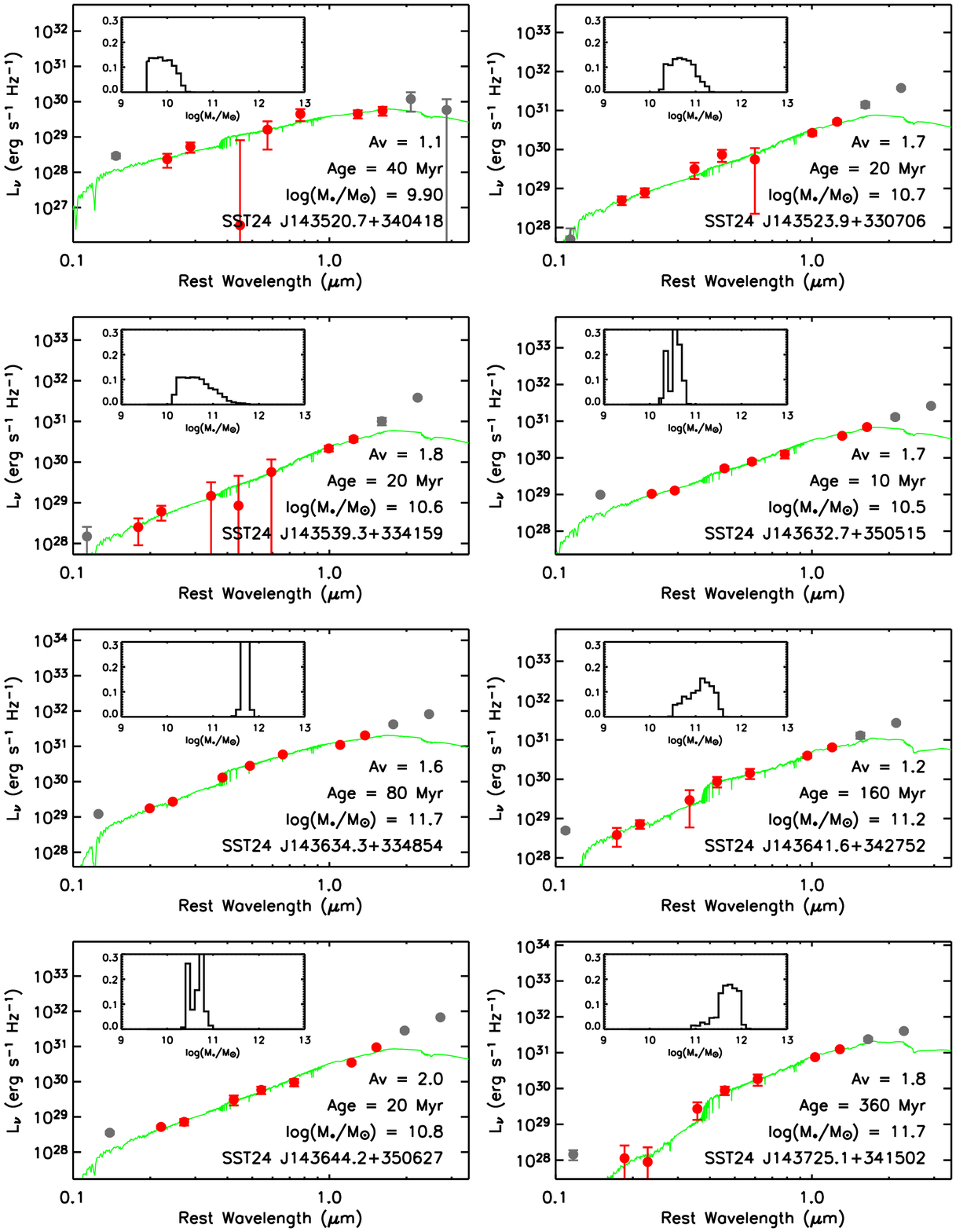}
\end{center}
\caption{Continued.}
\addtocounter{figure}{-1}
\end{figure}

\begin{figure}[!tp] 
\begin{center}
\includegraphics[height=0.8\textheight]{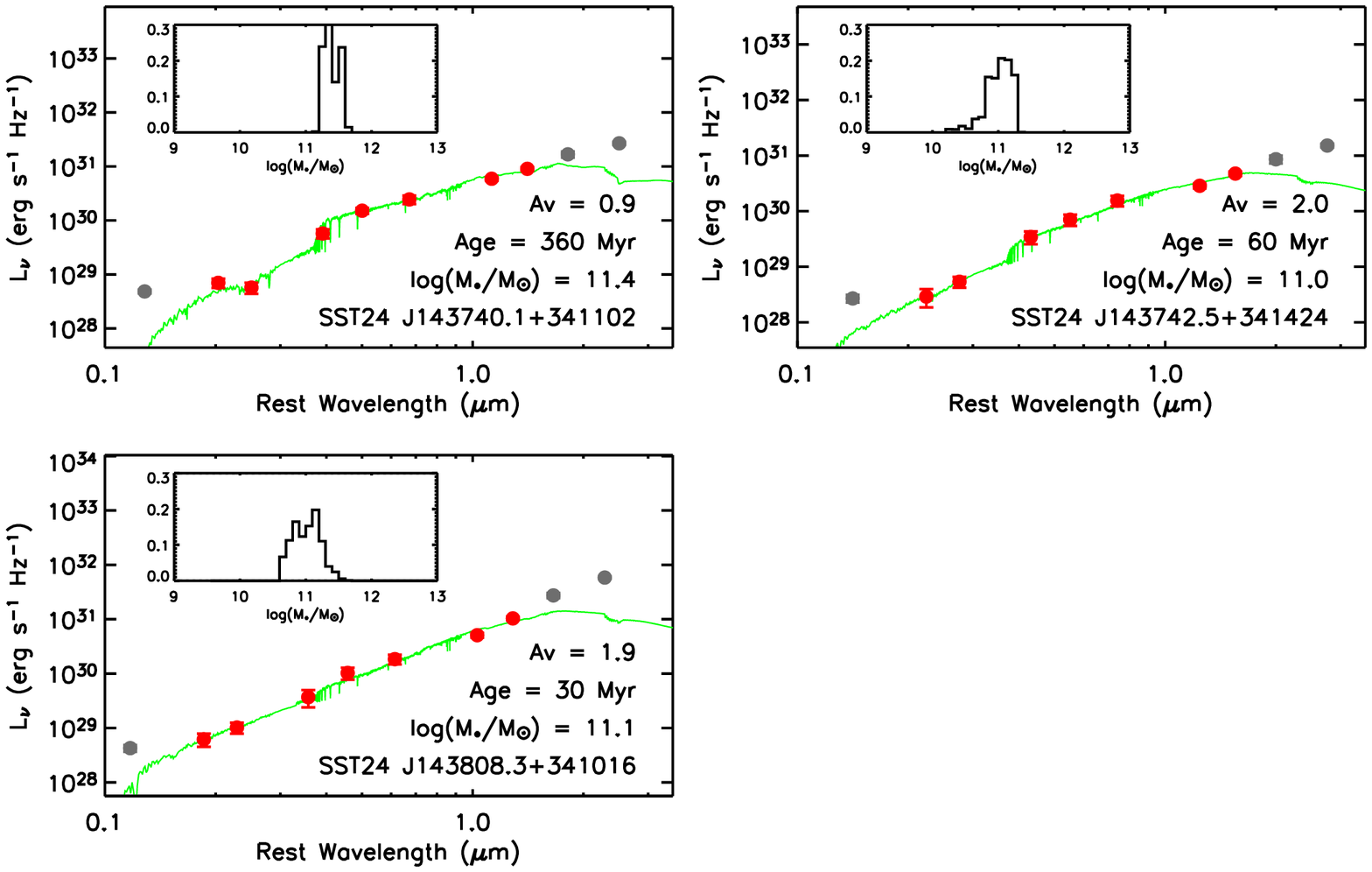}
\end{center}
\caption{Continued.}
\end{figure}

\clearpage


\begin{figure}[!tp] 
\begin{center}
\includegraphics[height=0.8\textheight]{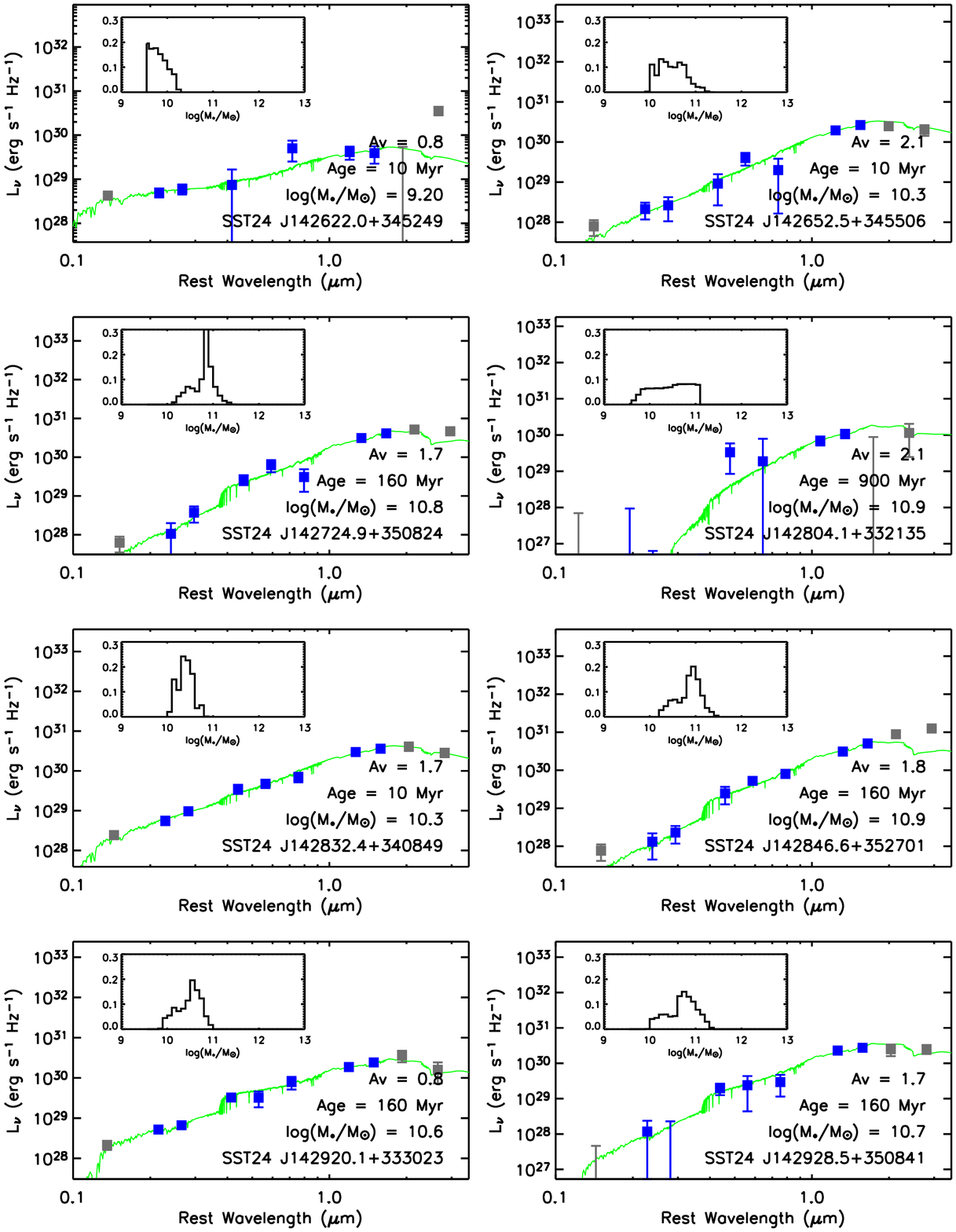}
\end{center}

\caption{ Same as Figure~\ref{fig:pldogsed}, but for bump DOGs.
\label{fig:bumpdogsed}}

\addtocounter{figure}{-1}
\end{figure}

\begin{figure}[!tp] 
\begin{center}
\includegraphics[height=0.8\textheight]{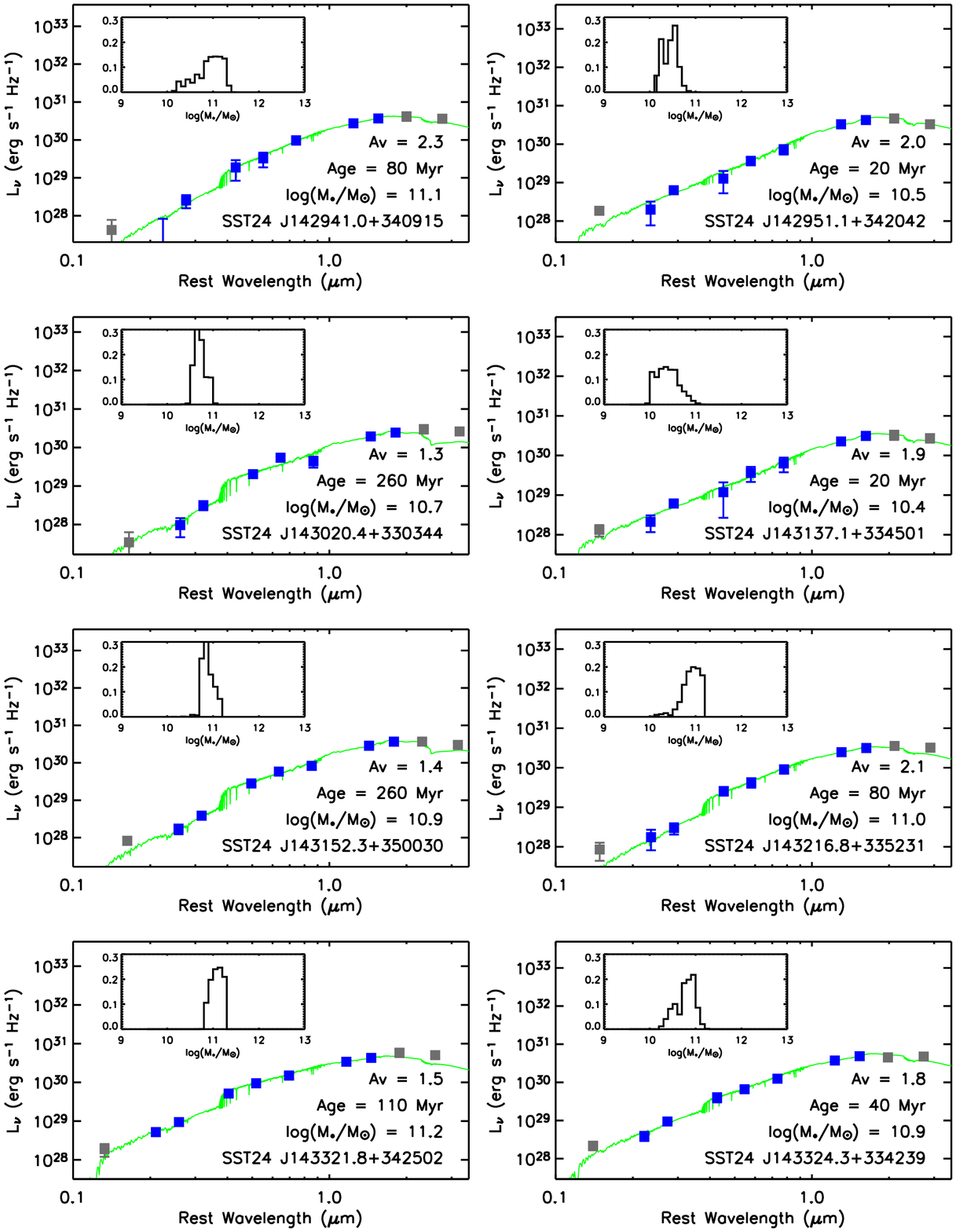}
\end{center}
\caption{Continued.}
\addtocounter{figure}{-1}
\end{figure}

\begin{figure}[!tp] 
\begin{center}
\includegraphics[height=0.8\textheight]{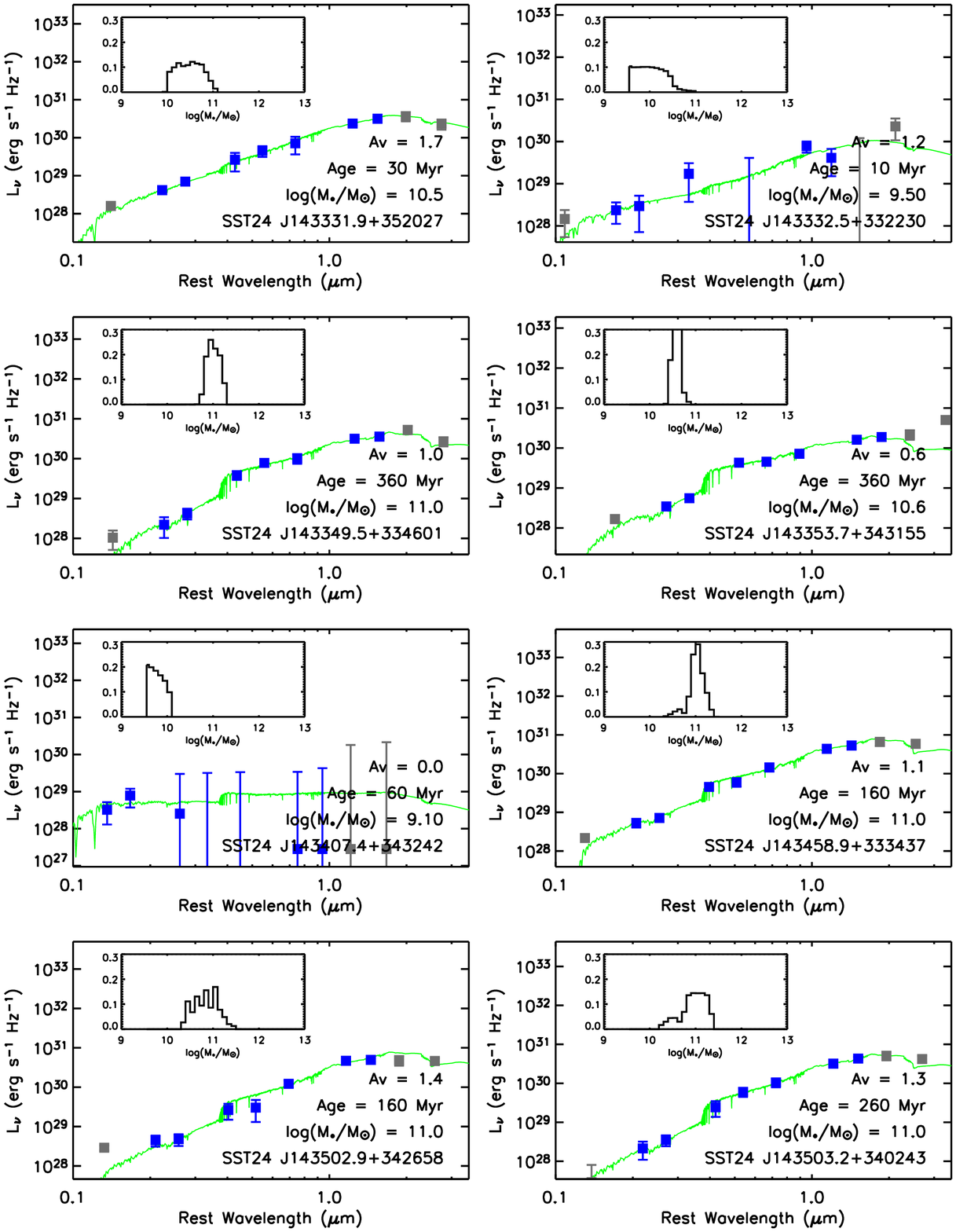}
\end{center}
\caption{Continued.}
\addtocounter{figure}{-1}
\end{figure}

\begin{figure}[!tp] 
\begin{center}
\includegraphics[height=0.8\textheight]{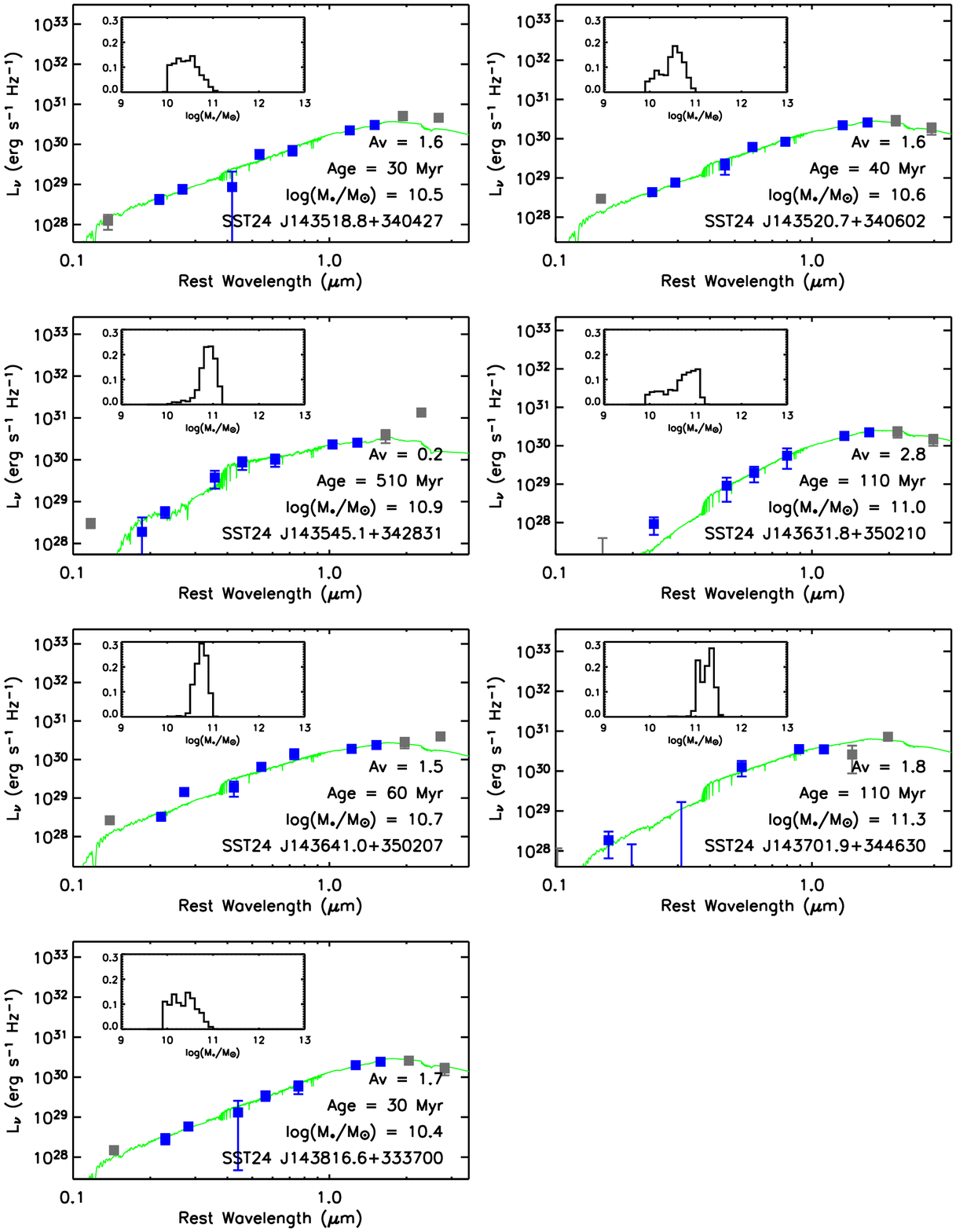}
\end{center}
\caption{Continued.}
\end{figure}

\clearpage


\begin{figure}[!tp] 
\begin{center}
\includegraphics[height=0.8\textheight]{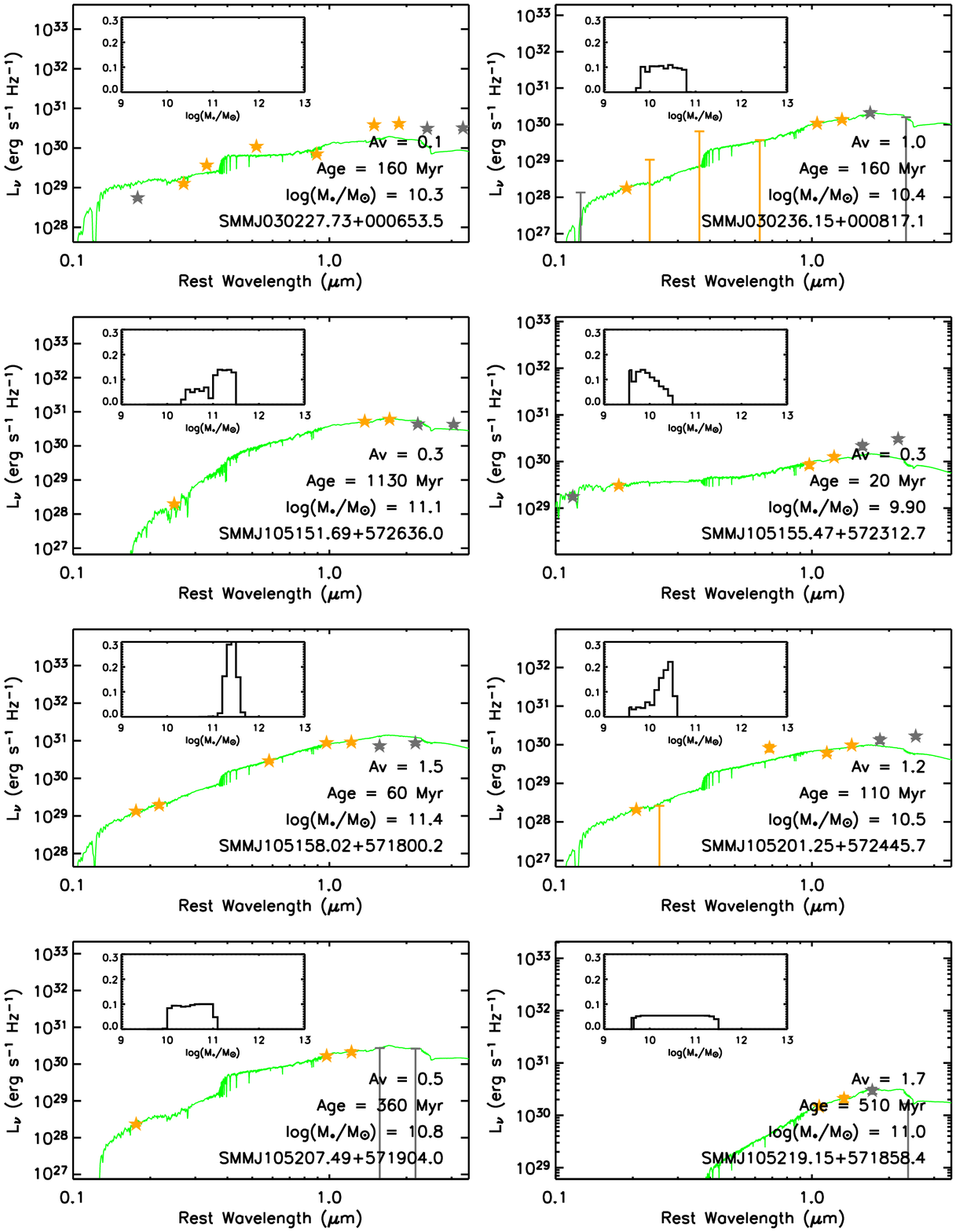}
\end{center}

\caption{ Same as Figure~\ref{fig:pldogsed}, but for SMGs.
\label{fig:smgsed}}
\addtocounter{figure}{-1}
\end{figure}

\begin{figure}[!tp] 
\begin{center}
\includegraphics[height=0.8\textheight]{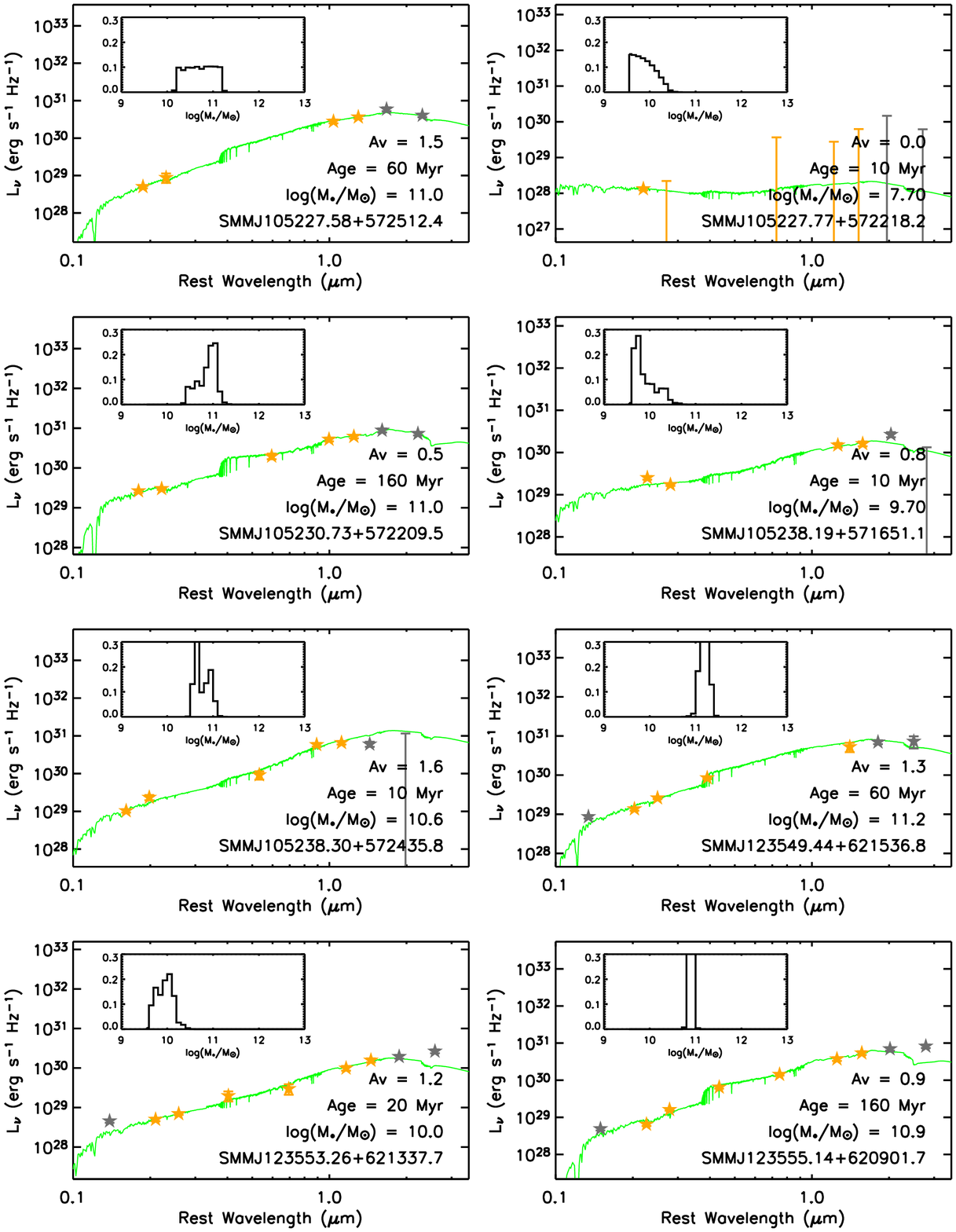}
\end{center}
\caption{Continued.}
\addtocounter{figure}{-1}
\end{figure}

\begin{figure}[!tp] 
\begin{center}
\includegraphics[height=0.8\textheight]{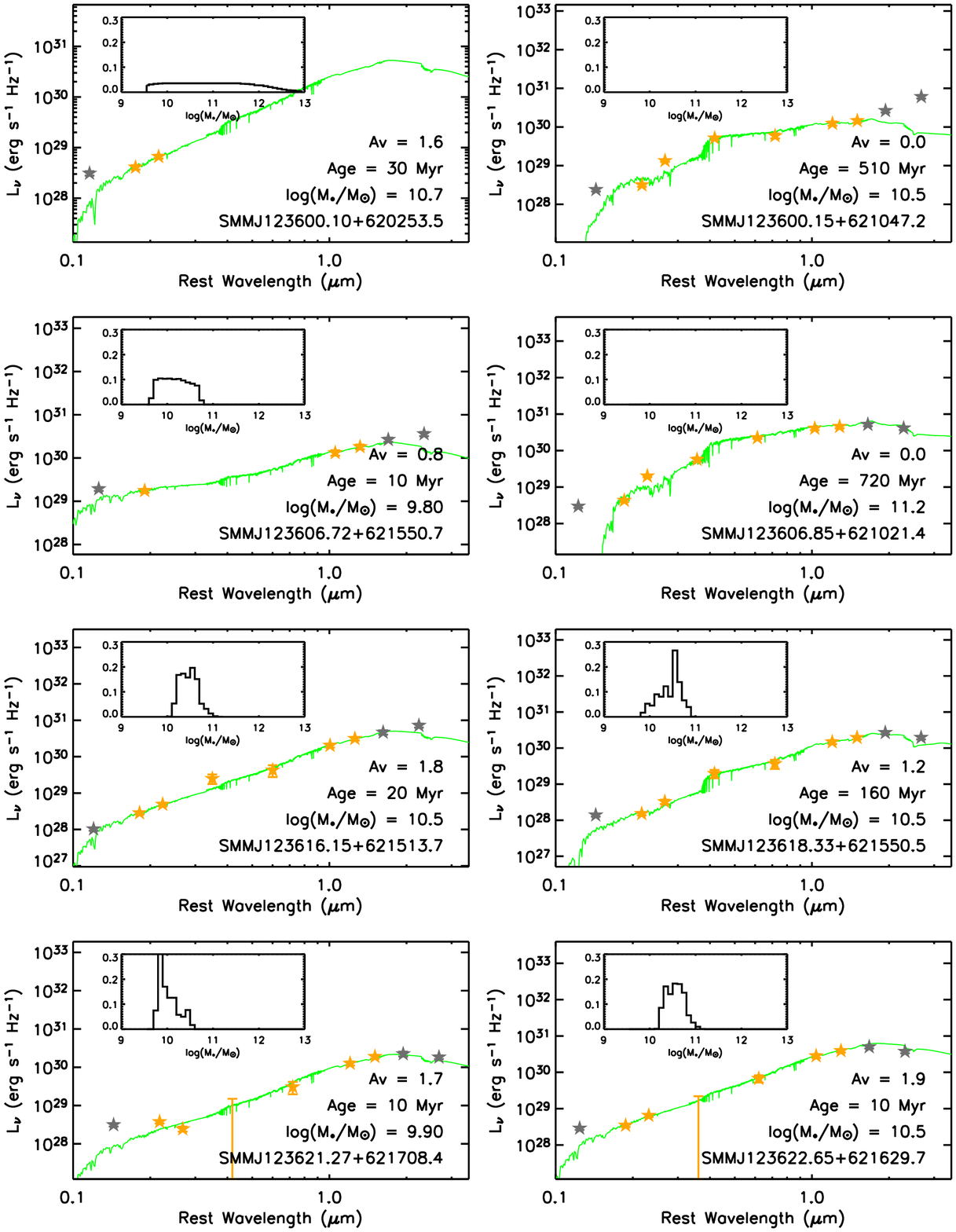}
\end{center}
\caption{Continued.}
\addtocounter{figure}{-1}
\end{figure}

\begin{figure}[!tp] 
\begin{center}
\includegraphics[height=0.8\textheight]{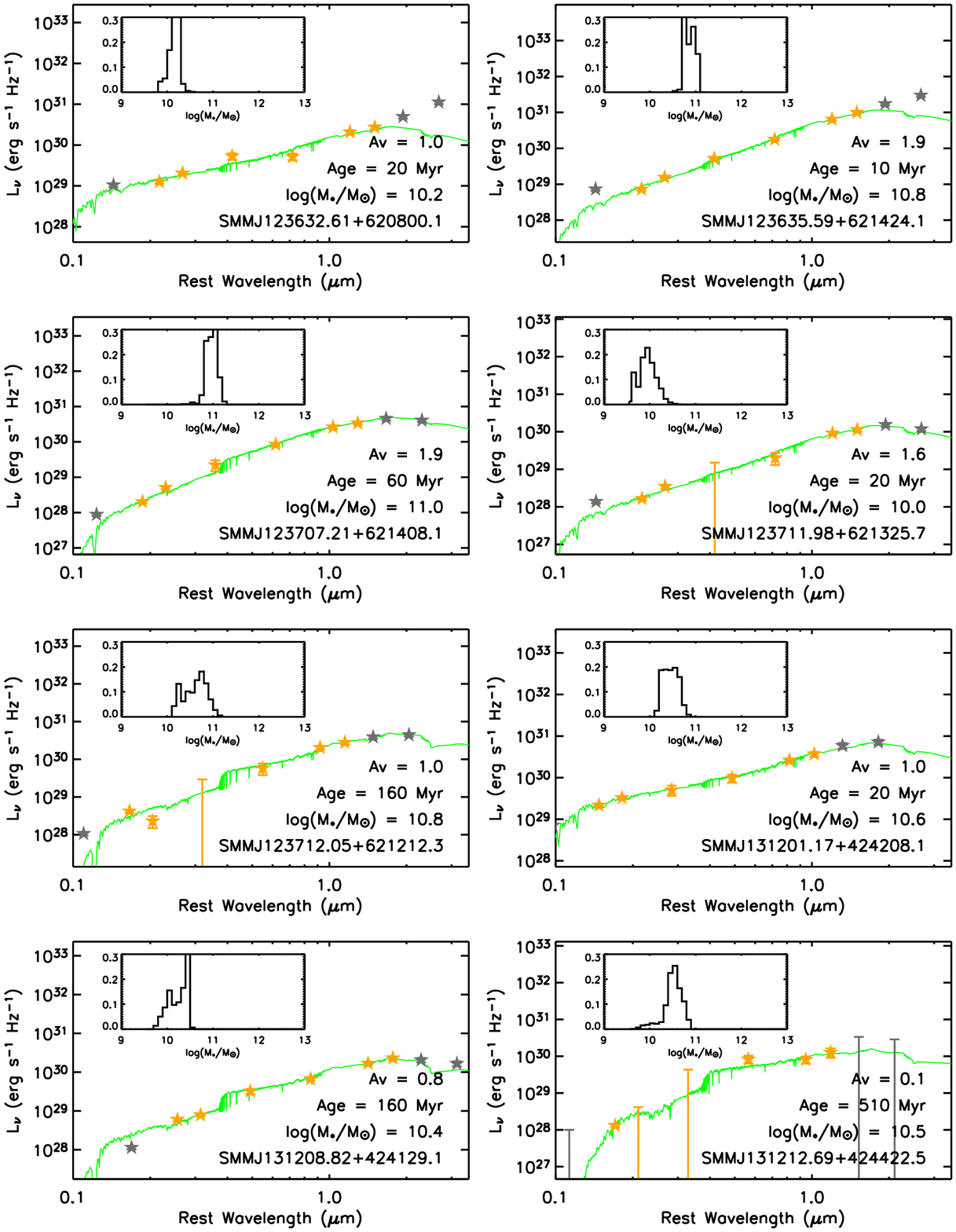}
\end{center}
\caption{Continued.}
\addtocounter{figure}{-1}
\end{figure}

\begin{figure}[!tp] 
\begin{center}
\includegraphics[height=0.8\textheight]{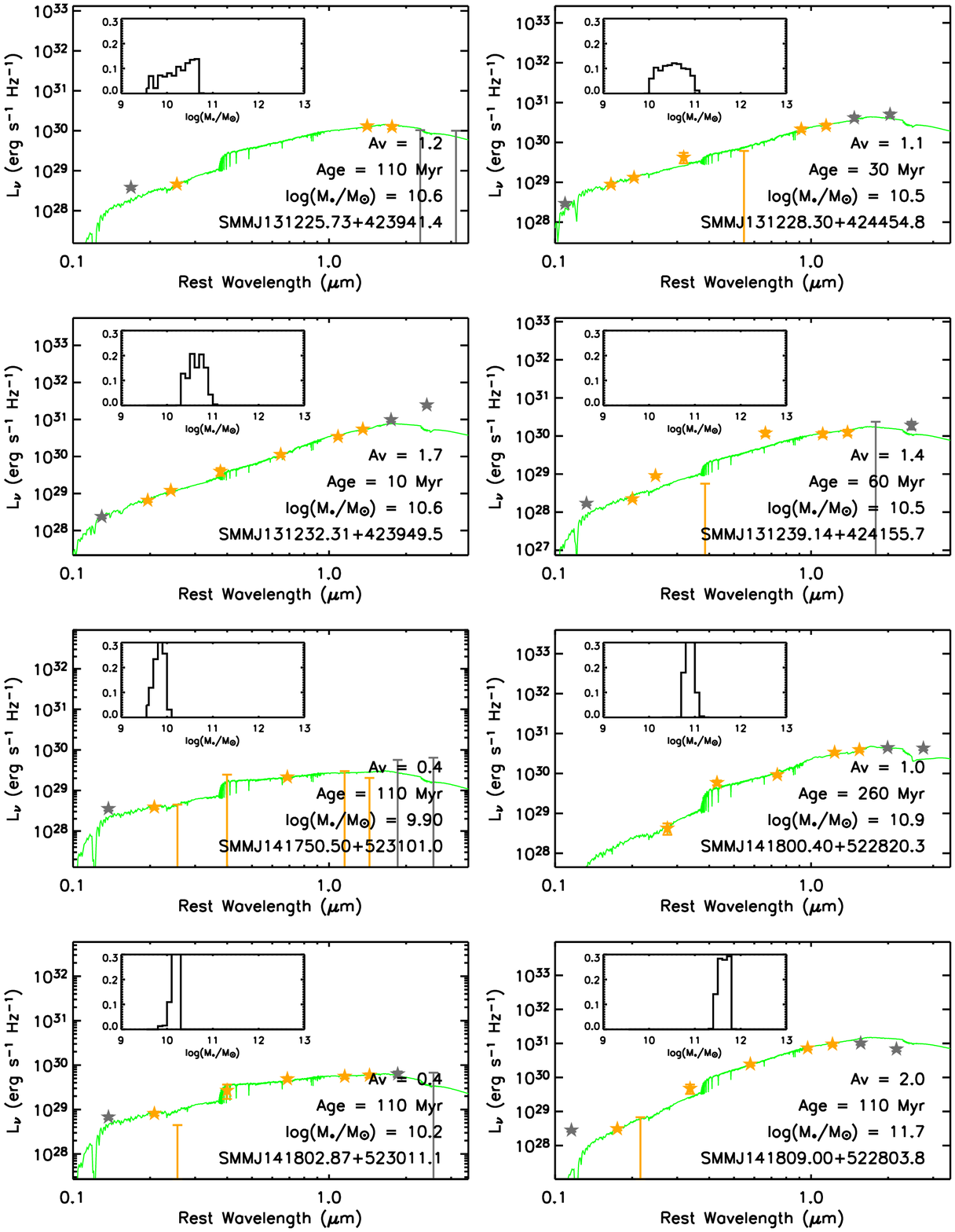}
\end{center}
\caption{Continued.}
\addtocounter{figure}{-1}
\end{figure}

\begin{figure}[!tp] 
\begin{center}
\includegraphics[height=0.8\textheight]{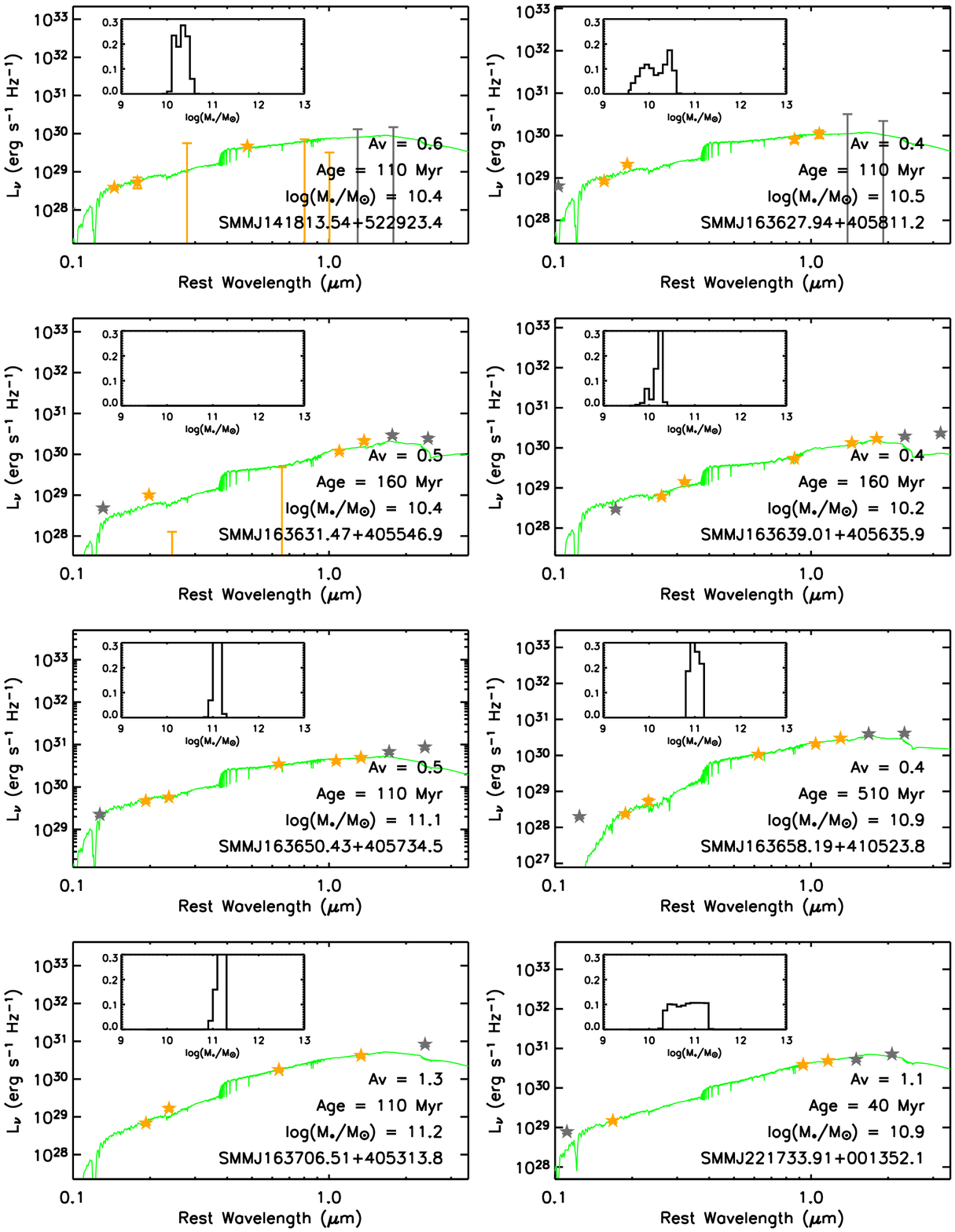}
\end{center}
\caption{Continued.}
\addtocounter{figure}{-1}
\end{figure}

\begin{figure}[!tp] 
\begin{center}
\includegraphics[height=0.8\textheight]{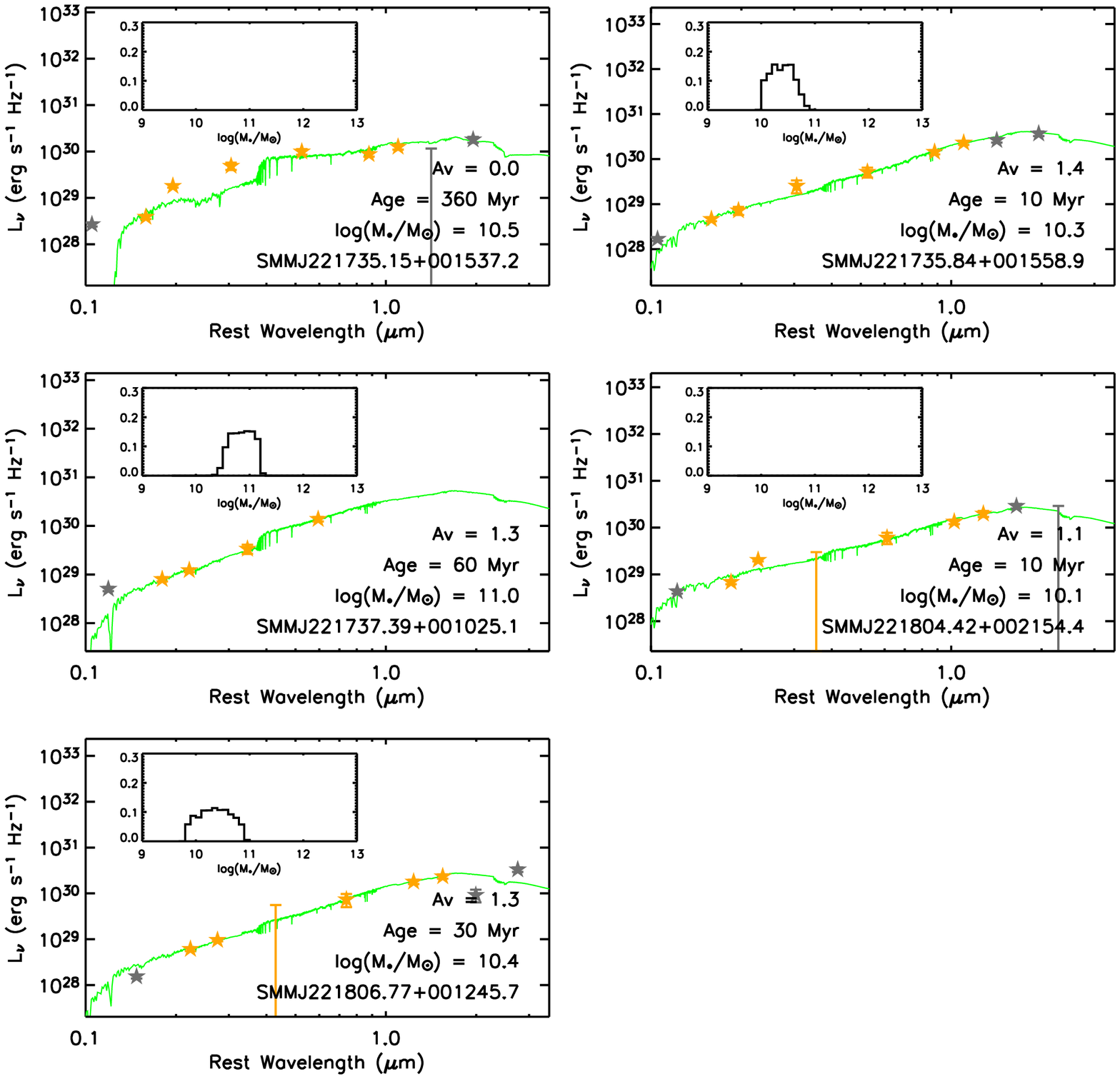}
\end{center}
\caption{Continued.}
\end{figure}
 
 \end{document}